\documentclass[aps,pra,amsmath,amssymb,tightenlines,epsfig,floatfix, twocolumn, longbibliography]{revtex4-1}
\usepackage{graphicx}
\usepackage{dcolumn}	
\usepackage{bm}			
\usepackage{amsfonts}
\usepackage{xspace}
\usepackage{color}
\usepackage{epstopdf}
\usepackage{multirow}

\begin{document}

\newcommand{\psihat}{\ensuremath{\hat{\psi}}\xspace}
\newcommand{\psihatd}{\ensuremath{\hat{\psi}^{\dagger}}\xspace}
\newcommand{\ahat}{\ensuremath{\hat{a}}\xspace}
\newcommand{\Ham}{\ensuremath{\mathcal{H}}\xspace}
\newcommand{\ahatd}{\ensuremath{\hat{a}^{\dagger}}\xspace}
\newcommand{\bhat}{\ensuremath{\hat{b}}\xspace}
\newcommand{\bhatd}{\ensuremath{\hat{b}^{\dagger}}\xspace}
\newcommand{\boldr}{\ensuremath{\mathbf{r}}\xspace}
\newcommand{\dr}{\ensuremath{\,d^3\mathbf{r}}\xspace}
\newcommand{\tr}{\ensuremath{\,\mathrm{Tr}}\xspace}
\newcommand{\dk}{\ensuremath{\,d^3\mathbf{k}}\xspace}
\newcommand{\etal}{\emph{et al.\/}\xspace}
\newcommand{\ie}{i.e.}
\newcommand{\eq}[1]{Eq.~(\ref{#1})\xspace}
\newcommand{\fig}[1]{Fig.~\ref{#1}\xspace}
\newcommand{\abs}[1]{\left| #1 \right|}
\newcommand{\proj}[2]{\left| #1 \rangle\langle #2\right| \xspace}
\newcommand{\Qhat}{\ensuremath{\hat{Q}}\xspace}
\newcommand{\Qhatd}{\ensuremath{\hat{Q}^\dag}\xspace}
\newcommand{\phihatd}{\ensuremath{\hat{\phi}^{\dagger}}\xspace}
\newcommand{\phihat}{\ensuremath{\hat{\phi}}\xspace}
\newcommand{\boldk}{\ensuremath{\mathbf{k}}\xspace}
\newcommand{\boldp}{\ensuremath{\mathbf{p}}\xspace}
\newcommand{\boldsigma}{\ensuremath{\boldsymbol\sigma}\xspace}
\newcommand{\boldalpha}{\ensuremath{\boldsymbol\alpha}\xspace}
\newcommand{\grad}{\ensuremath{\boldsymbol\nabla}\xspace}
\newcommand{\parti}[2]{\frac{ \partial #1}{\partial #2} \xspace}
 \newcommand{\vs}[1]{\ensuremath{\boldsymbol{#1}}\xspace}
\renewcommand{\v}[1]{\ensuremath{\mathbf{#1}}\xspace}
\newcommand{\Psihat}{\ensuremath{\hat{\Psi}}\xspace}
\newcommand{\Psihatd}{\ensuremath{\hat{\Psi}^{\dagger}}\xspace}
\newcommand{\Vhatd}{\ensuremath{\hat{V}^{\dagger}}\xspace}
\newcommand{\Xhat}{\ensuremath{\hat{X}}\xspace}
\newcommand{\Xhatd}{\ensuremath{\hat{X}^{\dag}}\xspace}
\newcommand{\Yhat}{\ensuremath{\hat{Y}}\xspace}
\newcommand{\Jhat}{\ensuremath{\hat{J}}\xspace}
\newcommand{\Yhatd}{\ensuremath{\hat{Y}^{\dag}}\xspace}
\newcommand{\jhat}{\ensuremath{\hat{J}}\xspace}
\newcommand{\lhat}{\ensuremath{\hat{L}}\xspace}
\newcommand{\Nhat}{\ensuremath{\hat{N}}\xspace}
\newcommand{\rhohat}{\ensuremath{\hat{\rho}}\xspace}
\newcommand{\ddt}{\ensuremath{\frac{d}{dt}}\xspace}
\newcommand{\nset}{\ensuremath{n_1, n_2,\dots, n_k}\xspace}
\newcommand{\notes}[1]{{\color{blue}#1}}
\newcommand{\cmc}[1]{{\color{red}#1}}
\newcommand{\sah}[1]{{\color{magenta}#1}}

\title{Quantum metrology with mixed states: When recovering lost information is better than never losing it}
\author{Simon A.~Haine}
\affiliation{School of Mathematics and Physics,  University of Queensland, Brisbane, Queensland, 4072, Australia}
\author{Stuart S.~Szigeti}
\affiliation{School of Mathematics and Physics,  University of Queensland, Brisbane, Queensland, 4072, Australia}
\affiliation{ARC Centre for Engineered Quantum Systems, University of Queensland, Brisbane, Queensland 4072, Australia}

\begin{abstract}
Quantum-enhanced metrology can be achieved by entangling a probe with an auxiliary system, passing the probe through an interferometer, and subsequently making measurements on both the probe and auxiliary system. Conceptually, this corresponds to performing metrology with the purification of a (mixed) probe state. We demonstrate via the quantum Fisher information how to design mixed states whose purifications are an excellent metrological resource. In particular, we give examples of mixed states with purifications that allow (near) Heisenberg-limited metrology and provide examples of entangling Hamiltonians that can generate these states. Finally, we present the optimal measurement and parameter-estimation procedure required to realize these sensitivities (i.e., that saturate the quantum Cram\'er-Rao bound). Since pure states of comparable metrological usefulness are typically challenging to generate, it may prove easier to use this approach of entanglement and measurement of an auxiliary system. An example where this may be the case is atom interferometry, where entanglement with optical systems is potentially easier to engineer than the atomic interactions required to produce nonclassical atomic states. 

\end{abstract}

\maketitle

\section{Introduction}
There is currently great interest in quantum metrology: the science of estimating a classical parameter $\phi$ with a quantum probe at a higher precision than is possible with a classical probe of identical particle flux. 
Given a fixed number of particles, $N$, the ultimate limit to the sensitivity is the Heisenberg limit $\Delta \phi = 1/N$ \cite{Holland:1993, Giovannetti:2006}. Na\"ively, the choice of probe state is a solved problem;  for instance, symmetric Dicke states \cite{Holland:1993, Lucke:2011} and spin-cat states \cite{Sanders:1995, Pezze:2013} input into a Mach-Zehnder (MZ) interferometer yield sensitivities of $\sqrt{2}/N$ and $1/N$, respectively. However, in practice achieving quantum-enhanced sensitivities is a significant challenge. This is due to both the deleterious effect of losses \cite{Demkowicz-Dobrzanski:2012} and the challenges associated with preparing nonclassical states with an appreciable number of particles \cite{Leibfried:2005, Wieczorek:2009, Gao:2010, Vlastakis:2013, Signoles:2014}. For example, protocols for generating a spin-cat state commonly require a large Kerr nonlinearity, which is either unavailable (e.g. in optical systems \cite{Jeong:2004}), difficult to engineer (e.g. in microwave cavities \cite{Rebic:2009, Lu:2013}), or is incompatible with the efficient operation of the metrological device (as in atom interferometers \cite{Debs:2011, Szigeti:2012, Robins:2013}).

\begin{figure}
\includegraphics[width=0.8\columnwidth]{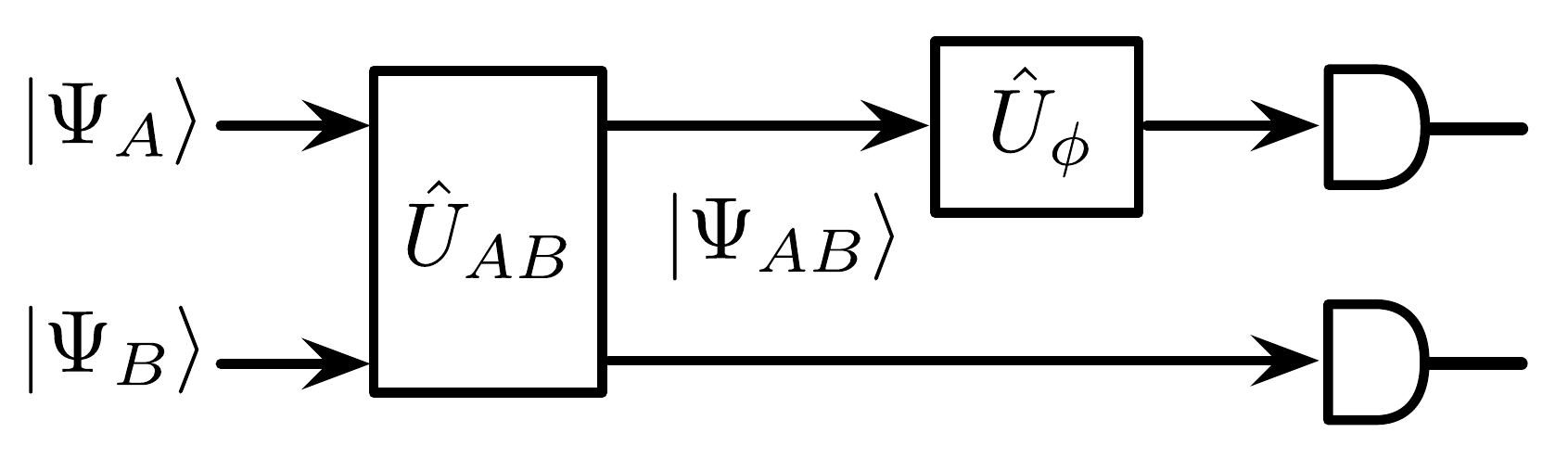}
\caption{ The unitary $\hat{U}_{AB} = \exp(-i \hat{\mathcal{H}}_{AB} t / \hbar)$ entangles system $A$ (probe) with system $B$ (auxiliary) before system $A$ passes through a measurement device described by $\hat{U}_\phi = \exp (-i \phi \hat{G}_A )$. If measurements are restricted to system $A$, then the QFI for an estimate of $\phi$ is $\mathcal{F}_A = \mathcal{F} [\hat{G}_A, \hat{\rho}_A]$, where $\hat{\rho}_A = \text{Tr}_B\left\{ |\Psi_{AB} \rangle \langle \Psi_{AB} |\right\}$. If measurements on both systems are permitted, then the QFI is $\mathcal{F}_{AB} = \mathcal{F} [\hat{G}_A, |\Psi_{AB}\rangle] = 4 \text{Var}(\hat{G}_A)_{\hat{\rho}_A} \geq \mathcal{F}_A$.}
\label{scheme}
\end{figure}

In this paper, we present an alternative route to quantum-enhanced metrology based on purifications of mixed states. Physically, this involves entangling the probe with an auxiliary system before the probe is affected by $\phi$, making measurements on \emph{both} the probe and auxiliary system, and subsequently using correlations between the two measurement outcomes in order to reduce the uncertainty in the estimated parameter (see Fig.~\ref{scheme}). This approach is advantageous in cases where it is easier to entangle the probe system with another system, rather than directly create highly entangled states of the probe system itself. An example of this is atom interferometry; although quantum squeezing can be produced in atomic systems via atomic interactions \cite{Kheruntsyan:2005b, Pu:2000, Gross:2011, Jaskula:2010, Bucker:2011, Haine:2011, Lewis-Swan:2014, Kitagawa:1993, Sorensen:2002, Gross:2010, Riedel:2010, Johnsson:2007a, Haine:2009, Haine:2014}, the technical requirements of high sensitivity, path separated atom interferometers are better suited to enhancement via entanglement with an optical system \cite{Jing:2000, Fleischhauer:2002b, Haine:2005, Haine:2005b, Haine:2006b, Hammerer:2010} and information recycling \cite{Haine:2013, Szigeti:2014b, Tonekaboni:2015, Haine:2015}.

The structure of this paper is as follows. In Sec.~\ref{secII} we introduce in detail the central idea of this paper: that purifications of mixed states can possess a large quantum Fisher information (QFI), and therefore represent an excellent resource for quantum metrology. In Sec.~\ref{secIII} we specialize to an $N$-boson probe state and Mach-Zehnder (MZ) interferometer, and show how to engineer purifications that yield sensitivities at and near the Heisenberg limit. Finally, in Sec.~\ref{secIV} we present optimal measurement schemes that allow these quantum-enhanced sensitivities to be achieved in practice.

\section{Quantum Fisher information for a purification}\label{secII}
We can determine the best sensitivity possible for any given metrology scheme via the QFI, $\mathcal{F}$, which places an absolute lower bound on the sensitivity, $\Delta \phi \geq 1/\sqrt{\mathcal{F}}$, called the quantum Cram{\'e}r-Rao bound (QCRB) \cite{Braunstein:1994, Paris:2009, Demkowicz-Dobrzanski:2014, Toth:2014}. This bound is independent of the choice of measurement and parameter estimation procedure, and depends only on the input state. Explicitly, if a state $\hat{\rho}_A$ is input into a metrological device described by the unitary operator $\hat{U}_\phi = \exp (- i \phi \hat{G}_A)$, then the QFI is
\begin{equation}
	\mathcal{F}_A \equiv \mathcal{F}[\hat{G}_A, \hat{\rho}_A] = 2 \sum_{i,j} \frac{(\lambda_i-\lambda_j)^2}{\lambda_i + \lambda_j}| \langle e_i| \hat{G}_A |e_j\rangle|^2, \label{QFI_mixed}
\end{equation}
where $\lambda_i$ and $|e_i\rangle$ are the eigenvalues and eigenvectors of $\hat{\rho}_A$, respectively. If $\hat{\rho}_A$ is pure, then Eq.~(\ref{QFI_mixed}) reduces to $\mathcal{F}_A = 4 \text{Var}(\hat{G}_A)$. 

A na\"ive consideration of the pure state QFI suggests that engineering input states with a large variance in $\hat{G}_A$ is an excellent strategy for achieving a high precision estimate of $\phi$. However, there are many operations on $\hat{\rho}_A$ that increase $\text{Var}(\hat{G}_A)$ at the expense of also \emph{decreasing} the purity $\gamma = \text{Tr}\left\{ \hat{\rho}_A^2 \right\}$. Since the QFI is convex in the state, any process that mixes the state typically decreases the QFI. Consequently, any improvement due to a larger $\text{Var}(\hat{G}_A)$ is usually overwhelmed by reductions in the QFI due to mixing. 

In order to concretely demonstrate this point, we specialize to an $N$-boson state input into a MZ interferometer. As discussed in \cite{Yurke:1986}, this system is conveniently described by the SU(2) Lie algebra $[ \hat{J}_i, \hat{J}_j] = i \epsilon_{ijk} \hat{J}_k$, where $\epsilon_{ijk}$ is the Levi-Civita symbol, for $i = x, y, z$. A MZ interferometer is characterized by $\hat{G}_A = \hat{J}_y$, therefore, for pure states, a large QFI requires a large $\text{Var}(\hat{J}_y)$. 

\begin{figure}
\includegraphics[width=1\columnwidth]{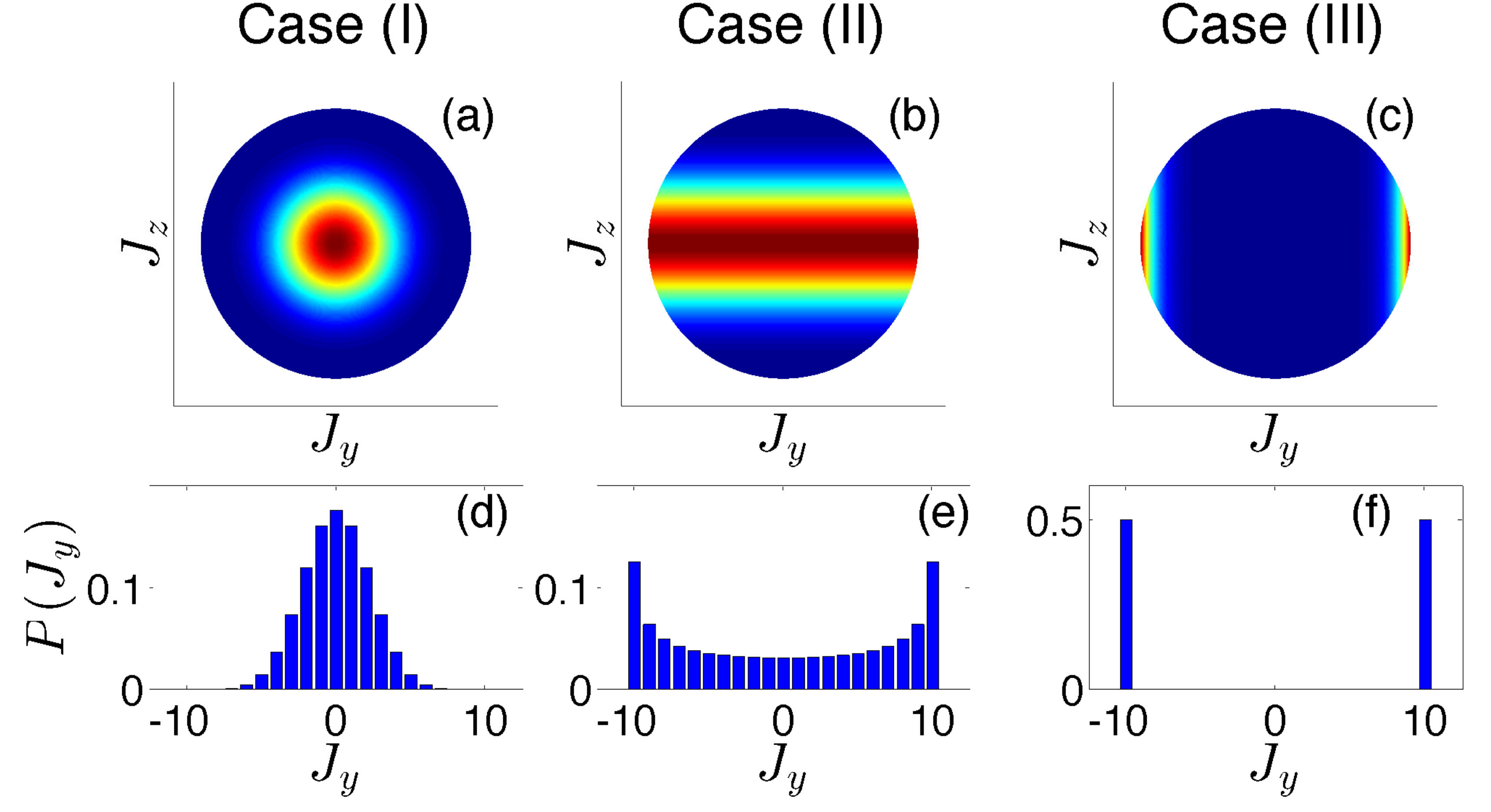}
\caption{(Color online) Husimi-$\mathcal{Q}$ function for Case (I) (a), Case (II) (b), and Case (III) (c). The projection in the $\hat{J}_y$ basis, $P(J_y)$ is shown for Case (I) (d), Case (II) (e), and Case~(III) (f). $N=20$ for all frames. 
}
\label{fig_Q}
\end{figure}

Without loss of generality, we restrict ourselves to the class of input states
\begin{equation}
	\hat{\rho}_A = \int_0^{2\pi} d\varphi \, \mathcal{P}(\varphi) |\alpha(\tfrac{\pi}{2}, \varphi)\rangle \langle \alpha(\tfrac{\pi}{2}, \varphi) |. \label{rho_P_fun}
\end{equation}
Here $|\alpha(\theta, \varphi)\rangle = \exp (-i \varphi \jhat_z) \exp (-i \theta \jhat_y) | j, j \rangle$ are spin coherent states, where $|j, m\rangle$ are Dicke states with total angular momentum $j = N/2$ and $\hat{J}_z$ projection $m$. We focus on the following three states in class~(\ref{rho_P_fun}), which are in order of increasing $\text{Var}(\hat{J}_y)$:
\begin{subequations}
\label{P_cases}
\begin{flalign}
	&\text{Case (I):}	&  &\mathcal{P}(\varphi) = \delta(\varphi),  & \label{case_1}\\
	&\text{Case (II):} 	&  &\mathcal{P}(\varphi) = \frac{1}{2\pi}, & \label{case_2}\\
	&\text{Case (III):} 	&  &\mathcal{P}(\varphi) = \frac{1}{2}\left[\delta \left(\varphi-\tfrac{\pi}{2}\right) + \delta \left(\varphi+ \tfrac{\pi}{2} \right) \right]. & \label{case_3}
\end{flalign}
\end{subequations}
These states can be conveniently visualized by plotting the Husimi-$\mathcal{Q}$ function \cite{Arecchi:1972, Agarwal:1998}
\begin{equation}
	\mathcal{Q}(\theta, \varphi) = \frac{2j+1}{4\pi} \langle \alpha (\theta, \varphi) | \rhohat_A |\alpha(\theta, \varphi)\rangle \, ,
\end{equation}
and the $\hat{J}_y$ projection of the state, $P(J_y) = \langle J_y | \hat{\rho}_A| J_y \rangle$, where $\jhat_y|J_y\rangle = J_y|J_y\rangle$ (see \fig{fig_Q}). 

None of these states yield sensitivities that surpass the standard quantum limit (SQL), $\Delta \phi = 1 / \sqrt{N}$. In Case~(I), $\hat{\rho}_A$ is a pure spin coherent state, $|\alpha(\pi/2,0)\rangle$, with $\mathcal{F}_A = 4\text{Var}(\jhat_y) = N$. In Case~(II), $\hat{\rho}_A$ is an incoherent mixture of Dicke states (i.e. it contains no off-diagonal terms in the $|j, m\rangle$ basis). Although $4\text{Var}(\jhat_y) = N(N+1)/2$ is much larger than for Case~(I), the QFI is only $\mathcal{F}_A = N/2$. Finally, Case~(III) is an incoherent mixture of maximal and minimal $\hat{J}_y$ eigenstates with $4\text{Var}(\jhat_y) = N^2$, which is the maximum possible value in SU(2). However, since the state is mixed the QFI is significantly less than this, with $\mathcal{F}_A = N/2$.

However, suppose the mixing in $\hat{\rho}_A$ arises from entanglement with an auxiliary system $B$ before system $A$ passes through the metrological device (see Fig.~\ref{scheme}). Specifically, for an input pure state $| \Psi_{AB} \rangle$ of a composite system $A \otimes B$, where $\hat{\rho}_A = \text{Tr}_B\left\{ |\Psi_{AB} \rangle \langle \Psi_{AB} |\right\}$, the QFI is
\begin{align}
	\mathcal{F}_{AB} 	&\equiv \mathcal{F}[\hat{G}_A, | \Psi_{AB} \rangle] \notag \\
					&= 4\left( \langle \Psi_{AB} | \hat{G}_A^2 |\Psi_{AB} \rangle - \langle \Psi_{AB}| \hat{G}_A |\Psi_{AB}\rangle^2\right) \nonumber \\ 
					&= 4\left( \text{Tr}_A \left[ \hat{G}_A^2 \hat{\rho}_A\right] - \text{Tr}_A \left[ \hat{G}_A \hat{\rho}_A\right]^2 \right) \nonumber \\ 
					&\equiv 4\text{Var}(\hat{G}_A)_{\hat{\rho}_A}.  \label{F_AB_rho}
\end{align}
Consequently, for a purification of $\hat{\rho}_A$ the QFI only depends on the variance in $\hat{G}_A$ of $\hat{\rho}_A$ \cite{Haine:2015, Mondal:2015}. Our na\"ive strategy of preparing a state with large $\text{Var}(\hat{G}_A)$ \emph{irrespective of its purity} is now an excellent approach. Indeed, in this situation the states in Cases (I)-(III) are now also arranged in order of increasing QFI, with Case~(II) and Case~(III) attaining a QFI of $N(N+1)/2$ and $N^2$, respectively. It is interesting to note that the QFI for Case~(III) is the maximum allowable for $N$ particles in SU(2) \cite{Pezze:2009}, and is usually obtained via the difficult to generate spin-cat state, which is a macroscopic superposition, rather than a classical mixture, of spin coherent states. Note also that $\mathcal{F}_{AB}$ is independent of any particular purification, and convexity implies that $\mathcal{F}_{AB} \geq \mathcal{F}_A$. That is, in principle any purification of $\hat{\rho}_A$ is capable of achieving sensitivities at least as good as, and usually much better than, $\hat{\rho}_A$ itself.

Quantum metrology with purifications is not simply a mathematical `trick'; physically, a purification corresponds to entangling the probe system $A$ with some auxiliary system $B$, and permitting measurements on both systems \footnote{If we restricted measurements to system $A$, then this is equivalent to tracing out system $B$, in which case the QFI is $\mathcal{F}_A$.}. Therefore, the practical utility of our proposal depends crucially on the existence of an entangling Hamiltonian that can prepare $\hat{\rho}_A$ in a state with large $\text{Var}(\hat{G}_A)_{\hat{\rho}_A}$. 

For the three cases described by Eq.~(\ref{rho_P_fun}) and Eqs.~(\ref{P_cases}), a purification of $\hat{\rho}_A$ can be written as 
\begin{equation}
	|\Psi_{AB}\rangle = \sum_{m=-j}^j c_m|j, m\rangle \otimes |B_m\rangle, \label{Psi_AB}
\end{equation}
with Case~(I) corresponding to $\langle B_m| B_n\rangle = 1$, Case~(II) corresponding to  $\langle B_m| B_n\rangle = \delta_{n,m}$, and Case~(III) corresponding to $\langle B_m| B_n\rangle = 1(0)$ for $|n-m|$ even (odd). In the following section, we present a simple scheme that converts a shot-noise limited spin coherent state [such as Case~(I)] to the enhanced QFI purifications of Cases~(II) and (III).

\section{Example entangling dynamics leading to increased QFI}\label{secIII}
Consider again the $N$-boson probe state (system $A$) input into a MZ interferometer (i.e. $\hat{G}_A = \hat{J}_y$). The QFI for a purification of $\rhohat_A$ can be written as
\begin{equation}
	\mathcal{F}_{AB} = 4\left(\langle \jhat_y^2\rangle - \langle \jhat_y\rangle^2\right) = \mathcal{F}_0 + \mathcal{F}_1 + \mathcal{F}_2,\label{QFI_Fs}
\end{equation} 
with
\begin{subequations}
\begin{eqnarray}
\mathcal{F}_0 &=& \tfrac{N}{2}(N+2) - 2\langle \jhat_z^2\rangle, \\
\mathcal{F}_1 &=& -\langle i(\jhat_+ - \jhat_-)\rangle^2, \\
\mathcal{F}_2 &=& -\langle \jhat_+^2 + \jhat_-^2\rangle ,
\end{eqnarray}
\end{subequations}
where $\hat{J}_\pm = \hat{J}_x \pm i\hat{J}_y$. Note that $\mathcal{F}_{0}$, $\mathcal{F}_{1}$, and $\mathcal{F}_{2}$ depend only on the matrix elements of $\hat{\rho}_A$ with $|n-m|$ equal to $0$, $1$, and $2$, in the $\hat{J}_z$ basis; writing $\mathcal{F}_{AB}$ in this form is very convenient for what follows. 

Before the interferometer, we assume the probe is coupled to some auxiliary system $B$ via the Hamiltonian
\begin{equation}
	\hat{\mathcal{H}}_{AB} = \hbar g \Jhat_z \hat{\Ham}_B \, . \label{ham}
\end{equation}
When system $B$ is a photon field and $\hat{\Ham}_B$ is proportional to the number of photons in the field, then $\hat{\mathcal{H}}_{AB}$ describes the weak probing of the population difference of an ensemble of two-level atoms with far-detuned light \cite{Wasilewski:2010, Hammerer:2010, Schleier-Smith:2010, Chen:2011, Szigeti:2009, Szigeti:2010, vanderbruggen2011spin, bernon2011heterodyne, Brahms:2012, Bohnet:2014, Cox:2015}, or dispersive coupling between a microwave cavity and a superconducting qubit \cite{Wallraff:2004, Schuster:2005, Haigh:2015}. We will explore this specific case shortly, however, for now we keep $\hat{\Ham}_B$ completely general. If the initial system state is a product state $|\Psi_{AB}(0)\rangle = |\Psi_A\rangle\otimes |\Psi_B\rangle$, after some evolution time the state of the system will be given by \eq{Psi_AB} with $c_m = \langle m | \Psi_A\rangle$ and 
 $|B_m\rangle = \exp(-i m g t \hat{\Ham}_B) |\Psi_B\rangle$. The reduced density operator of system $A$ is then
\begin{equation}
	\hat{\rho}_A = \sum_{n, m = -j}^j c_n c_m^* \mathcal{C}_{n-m} |j, n\rangle\langle j, m|, \label{rho_spin_coupling}
\end{equation}
where the coherence of system $A$ is determined via
\begin{equation}
	\mathcal{C}_{n-m} = \langle B_m |B_n \rangle = \langle \Psi_B| e^{-i (n-m) g t \hat{\Ham}_B} |\Psi_B\rangle \, . \label{defn_C}
\end{equation}
When $\mathcal{C}_{n-m} = 1$, the system remains separable and system $A$ is a pure state, whereas if $\mathcal{C}_{n-m} = \delta_{n, m}$ then $\hat{\rho}_A$ is an incoherent mixture of Dicke states.

Using Eq.~(\ref{rho_spin_coupling}), $\mathcal{F}_0$, $\mathcal{F}_1$, and $\mathcal{F}_2$ can be written as
\begin{subequations}
\begin{eqnarray}
\mathcal{F}_0 &=& \tfrac{N}{2}(N+2) - 2\langle \jhat_z^2\rangle, \\
\mathcal{F}_1 &=& -\langle i(\mathcal{C}_1\jhat_+ - \mathcal{C}_1^*\jhat_-)\rangle^2, \\
\mathcal{F}_2 &=& -\left(\mathcal{C}_2\langle \jhat_+^2\rangle + \mathcal{C}_2^*\langle \jhat_-^2\rangle\right) ,
\end{eqnarray}
\end{subequations}
where the above expectation values are calculated with respect to $|\Psi_A\rangle$. The effect of the entanglement between systems $A$ and $B$ is entirely encoded in the coherences $\mathcal{C}_1$ and $\mathcal{C}_2$; coherences greater than 2nd order do not affect the QFI.

Let us consider the effect on the QFI of each term in Eq.~(\ref{QFI_Fs}). $\mathcal{F}_0$ is independent of the entanglement between systems $A$ and $B$, and will be of order $N^2/2$ if $| \Psi_A\rangle$ has $\langle \hat{J}_z^2 \rangle \sim N$ (e.g. the spin coherent state $| \alpha(\pi/2, \varphi)\rangle$ has $\langle \jhat_z^2\rangle = N/4$). This suggests that a sufficient condition for Heisenberg scaling is $\mathcal{F}_1 \sim \mathcal{F}_2 \sim \langle \hat{J}_z^2 \rangle \sim N$. In fact, since $\mathcal{F}_1 \leq 0$, the maximum QFI state must necessarily have $\mathcal{C}_1 = 0$. In contrast, $\mathcal{F}_2$ can be positive or negative, in which case a state with $\mathcal{C}_2 = 0$ and another state with $\mathcal{C}_2 = 1$ and $\mathcal{F}_2 \sim + N^2 /2$ might both be capable of (near) Heisenberg-limited metrology. We consider examples of both states below.

\subsection{Case~(II): Example dynamics yielding $\mathcal{F}_{AB} \simeq N^2/2$} \label{dynamics_case_2}
To concretely illustrate the increased QFI a purification of $\hat{\rho}_A$ can provide, we assume system $B$ is a single bosonic mode, described by annihilation operator $\hat{b}$, and take $\hat{\Ham}_B = \bhatd\bhat$ such that
\begin{equation}
	\hat{\mathcal{H}}_{AB} = \hbar g \Jhat_z \bhatd\bhat \, . \label{ham_coh}
\end{equation}
If the initial state of system $B$ is a Glauber coherent state $|\beta \rangle$ \cite{Walls:2008}, then the coherences described by Eq.~(\ref{defn_C}) simplify to
\begin{equation}
	\mathcal{C}_{n-m} = \exp\left[ -|\beta|^2\left(1- e^{-i(n-m)gt} \right)\right]. \label{specific_C}
\end{equation}
$|\mathcal{C}_{n-m}|^2$ decays on a timescale $g t \sim [(n-m)|\beta|^2]^{-1}$. Although the non-orthogonality of $\langle \beta | \beta e^{i\theta}\rangle$ ensures that $\mathcal{C}_{n-m}$ never actually reaches zero, it becomes very small for even modest values of $|\beta|^2$.  

If the initial condition of system $A$ is $|\Psi_A\rangle = |\alpha(\theta, \phi)\rangle$, then $\mathcal{F}_{AB}$ has the simple analytic form given by (see Appendix~\ref{sec_appendix})
\begin{subequations}
\label{Fs_for_example}
\begin{align}
	\mathcal{F}_0	&= N\left( 1 + \tfrac{(N-1)}{2}\sin^2\theta \right), \\
	\mathcal{F}_1	&= - N^2 \sin^2 \theta \sin^2\left( |\beta|^2 \sin\left( g t \right) + \phi \right) e^{- 4 |\beta|^2 \sin^2\left( g t / 2 \right)}, \\
	\mathcal{F}_2	&= \tfrac{N(1-N)}{2}\sin^2\theta \cos \left( |\beta|^2 \sin\left( 2 g t \right) + 2\phi \right) e^{- 2 |\beta|^2 \sin^2\left( g t \right)}.
\end{align}
\end{subequations}
In contrast, calculating $\mathcal{F}_A$ via Eq.~(\ref{QFI_mixed}) requires the diagonalization of $\hat{\rho}_A$, which must be performed numerically. 

\begin{figure}
\includegraphics[width=1\columnwidth]{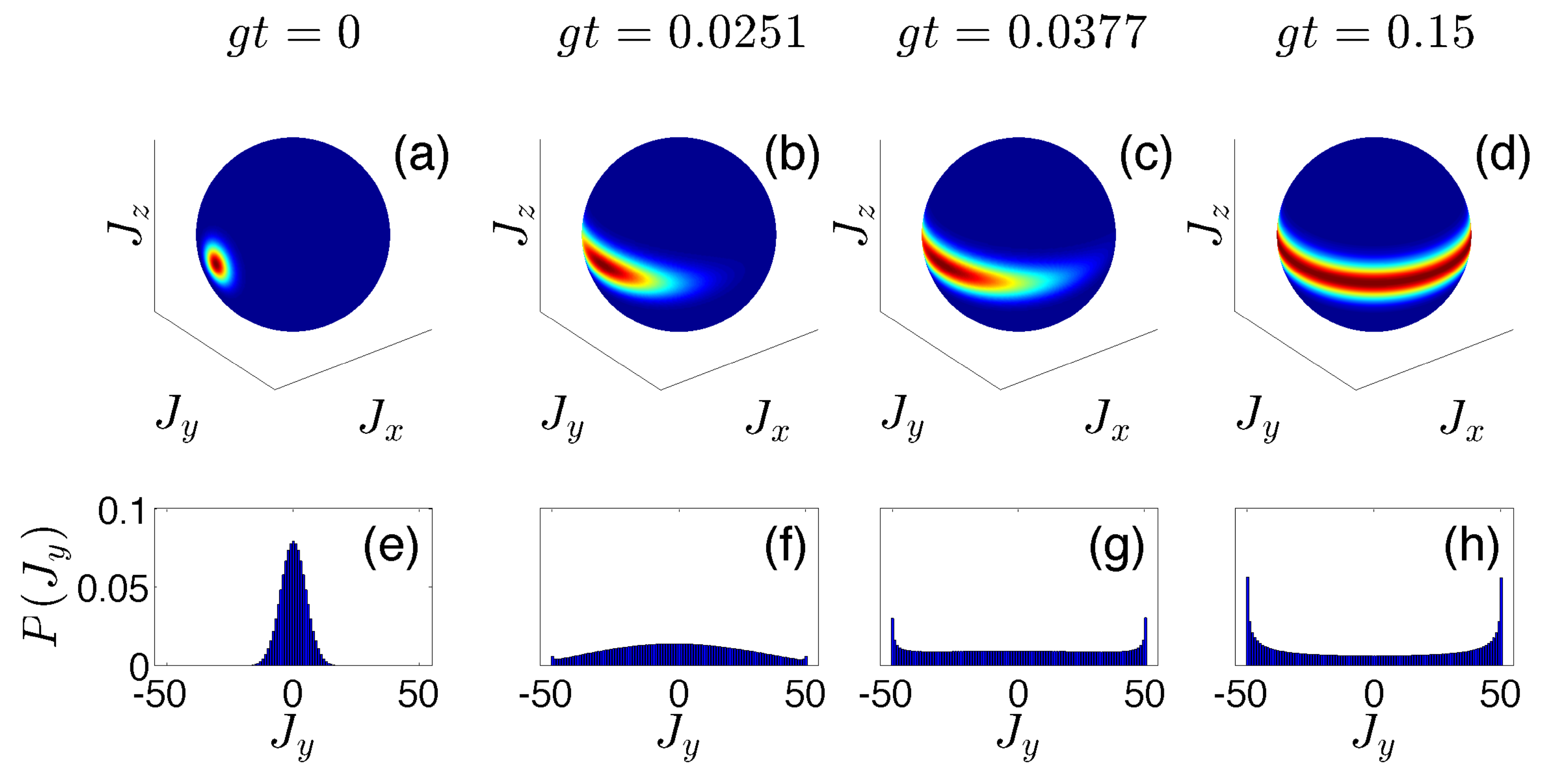}
\caption{(Color online) Time snapshots of the Husimi-$\mathcal{Q}$ function and $\jhat_y$ projection illustrating the evolution of a maximal $\hat{J}_x$ eigenstate under entangling Hamiltonian \eq{ham_coh}. The snapshots were chosen to correspond to times when the rotation around the $J_z$ axis is such that $\langle \hat{J}_y\rangle =0$, which roughly corresponds to the local maxima of $\mathcal{F}_{AB}$ in Fig.~\ref{fig2}(c). (Parameters: $N=100$, $|\beta|^2 = 500$).
}
\label{figBS1}
\end{figure}

We first demonstrate the effect of vanishing 1st and 2nd order coherence on $\mathcal{F}_{AB}$ by preparing system $A$ in the maximal $\hat{J}_x$ eigenstate, $|\alpha(\pi/2, 0)\rangle$, with $N=100$, and a Glauber coherent state for system $B$ with average particle number $|\beta|^2 =500$. The initial state for system $A$ is precisely Case~(I) [see Eq.~(\ref{case_1})], and has a QFI of $N$. As shown in Fig.~\ref{figBS1}, under the evolution of \eq{ham_coh}, $\rhohat_A$ tends towards an incoherent mixture of Dicke states [Case~(II)], with the corresponding broadening of the $P(J_y)$ distribution.    

\begin{figure}
\includegraphics[width=1\columnwidth]{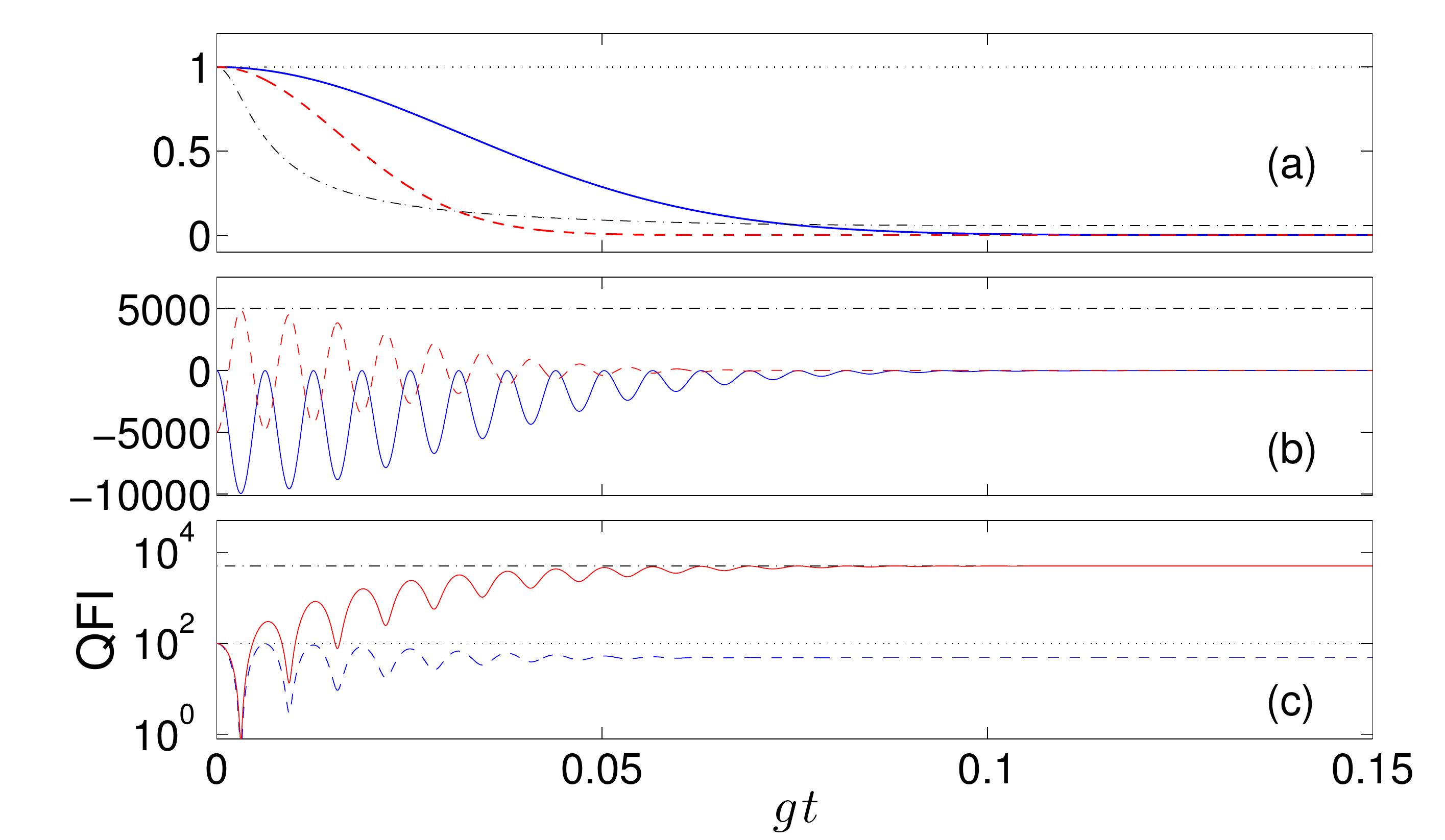}
\caption{(Color online) Evolution of a maximal $\hat{J}_x$ eigenstate under entangling Hamiltonian \eq{ham_coh}. (a) Coherences $|\mathcal{C}_1|^2$ (blue solid line), $|\mathcal{C}_2|^2$ (red dashed line) and purity $\gamma$ (black dot-dashed line). (b): Three components of $\mathcal{F}_{AB}$ [see Eq.~(\ref{QFI_Fs})]: $\mathcal{F}_0$ (black dot-dashed line), $\mathcal{F}_1$ (blue solid line), and $\mathcal{F}_2$ (red dashed line). (c) QFI for $\hat{\rho}_A$, $\mathcal{F}_A$ (blue dashed line), and a purification of $\hat{\rho}_A$, $\mathcal{F}_{AB}$ (red solid line). For reference, we have included $N$ (black dotted line) and $\mathcal{F}_0 = N(N+1)/2 \approx N^2 / 2$ (black dot-dashed line), which correspond to phase sensitivities at the SQL and $\sqrt{2} \times$Heisenberg limit, respectively. (Parameters: $N=100$, $|\beta|^2 = 500$).}
\label{fig2}
\end{figure}

Figure~\ref{fig2}(a) shows that both coherences $\mathcal{C}_1$ and $\mathcal{C}_2$ rapidly approach zero, which causes $\mathcal{F}_1$ and $\mathcal{F}_2$ to vanish [see Fig.~\ref{fig2}(b)]. Consequently, $\mathcal{F}_{AB}$ approaches $\mathcal{F}_0 = N(N+1)/2$, which allows a phase sensitivity of approximately $\sqrt{2} \times$Heisenberg limit [see Fig.~\ref{fig2}(c)]. In contrast, the effect of the mixing causes the QFI of $\hat{\rho}_A$ itself to decrease from $N$ to $\mathcal{F}_A = N/2$, with $\mathcal{F}_A \leq N$ for all $t$. This remains true even if the if $G_A$ is rotated to lie in any arbitrary direction on the Bloch sphere. 

The oscillations in $\mathcal{F}_1$ and $\mathcal{F}_2$ (and consequently $\mathcal{F}_A$ and $\mathcal{F}_{AB}$) before the plateau are due to the complex rotation of $\mathcal{C}_1$ and $\mathcal{C}_2$, which causes rotations of $\hat{\rho}_A$ around the $J_z$ axis before being overwhelmed by the overall decay in magnitude. Furthermore, although the purity of the state also decays, it never vanishes, thereby illustrating that it is not the entanglement \emph{per se} that is causing the QFI enhancement for a purification of $\hat{\rho}_A$.

\subsection{Case~(III): Example dynamics yielding $\mathcal{F}_{AB} = N^2 $} \label{dynamics_case_3}
\begin{figure}
\includegraphics[width=1\columnwidth]{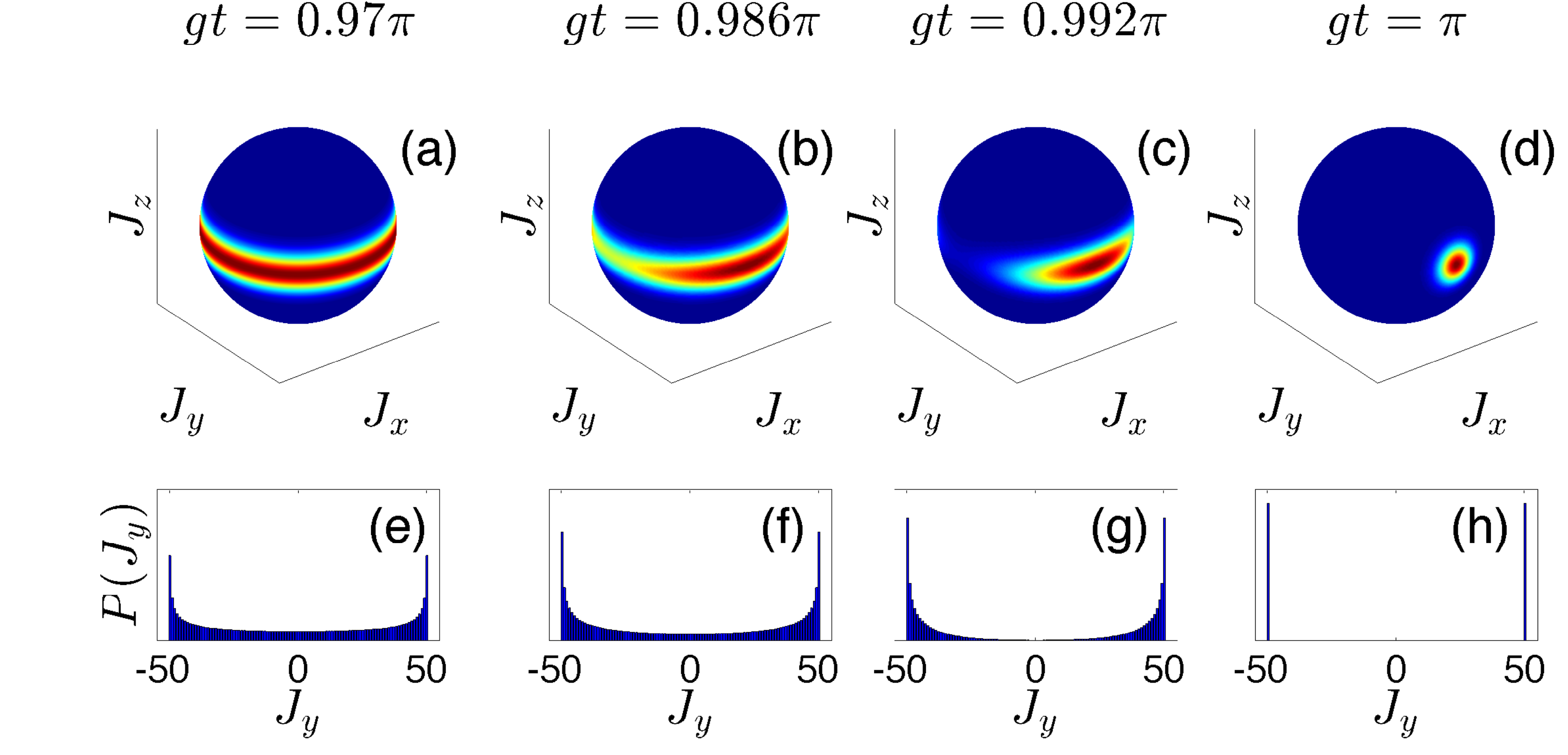}
\caption{(Color online) Husimi-$\mathcal{Q}$ function and $\jhat_y$ projection for  an initial state $|\Psi_{AB}(0)\rangle = |\alpha(\pi/2, \pi/2)\rangle \otimes |\beta\rangle$ under the evolution of \eq{ham_coh} for different values of $gt$. The $\mathcal{Q}$ function is symmetric about reflection of the $J_y$ axis, resulting in part of the function being hidden from view on the reverse side of the sphere. (Parameters: $N=100$, $|\beta|^2 = 500$).  
}
\label{figBS2}
\end{figure}

At $gt = \pi$, there is a revival in $|\mathcal{C}_n|^2$ for $n$ even, but not for $n$ odd. Figure~\ref{figBS2} shows the Husimi-$\mathcal{Q}$ function under the evolution of \eq{ham_coh} for times close to $gt=\pi$, when the initial state of system $A$ is the maximal $\hat{J}_y$ eigenstate $|\alpha(\pi/2, \pi/2)\rangle$. 

\begin{figure}
\includegraphics[width=1\columnwidth]{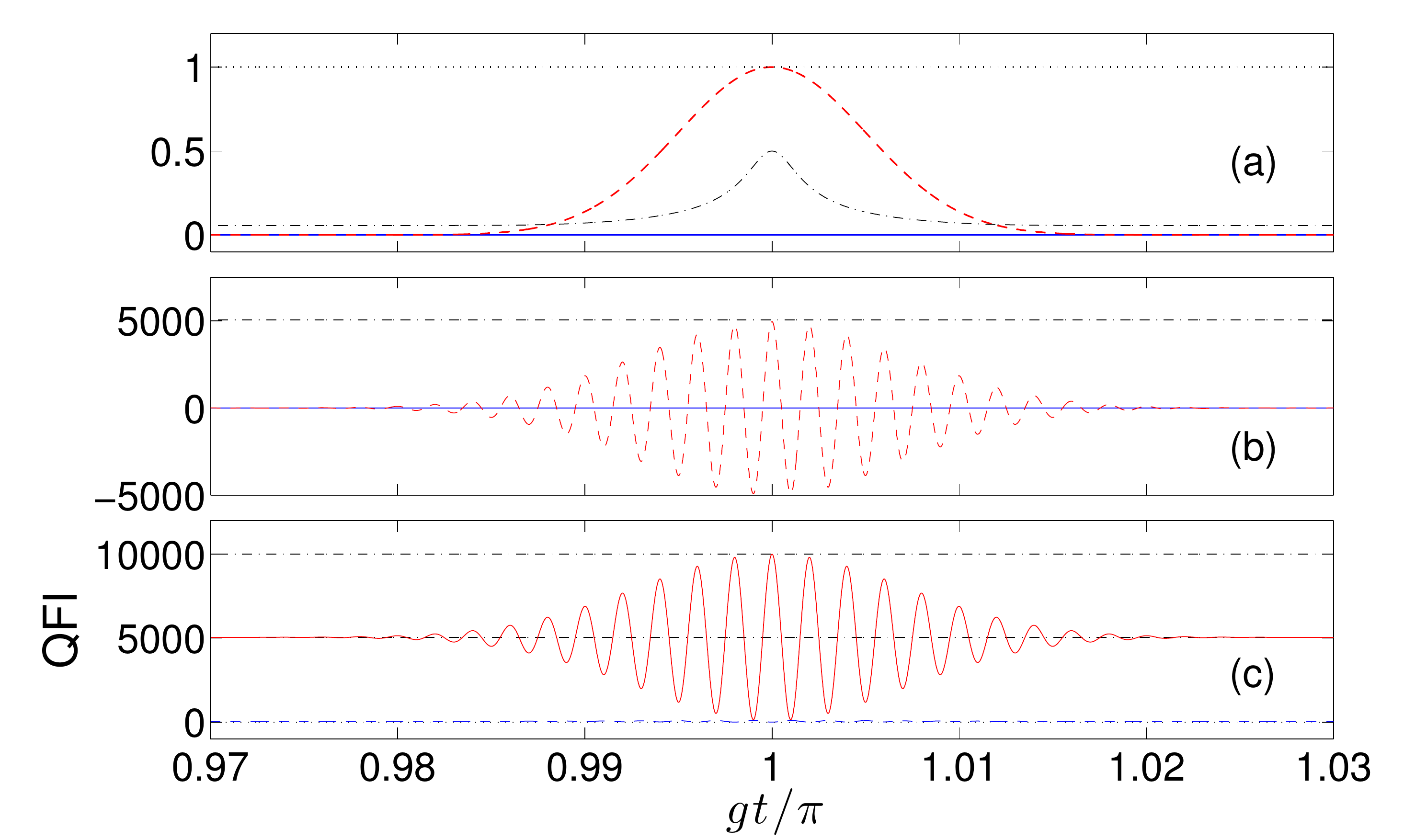}
\caption{(Color online) Evolution of a maximal $\hat{J}_y$ eigenstate near $gt = \pi$ under entangling Hamiltonian \eq{ham_coh}. (a) $|\mathcal{C}_1|^2$ (blue solid line), $|\mathcal{C}_2|^2$ (red dashed line), and $\gamma$ (black dot-dashed line). (b):  $\mathcal{F}_0$ (black dot-dashed line), $\mathcal{F}_1$ (blue solid line), and $\mathcal{F}_2$ (red dashed line). (c) $\mathcal{F}_A$ (blue dashed line) and $\mathcal{F}_{AB}$ (red solid line). For comparison, we have included $\mathcal{F}_0 = N(N+1)/2 \approx N^2 / 2$ and the Heisenberg limit $N^2$ (black dot-dashed lines); $\mathcal{F}_A \leq N$ for all $t$. (Parameters: $N=100$, $|\beta|^2 = 500$).  
}
\label{pseudo_cat_plots}
\end{figure}

The QFI is initially zero, but the decay of $\mathcal{C}_1$ and $\mathcal{C}_2$ rapidly increases to $\mathcal{F}_{AB} = \mathcal{F}_0 = N(N+1)/2$ as in the previous example. As $g t \rightarrow \pi$,  the revival of $|\mathcal{C}_2|^2$ causes $\mathcal{F}_{AB}$ to briefly \emph{increase} to $N^2$ (see Fig.~\ref{pseudo_cat_plots}). This is the Heisenberg limit, which is the QFI of a (pure) spin-cat state and the maximum QFI for SU(2) \cite{Pezze:2009}. 
At $gt=\pi$, $\rhohat_A$ is identical to a classical mixture of $|\alpha(\pi/2, \pi/2)\rangle$ and $|\alpha(\pi/2, -\pi/2)\rangle$, however, its $\mathcal{Q}$-function is similar to that of a spin-cat state, and purifications of it behave as a spin-cat state for metrological purposes. For these reasons, we call this state a \emph{pseudo-spin-cat state}.

\subsection{Example dynamics for particle-exchange Hamiltonian} \label{sec_particle_exchange}
In the previous two examples the $\hat{J}_z$ projection was a conserved quantity, so any entanglement between systems $A$ and $B$ can only degrade the coherence in the $\hat{J}_z$ basis of system $A$ (ultimately resulting in an enhanced QFI). The situation is more complicated when considering a Hamiltonian that does not conserve the $\hat{J}_z$ projection, such as when a spin flip in system $A$ is correlated with the creation or annihilation of a quantum in system $B$. Here, we encounter scenarios where the interaction can either create or destroy coherences in the $\hat{J}_z$ basis of system $A$, and although a significant QFI enhancement is still possible, it depends upon the initial state of system $B$. 

As a concrete illustration, consider the particle-exchange Hamiltonian
\begin{equation}
	\hat{\mathcal{H}}_\pm = \hbar g \left(\jhat_\pm \bhatd + \jhat_\mp \bhat \right), \label{Hpm}
\end{equation}
and assume that system $A$ is initially prepared in the maximal $\hat{J}_z$ eigenstate $| \Psi_A\rangle = |\alpha(0, 0)\rangle = |j,j\rangle$ (N.B. this has $\text{Var}(\hat{J}_y) = N/4$, and therefore a QFI of $N$). Then $\hat{\mathcal{H}}_-$ and $\hat{\mathcal{H}}_+$ physically correspond to Raman superradiance \cite{Moore:2000, Haine:2013} and quantum state transfer \cite{Haine:2005, Haine:2005b, Haine:2006b, Hammerer:2010, Szigeti:2014b, Tonekaboni:2015} processes, respectively. After some period of evolution, the combined state of systems $A \otimes B$ takes the form of Eq.~(\ref{Psi_AB}).

First, consider the case when the initial state for system $B$ is a large amplitude coherent state (i.e. $|\Psi_B\rangle = |\beta\rangle$). Here the addition/removal of a quantum to/from system $B$ has a minimal effect on the state and the system remains approximately separable, since $| \langle B_n | B_m\rangle |^2 \approx 1$. 	It is therefore reasonable to make the undepleted pump approximation $\bhat \rightarrow \beta$, such that $\hat{\Ham}_\pm \rightarrow \hbar g \beta \jhat_x$ (assuming $\beta$ is real). Hence, the effect of the interaction is simply a rotation around the $J_x$ axis, which can create coherence in the $\hat{J}_z$ basis, and so $\mathcal{F}_A = \mathcal{F}_{AB} \leq N$ for all time. 

\begin{figure}
\includegraphics[width=1\columnwidth]{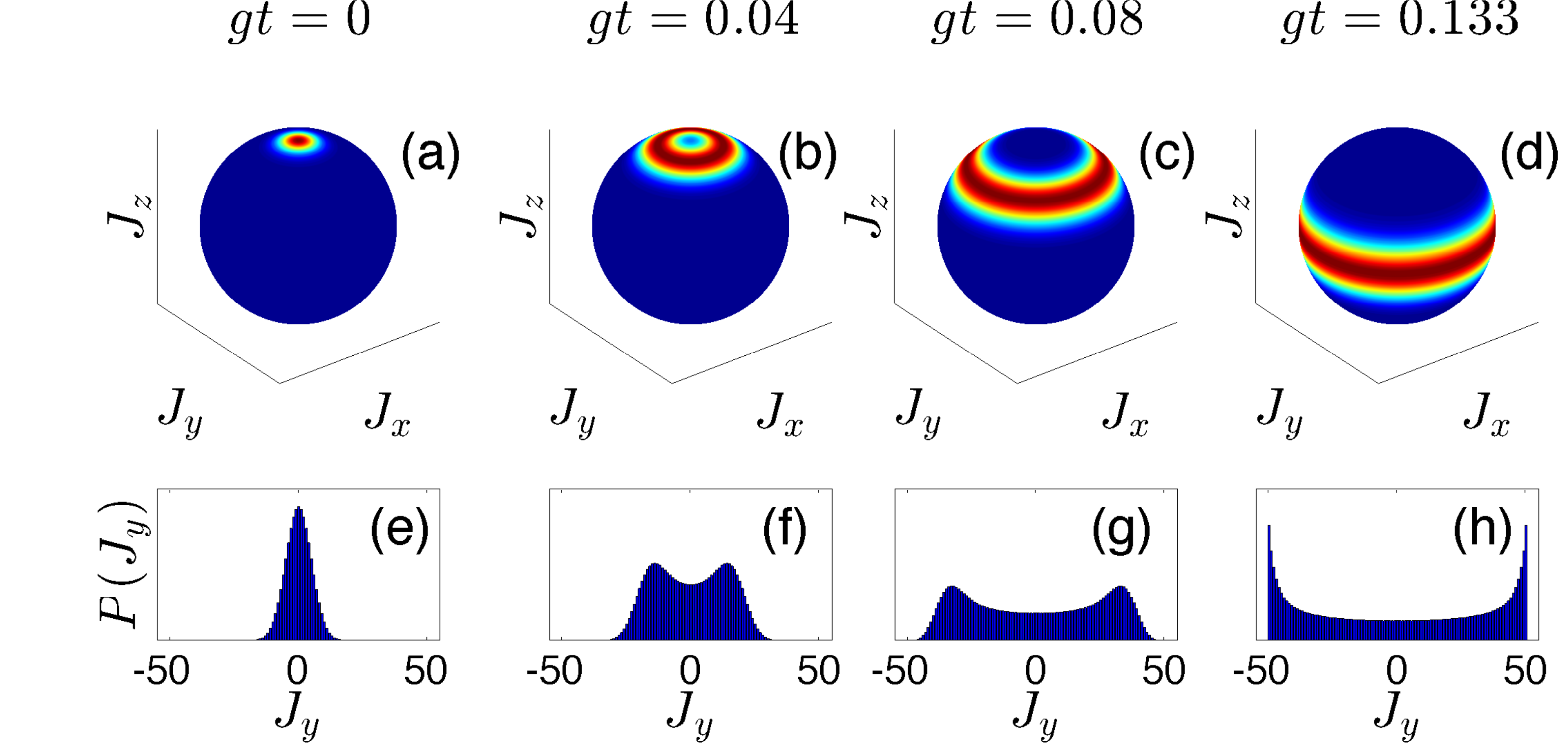}
\caption{(Color online) Husimi-$\mathcal{Q}$ function and $\jhat_y$ projection for  an initial state $|\Psi_{AB}(0)\rangle = |\alpha(0, 0)\rangle \otimes |N_B\rangle$ under the evolution of $\hat{\Ham}_-$ for different values of $gt$. (Parameters: $N=100$, $N_B = 20$).  
}
\label{figBS3}
\end{figure}

\begin{figure}
\includegraphics[width=1\columnwidth]{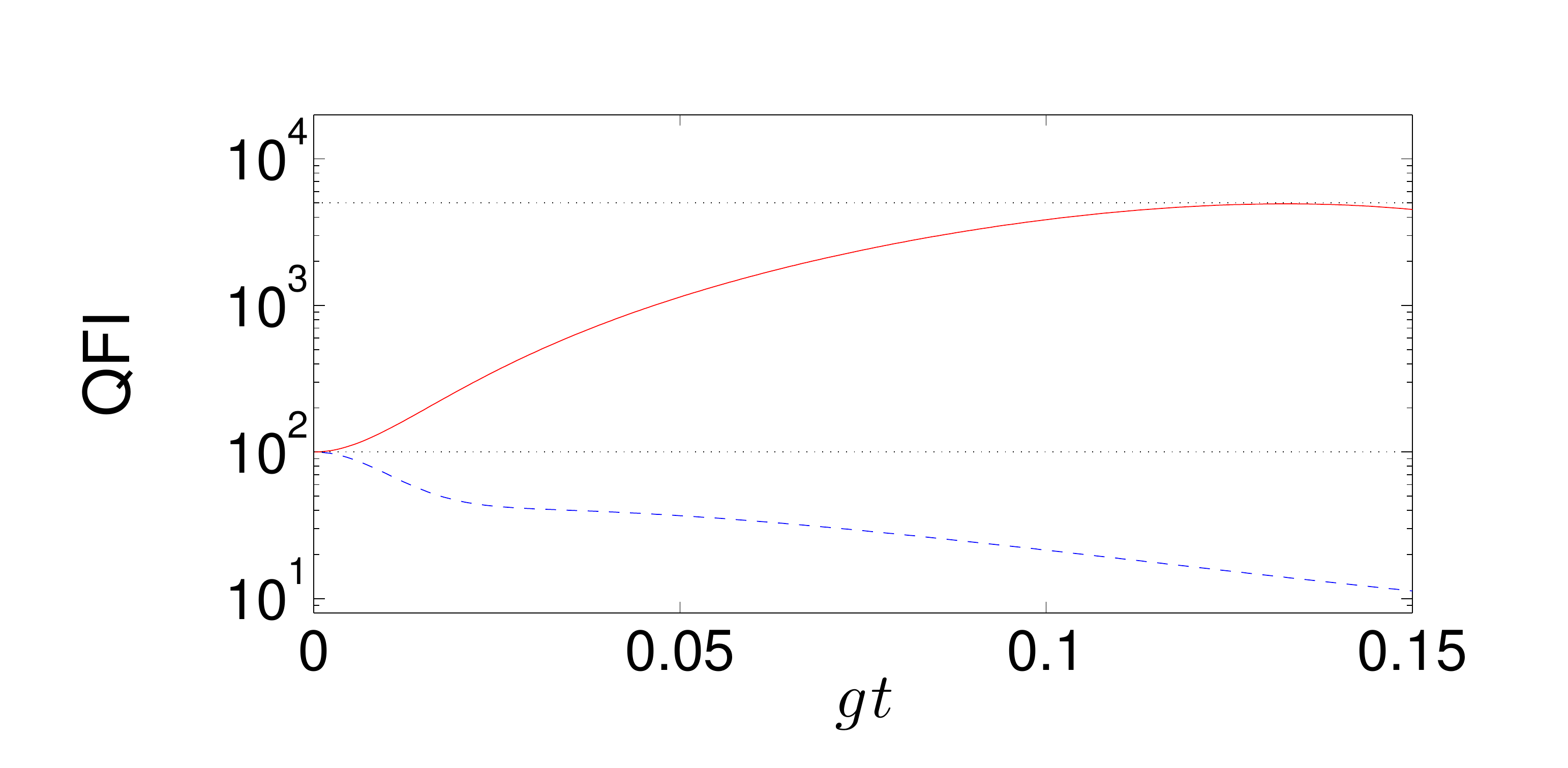}
\caption{(Color online) $\mathcal{F}_A$ (blue dashed line) and $\mathcal{F}_{AB}$ (red solid line) for an initial state $|\Psi_{AB}(0)\rangle = |\alpha (0,0)\rangle \otimes |N_B\rangle$ under the evolution of $\hat{\Ham}_-$. We have indicated $N$ and $N^2/2$ with black dotted lines for comparison. (Parameters: $N=100$, $N_B = 20$).  
}
\label{downconvert_fisher}
\end{figure}

In the opposite limit where the initial state of system $B$ is a Fock state with $N_B$ particles, $|\Psi_B\rangle = |N_B\rangle$, then
\begin{equation}
	|B_m\rangle = |N_B \pm (m-j)\rangle, \label{B_m_fock}
\end{equation}
and $\langle B_m | B_n\rangle = \delta_{n,m}$. This ensures that the first and second order coherences vanish, and $\mathcal{F}_1 = \mathcal{F}_2=0$ for all time. That is, as illustrated in Fig.~\ref{figBS3}, the state moves towards the equator and ultimately evolves to an incoherent Dicke mixture [i.e. Case~(II)]. As described in Sec.~\ref{dynamics_case_2}, and shown in Fig.~\ref{downconvert_fisher}, the QFI increases to a maximum of approximately $\mathcal{F}_{AB} \approx N^2 /2$. Although setting $N_B=0$ (i.e. a vacuum state) leads to a larger variance in $\jhat_z$, $\mathcal{F}_{AB}$ still reaches approximately $70 \%$ of $N^2 /2$.  

We therefore see that for the Hamiltonian~(\ref{Hpm}), a large QFI enhancement is achieved provided the initial state $|\Psi_B\rangle$ has small number fluctuations. Compare this to the Hamiltonian~(\ref{ham_coh}), where the choice $|\Psi_B\rangle = |N_B\rangle$ leads to no entanglement between systems $A$ and $B$, while in contrast an initial state with small phase fluctuations (and therefore large number fluctuations), such as a coherent state, causes rapid decoherence in $\rhohat_A$.

\section{Optimal measurement schemes} \label{secIV}
Although the QFI determines the optimum sensitivity for a given initial state, it is silent on the question of how to achieve this optimum. It is therefore important to identify a) which measurements to make on each system and b) a method of combining the outcomes of these measurements - which we refer to as a \emph{measurement signal} ($\hat{\mathcal{S}}$) - that saturates the QCRB. We do this below for purifications of the incoherent Dicke mixture [Case~(II)] and the pseudo-spin-cat state [Case~(III)].

\subsection{Optimal measurements for incoherent Dicke mixture [Case (II)]:} \label{opt_2}
It is worthwhile briefly recounting the optimal estimation procedure for a symmetric Dicke state $|j, 0 \rangle$ input into a MZ interferometer. A MZ interferometer rotates $\hat{J}_z$ according to $\jhat_z(\phi) = \hat{U}^\dag_\phi \jhat_z \hat{U}_\phi = \cos\phi \jhat_z - \sin \phi \jhat_x$. Since symmetric Dicke states satisfy $\langle \jhat_x \rangle = \langle \jhat_z\rangle = \langle \jhat_z^2\rangle =0$ and $\langle \jhat_x^2 \rangle = N(N+2)/8$, it is clear that the \emph{fluctuations} in $\jhat_z(\phi)$ contain the phase information, and therefore the quantity $\hat{\mathcal{S}} = [\jhat_z(\phi)]^2$ oscillates between $0$ and $N(N+2)/8$. It can be shown that at the operating point $\phi \to 0$, $\text{Var}(\hat{\mathcal{S}}) \to 0$, and the quantity $(\Delta \phi)^2 \rightarrow \text{Var}(\hat{\mathcal{S}})/ (\partial_\phi \langle \hat{\mathcal{S}}\rangle )^2 = 1/ \mathcal{F}_A$, and therefore the signal $\hat{\mathcal{S}}$ saturates the QCRB \cite{Bouyer:1997, Kim:1998, Lucke:2011}.

For an incoherent Dicke mixture, we have $\langle \jhat_x \rangle = \langle \jhat_y \rangle = \langle \jhat_z\rangle=0$, and $\langle \jhat_x^2 \rangle = N(N+1)/8$. Unfortunately, the non-zero variance in $\jhat_z$ (i.e. $\langle \jhat_z^2\rangle = N/4$) implies that $\text{Var}(\hat{\mathcal{S}}) \gg 0 $ for all $\phi$, and the signal no longer saturates the QCRB. However, since the states $|B_m\rangle$ in the purification Eq.~(\ref{Psi_AB}) are orthonormal, a projective measurement of some system $B$ operator diagonal in the $|B_m \rangle$ basis projects system $A$ into a $\hat{J}_z$ eigenstate (i.e. a Dicke state). That is, these measurement outcomes on system $B$ are correlated with $\hat{J}_z$ measurement outcomes on system $A$. Therefore, subtracting both measurements yields a quantity with very little quantum noise. 

More precisely, if we can construct an operator $\hat{\mathcal{S}}_B$ on system $B$ that is correlated with $\hat{J}_z$ measurements on system $A$ (i.e. $\hat{\mathcal{S}}_B |\Psi_{AB}\rangle = \hat{J}_z |\Psi_{AB}\rangle$), then we can construct the quantity $\hat{\mathcal{S}}_0 = \hat{J}_z - \hat{\mathcal{S}}_B$ which has the property $\langle \hat{\mathcal{S}}_0 \rangle = \langle \hat{\mathcal{S}}_0^2 \rangle =0$.  This motivates the signal choice  
\begin{equation}
	\hat{\mathcal{S}} = \left(\hat{U}_\phi^\dag \hat{\mathcal{S}}_0 \hat{U}_\phi\right)^2 = ( \cos \phi \hat{J}_z - \sin \phi \hat{J}_x - \hat{\mathcal{S}}_B )^2. \label{sig_Dicke_mix}
\end{equation} 
Using $\hat{\mathcal{S}}_B |\Psi_{AB}\rangle = \hat{J}_z |\Psi_{AB}\rangle$ and the fact that non-$\jhat_z$ conserving terms vanish due to the absence of off-diagonal terms in the $\jhat_z$ representation of $\rhohat_A$, (e.g. expectation values with an odd power of $\hat{J}_x$ vanish), we can show that
\begin{subequations}
\label{expectations_optimal_S}
\begin{align}
\langle \hat{\mathcal{S}} \rangle &= \langle \hat{J}_z^2\rangle\left(\cos \phi -1\right)^2 + \langle \hat{J}_x^2\rangle \sin^2 \phi,\\
\langle \hat{\mathcal{S}}^2\rangle &= \langle  \jhat_z^4\rangle (\cos \phi-1)^4 + \langle  \jhat_x^4\rangle \sin^4\! \phi \notag  \\
&+ \langle  \jhat_z^2 \jhat_x^2 +\jhat_x^2 \jhat_z^2+ 4  \jhat_z \jhat_x \jhat_x \jhat_z\rangle \sin^2\!\phi (\cos\phi-1)^2\notag\\
&+  2 i \langle   (\jhat_z \jhat_x \jhat_y - \jhat_y \jhat_x \jhat_z)\rangle \sin^2\!\phi \cos\phi (\cos\phi-1) \notag \\ 
&+ \langle   \jhat_y^2\rangle \cos^2\!\phi \sin^2\!\phi.
\end{align}
\end{subequations}
Note that the above expectation values can be taken with respect to $|\Psi_{AB}\rangle$ or $\hat{\rho}_A$. The best sensitivity occurs at small displacements around $\phi=0$. Taking the limit as $\phi \rightarrow 0$ and noting that $\langle \jhat_x^2\rangle = \langle \jhat_y^2\rangle $ gives
\begin{equation}
	(\Delta \phi)^2 = \left.\frac{\text{Var}(\hat{\mathcal{S}})}{\big(\partial_\phi\langle \hat{\mathcal{S}}\rangle\big)^2}\right|_{\phi=0} = \frac{1}{4\langle \jhat_x^2\rangle} = \frac{1}{4\langle \jhat_y^2\rangle} = \frac{1}{\mathcal{F}_{AB}}.
\end{equation}
This demonstrates that the signal \eq{sig_Dicke_mix} is optimal since it saturates the QCRB.

\subsection{Optimal measurements for pseudo-spin-cat state [Case (III)]:} \label{opt_3}
Pure spin-cat states have the maximum QFI possible for $N$ particles in SU(2), are eigenstates of the parity operator, and indeed parity measurements saturate the QCRB \cite{Bollinger:1996}. Pseudo-spin-cat states (case (III)) also have maximal QFI, and since $\langle B_n|B_m\rangle = 1(0)$ for $|n-m|$ even(odd), a projective measurement of system $B$ yields no information other than the parity of the $\jhat_z$ projection. This suggests that a measurement of parity could be optimal. 

In analogy with Case~(II), our aim is to construct an operator $\hat{\mathcal{S}}_0$ where the correlations between systems $A$ and $B$ lead to a reduction in $\text{Var}(\hat{\mathcal{S}}_0)$ and the system mimics a pure spin-cat state. Introducing the quantity 
\begin{equation}
	\hat{\mathcal{S}}_0 = \hat{\Pi}_A\hat{\mathcal{S}}_B \equiv \hat{\Pi}_A\hat{\Pi}_B, 
\end{equation}
where $\hat{\Pi}_{A(B)}$ is the parity operator for system $A$($B$), defined by $\hat{\Pi}_A|j, m \rangle = (-1)^m |j, m\rangle$ and $\hat{\Pi}_B|B_m\rangle = (-1)^m|B_m\rangle$, we see that pseudo-spin-cat states satisfy $\hat{\mathcal{S}}_0|\Psi_{AB}\rangle = |\Psi_{AB}\rangle$, and therefore $\text{Var}(\hat{\mathcal{S}}_0) = 0$. This motivates the signal choice $\hat{\mathcal{S}} = \hat{U}^\dag_\phi \hat{\mathcal{S}}_0 \hat{U}_\phi$. 

To calculate the sensitivity, we need to compute $\langle \hat{\mathcal{S}} \rangle$ and $\langle \hat{\mathcal{S}}^2 \rangle$. Trivially, $\langle \hat{\mathcal{S}}^2 \rangle = 1$ for all states. For $\phi \ll 1$, expanding $\hat{U}_\phi$ to second order in $\phi$ gives
\begin{align}
	\langle  \hat{\mathcal{S}} \rangle &\approx \langle (1 + i\phi \jhat_y  - \tfrac{1}{2}\phi^2\jhat_y^2) \hat{\mathcal{S}}_0 (1 - i\phi \jhat_y  - \tfrac{1}{2}\phi^2\jhat_y^2)\rangle \notag \\
	&= 1 + i\phi\left(\langle\jhat_y\hat{\mathcal{S}}_0\rangle - \langle\hat{\mathcal{S}}_0\jhat_y \rangle \right) \notag\\
&+ \phi^2\left[\langle\jhat_y\hat{\mathcal{S}}_0\jhat_y\rangle - \frac{1}{2}\left(\langle\jhat_y^2\hat{\mathcal{S}}_0\rangle + \langle \hat{\mathcal{S}}_0\jhat_y^2\rangle \right)\right] \notag \\
&+ \mathcal{O}(\phi^3),
\end{align}
The relation $\langle B_n | B_{n\pm 1}\rangle = 0$ ensures that terms linear in $\jhat_y$ go to zero:
\begin{align}
\langle \jhat_+\rangle &= \sum_{m,n} c_m c_{n}^* \langle j, n|\jhat_+| j, m \rangle \langle B_n| B_m\rangle \nonumber \\
				&\propto \sum_{n,m} c_m c_{n}^* \delta_{n, m+1} \langle B_n| B_m \rangle \nonumber \\
				&=   \sum_{m} c_{m} c_{m+1}^* \langle B_{m+1}| B_m\rangle =0 \, .
\end{align}
However, unlike Case~(II), the condition $\langle B_n | B_{n\pm 2}\rangle = 1$ preserves terms such as $\langle \jhat_+^2\rangle$. Noting that $\jhat_y$ flips the parity of any state in subsystem $A$ but not subsystem $B$: 
\begin{subequations}
\begin{align}
	\hat{\Pi}_A \jhat_y |\Psi_{AB} \rangle	&= - \jhat_y\hat{\Pi}_A |\Psi_{AB}\rangle, \\
	\hat{\Pi}_B \jhat_y |\Psi_{AB}\rangle &=  \jhat_y\hat{\Pi}_B |\Psi_{AB}\rangle, 
\end{align}
\end{subequations}
and using $\hat{\mathcal{S}}_0|\Psi_{AB}\rangle = |\Psi_{AB}\rangle$ gives
\begin{equation}
\langle \jhat_y \hat{\mathcal{S}}_0\jhat_y\rangle  = -\langle  \hat{\mathcal{S}}_0  \jhat_y^2 \rangle = -\langle  \jhat_y^2 \hat{\mathcal{S}}_0 \rangle = -\langle  \jhat_y^2 \rangle.
\end{equation}
Therefore
\begin{equation}
	\langle \hat{\mathcal{S}}\rangle = 1 - 2\phi^2 \langle \jhat_y^2\rangle + \mathcal{O}(\phi^3) \, .
\end{equation}
Since $\hat{\mathcal{S}}_0^2 =1$ implies that $\hat{\mathcal{S}}^2 = 1$, we obtain
\begin{equation}
	\text{Var}(\mathcal{\hat{\mathcal{S}}}) =  4 \phi^2 \langle \jhat_y^2\rangle + \mathcal{O}(\phi^4),
\end{equation}
and consequently
\begin{equation}
(\Delta \phi)^2 = \frac{\text{Var}(\mathcal{\hat{\mathcal{S}}})}{( \partial_\phi \langle \hat{\mathcal{S}}\rangle)^2 } = \frac{4 \phi^2 \langle \jhat_y^2\rangle}{16 \phi^2 \langle \jhat_y^2\rangle^2 }= \frac{1}{4 \langle \jhat_y^2\rangle} = \frac{1}{\mathcal{F}_{AB}} \, .
\end{equation} 
This demonstrates that the signal saturates the QCRB and is therefore optimal. 

The optimal estimation schemes presented in Secs.~\ref{opt_2} and \ref{opt_3} illustrate a somewhat counterintuitive fact: although the optimal measurement of system $B$ for a pseudo-spin-cat state provides \emph{less} information about system $A$ than for an incoherent Dicke mixture, the pseudo-spin-cat state yields the better (in fact \emph{best}) sensitivity.

\subsection{System $B$ observables that approximate optimal measurements} \label{secV}
We now turn to the explicit construction of physical observables that approximate $\hat{\mathcal{S}}_B$. In general, the choice of $\hat{\mathcal{S}}_B$ depends upon the specific purification of $\hat{\rho}_A$. Physically, the initial state of system $B$ and the entangling Hamiltonian matter. However, there is no guarantee that $\hat{\mathcal{S}}_B$ exists, and if it does there is no guarantee that a measurement of this observable can be made in practice. Nevertheless, as we show below, it may be possible to make a measurement of an observable that \emph{approximates} $\hat{\mathcal{S}}_B$, and can therefore give near-optimal sensitivities.

\subsubsection{Case~(II)}

To begin, consider the situation in Sec.~\ref{dynamics_case_2}: the evolution of the state $|\alpha(\pi/2,0)\rangle \otimes |\beta\rangle$ under the Hamiltonian~(\ref{ham_coh}). We require $\hat{\mathcal{S}}_B |\Psi_{AB}\rangle = \hat{J}_z |\Psi_{AB}\rangle$. After some evolution time $t$:
\begin{equation}
	|\Psi_{AB} \rangle = \sum_{m=-j}^j c_m |j,m\rangle |\beta e^{-i m g t}\rangle \, . \label{Psi_AB_beta}
\end{equation}
Clearly, the \emph{phase} of the coherent state is correlated with the $\hat{J}_z$ projection of system $B$. This can be extracted via a homodyne measurement of the phase quadrature $\Yhat_B = i(\bhat - \bhatd)$ \cite{Bachor:2004}. In fact, provided $ m gt  \ll 1$, phase quadrature measurements of $| \beta \exp(-i m gt)\rangle$ are linearly proportional to the $\hat{J}_z$ projection:
\begin{equation}
	\langle \beta e^{-i m gt} | \Yhat_B | \beta e^{-i m gt}\rangle =  2 \beta \sin \left( m gt\right) \approx 2 \beta m g t, \label{eqn_lin_jz}
\end{equation}
where without loss of generality we have taken $\beta$ to be real and positive. Consequently, the scaled phase quadrature
\begin{equation}
\hat{\mathcal{S}}_B = \frac{\Yhat_B}{ 2 \beta gt} \label{S_B_case_2}
\end{equation}
satisfies
\begin{subequations}
\label{eqn_no_flucts}
\begin{align}
	\langle \Psi_{AB} | \left(\jhat_z - \hat{\mathcal{S}}_B\right) | \Psi_{AB}\rangle &\approx 0, \\
	\langle \Psi_{AB} | \left(\jhat_z - \hat{\mathcal{S}}_B\right)^2 | \Psi_{AB}\rangle &\approx \frac{1}{(2 \beta gt)^2}, 
\end{align}
\end{subequations}
and so the fluctuations in $(\jhat_z - \hat{\mathcal{S}}_B)$ become arbitrarily small (and $\hat{\mathcal{S}}_B$ becomes perfectly correlated with $\hat{J}_z$) as $(\beta gt)^2$ becomes large.  
This suggests that Eqs.~(\ref{sig_Dicke_mix}) and (\ref{S_B_case_2}) should be a good approximation to an optimal measurement signal.

\begin{figure}
\includegraphics[width=1\columnwidth]{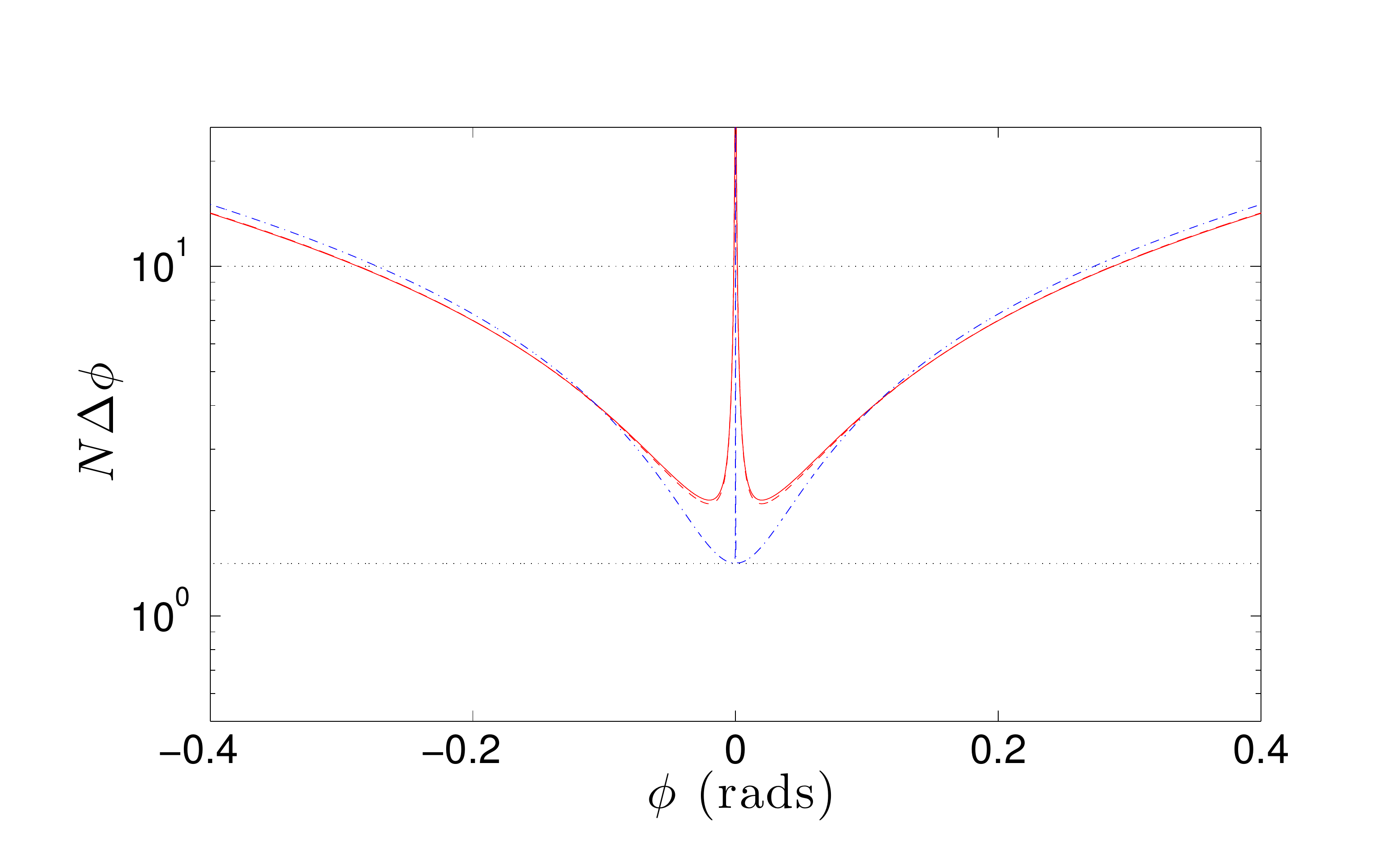}
\caption{(Color online) $\Delta \phi$ versus $\phi$ using \eq{s_opt_dicke} for a state of the form \eq{Psi_AB_beta}, with $N=100$, and $gt = 10^{-2}$. The blue dot-dashed line is with $|\beta|^2 = 10^6$ ($\beta gt = 10$), and the red solid line is for $|\beta|^2 = 10^4$ ($\beta gt =1$). The red dashed line shows the approximate expression for the sensitivity (\eq{sens_anal}) for $|\beta|^2 = 10^4$. For $|\beta|^2 = 10^6$, the numerical calculation and \eq{sens_anal} are identical. The upper and lower black dotted lines represent the standard quantum limit ($1/\sqrt{N}$), and $\sqrt{2}/N$ respectively. The divergence in $\Delta \phi$ close to $\phi=0$ in both cases is due to the imperfect correlations between $\hat{\mathcal{S}}_B$ and $\hat{J}_z$ leading to non-zero variance in $\hat{\mathcal{S}}$. If the correlations were perfect and $\text{Var}(\mathcal{\hat{\mathcal{S}}})|_{\phi=0}=0$, $\Delta \phi$ would reach exactly $1/\sqrt{\mathcal{F}_{AB}}$ at $\phi=0$.    
}
\label{fig_dicke_coherent}
\end{figure}

More precisely, assume that
\begin{equation}
	\frac{1}{\beta} \ll \frac{N g t}{2} \ll 1. \label{nice_cond}
\end{equation}
The first inequality ensures that $\langle \beta e^{-i n gt} | \beta e^{-i m gt}\rangle \approx \delta_{n,m}$ and so $\hat{\rho}_A$ is approximately an incoherent Dicke mixture, while the second inequality implies that we are in the linearized regime where Eq.~(\ref{eqn_lin_jz}) and Eqs.~(\ref{eqn_no_flucts}) hold. Then the signal
\begin{equation}
	\hat{\mathcal{S}}_\text{approx} = \left(\jhat_z \cos \phi - \jhat_x \sin \phi- \frac{\Yhat_B}{2 gt}\right)^2 \, , \label{s_opt_dicke}
\end{equation}
yields the sensitivity
\begin{align}
	(\Delta \phi)^2	&\approx \frac{1}{\big( \partial_\phi \langle \hat{\mathcal{S}} \rangle\big)^2} \Big\{ \text{Var}(\hat{\mathcal{S}}) - \frac{\sin^2(\phi/2)}{\beta^2}\langle \hat{J}_z^4\rangle + \frac{2}{(2 \beta g t)^4}\notag \\
				&+ \frac{4}{(2 \beta g t)^2} \left[ (\cos \phi - 1)^2 \langle \hat{J}_z^2 \rangle + \sin^2 \phi \langle \hat{J}_x^2 \rangle \right] \Big\}, \label{sens_anal}
\end{align}
where $\text{Var}(\hat{\mathcal{S}})$ and $\partial_\phi \langle \hat{\mathcal{S}} \rangle$ are given by the expectations~(\ref{expectations_optimal_S}) of the optimal signal Eq.~(\ref{sig_Dicke_mix}). Figure (\ref{fig_dicke_coherent}) shows \eq{sens_anal} compared to an exact numeric calculation.

Condition~(\ref{nice_cond}) typically ensures that the term proportional to $\langle \hat{J}_z^4\rangle$ is small in comparison to the term proportional to $1 / (2 \beta g t)^2$. We therefore see that our approximate signal $\hat{\mathcal{S}}_\text{approx}$ gives a sensitivity worse than the QCRB, and furthermore at an operating point $\phi \neq 0$. Nevertheless, $\Delta \phi$ approaches the QCRB at $\phi = 0$ as $\beta^2$ and $(2 \beta gt)^2$ approach infinity. Therefore, for a sufficiently large $\beta g t$, we can achieve near-optimal sensitivities close to $\phi = 0$. This is illustrated in Fig.~\ref{fig_dicke_coherent}. When $\beta gt = 10$, we find that $\Delta \phi$ is very close to the QCRB. In contrast, for $\beta gt =1$, the imperfect correlations between $\hat{J}_z$ and $\hat{\mathcal{S}}_b$ prevent the sensitivity from reaching the QCRB; nevertheless, the sensitivity is still below the SQL. Note that there is a slight deviation between Eq.~(\ref{sens_anal}) and the numerical calculation of the sensitivity using the state (\ref{Psi_AB_beta}). This is due to terms neglected by our approximations; in particular, the nonlinear terms ignored by linearizations such as Eq.~(\ref{eqn_lin_jz}) and those neglected terms that arise due to the small (but strictly non-zero) off-diagonal elements of $\hat{\rho}_A$.

\subsubsection{Case~(III)}

Now, consider the situation in Sec.~\ref{dynamics_case_3}: the evolution of the state $|\alpha(0,0)\rangle \otimes |\beta\rangle$ under the Hamiltonian~(\ref{ham_coh}) that at $gt=\pi$ approximately results in a pseudo-spin-cat state.

In order to find an operator that approximates $\hat{\mathcal{S}}_B = \hat{\Pi}_B$, we introduce the amplitude quadrature operator $\Xhat_B = (\bhat + \bhatd)$, and notice that
\begin{align}
	\langle \beta e^{-i m \pi} | \Xhat_B | \beta e^{-i m \pi}\rangle &=  2\beta (-1)^m \notag \\
												  &= 2 \beta \langle j,m| \hat{\Pi}_A | j, m\rangle \, .
\end{align} 
That is, amplitude quadrature measurements of $|\beta e^{-i m \pi} \rangle$ are proportional to parity measurements on system $B$, which are directly correlated with parity measurements on $\hat{J}_z$ eigenstates. Indeed, the quantity
\begin{equation}
	\hat{\mathcal{S}}_0 = \hat{\Pi}_A \frac{\Xhat_B}{ 2 \beta} \, ,
\end{equation}
has a variance $\text{Var}(\hat{\mathcal{S}}_0) = 1/ (2 \beta)^2$ that becomes vanishingly small as the amplitude of the coherent state is increased. We therefore expect the signal
\begin{equation}
	\hat{\mathcal{S}}_\text{approx} = \hat{U}_\phi^\dag \left(\hat{\Pi}_A \frac{\Xhat_B}{ 2 \beta}\right) \hat{U}_\phi \label{s_opt_cat}
\end{equation}
will be a good approximation to the optimal measurement $\hat{\mathcal{S}}$.  

\begin{figure}
\includegraphics[width=1\columnwidth]{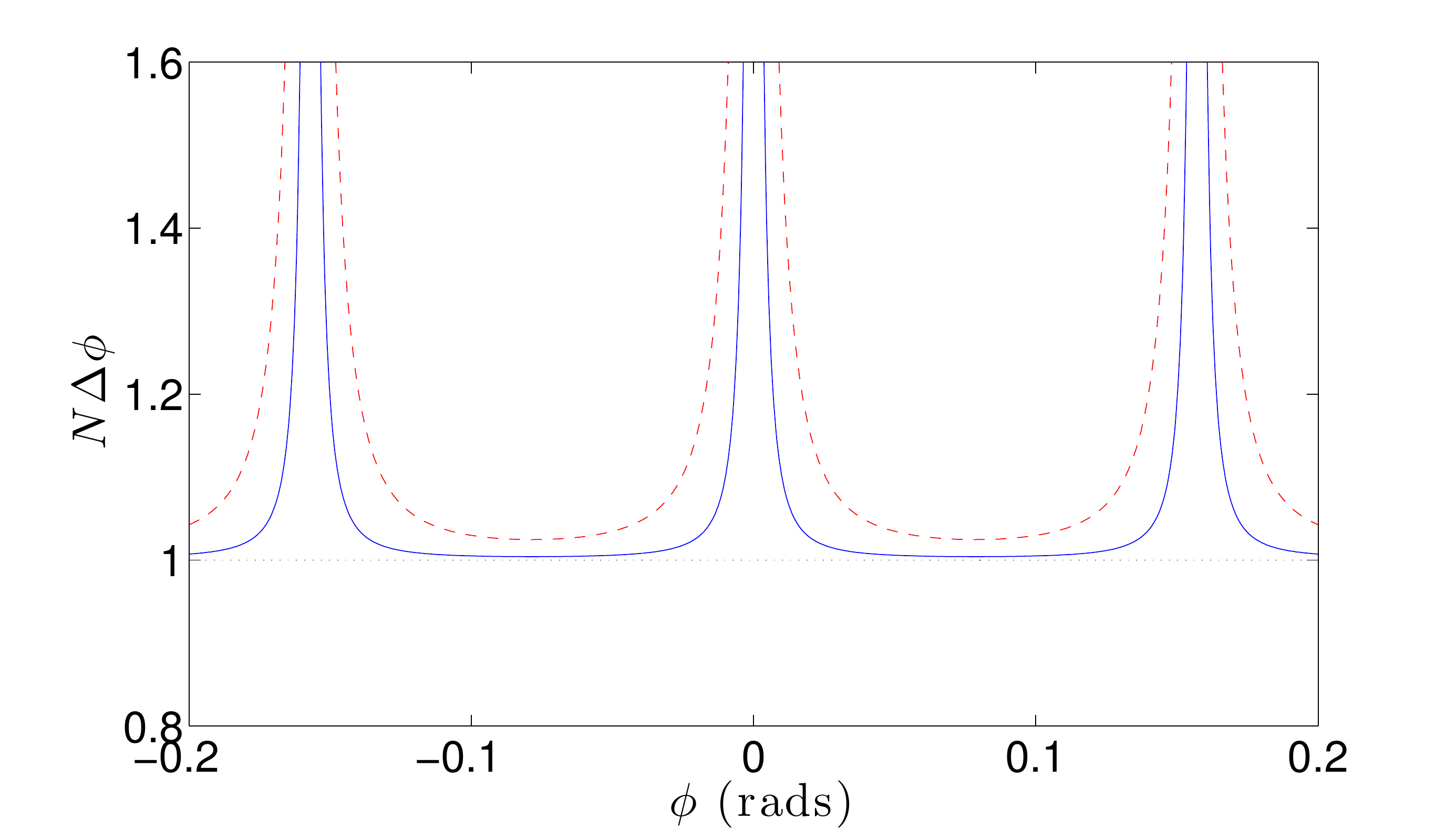}
\caption{(Color online) Phase sensitivity of the approximate signal Eq.~(\ref{s_opt_cat}) for a state of the form \eq{Psi_AB_beta} at $gt = \pi $ (i.e. a pseudo-spin-cat state) with $N=20$. The blue solid line and red dashed line are for $|\beta|^2 = 30$ and $|\beta|^2 = 5$, respectively. The black dotted line indicates the Heisenberg limit $\Delta \phi = 1/N$ (which is the QCRB). Note that the vertical axis is a linear scale.
}
\label{fig_deltaphi_cat}
\end{figure}

Figure~\ref{fig_deltaphi_cat} shows the sensitivity for a state of the form \eq{Psi_AB_beta} at $gt = \pi$ with $N=20$. When $|\beta |^2 = 30$, the sensitivity is very close to the Heisenberg limit, while for $|\beta|^2 = 5$ there is a slight degradation in the sensitivity due to imperfect correlations. In contrast to the approximate optimal measurement scheme for Case~(II), which requires a large amplitude coherent state, here the signal~(\ref{s_opt_cat}) is almost optimal even for small amplitude coherent states. This is because $\langle \beta e^{-i \pi} | \beta \rangle = \exp (-2 |\beta|^2)$ is approximately zero even for modest values of $\beta$. 

In situations where system $A$ is an ensemble of atoms, and system $B$ is an optical mode, it would be challenging to achieve the strong atom-light coupling regime required for $gt = \pi$. On the other hand, the choice of an initial coherent state for system $B$ ensures that the sensitivity is reasonably insensitive to losses in system $B$. In particular, since particle loss from a coherent state acts only to reduce the state's amplitude, provided the coherent state remains sufficiently large \emph{after} losses in order to satisfy the requirements for near-optimal measurements, near-Heisenberg-limited sensitivities should be obtainable. 

\subsubsection{Particle-exchange Hamiltonian}
Finally, for completeness we include the optimal measurement scheme for the state attained after evolving the product state $|\alpha(0,0)\rangle \otimes |N_B \rangle$ under the Hamiltonian~(\ref{Hpm}) (see Sec.~\ref{sec_particle_exchange}). The optimal measurement signal is simply Eq.~(\ref{sig_Dicke_mix}) with
\begin{equation}
	\hat{\mathcal{S}}_B = \frac{N}{2} \pm \left( N_b - \bhatd\bhat \right). \label{SB_fock}
\end{equation}
This choice of $\hat{\mathcal{S}}_B$ can be constructed by counting the number of particles in system $B$, and it satisfies $\hat{\mathcal{S}}_B|\Psi_{AB}\rangle = \jhat_z|\Psi_{AB}\rangle$ as required. As \emph{any} entangling Hamiltonian of the form
\begin{equation}
\hat{\mathcal{H}}_\pm = \sum_{k} \mathcal{A}_k \left(\jhat_\pm \bhatd + \jhat_\mp \bhat\right)^k + \sum_{j,k} \mathcal{B}_{j,k} \jhat_z^k \left(\bhatd \bhat \right)^j
\end{equation}
will lead to a state of the form \eq{Psi_AB}, with $|B_m\rangle$ given by \eq{B_m_fock} (assuming an initial state $|\Psi_A\rangle = |\alpha(0,0)\rangle$,  $|\Psi_B\rangle = |N_B\rangle$), \eq{SB_fock} also transfers to these systems. 

\section{Discussion and Conclusions} 

We have shown that purifications of mixed states represent an excellent resource for quantum metrology. In particular, we showed that if probe system $A$ and auxiliary system $B$ are entangled such that the 1st and 2nd order coherences of system $A$ vanish, then near-Heisenberg-limited sensitivities can be achieved provided measurements on both systems $A$ and $B$ are allowed. Although we focused on the situation where this entanglement is generated via a few specific Hamiltonians, our conclusions hold irrespective of the specific entanglement generation scheme.

While preparing this paper, we also numerically examined the effect of decoherence on the sensitivity of our metrological schemes. In particular, we found that the effect of particle loss, spin flips, and phase diffusion on purifications of the pseudo-spin-cat state from Fig.~\ref{fig_Q} was identical to that of a pure spin-cat state.  

Although these purified states are no more or less robust to decoherence than other nonclassical pure states, there are situations where they are easier to generate. The example we are most familiar with is atom interferometry, where atom-light entanglement and information recycling is more compatible with the requirements of high precision atom interferometry than the preparation of nonclassical atomic states via interatomic interactions \cite{Szigeti:2014b}. However, controlled interactions are routinely engineered between atoms and light \cite{Bloch:2008, Cronin:2009, Hammerer:2010, Li:2013}, superconducting circuits and microwaves \cite{Paik:2011, Xiang:2013}, light and mechanical systems \cite{Aspelmeyer:2014}, and ions and light \cite{Leibfried:2003, Wineland:2013, Stute:2013}. Given that high efficiency detection is available in all these systems \cite{Rowe:2001, Smith:2012, Guerlin:2007, Groen:2013, Murch:2013, Cohen:2013}, the application of our proposal to a range of metrological platforms is plausible in the near term. 

It is important to note that although the QFI approaches the Heisenberg limit ($F_{AB}=N^2$, in Case (III), for example), this is not the \emph{true} Heisenberg limit, as $N$ refers only to the number of particles in system $A$ (which pass through the interferometer), rather than the \emph{total} number of particles $N_t$ in system $A$ \emph{and} system $B$. However, there are some situations where the number of particles in system $A$ is by far the more valuable resource, which is why it makes sense to report the QFI in terms of $N$ rather than $N_t$. For example, consider the case of inertial sensing with atom interferometry, where system $A$ is atoms, and system $B$ is photons. The atoms are sensitive to inertial phase shift, but it is difficult to arbitrarily increase the atomic flux. However, a gain can be achieved by adding some number of photons to the system, which are comparatively `cheep' compared to atoms. 

Finally, we note that not all quantum systems are created equal; certain quantum information protocols, such as quantum error correction \cite{Kessler:2014} and no-knowledge feedback \cite{Szigeti:2014}, are better suited to some platforms than others. Our proposal allows an experimenter to both perform quantum-enhanced metrology and take advantage of any additional benefits a hybrid quantum system provides. 

\section{Acknowledgements} 

We would like to acknowledge useful discussions with Carlton Caves, Joel Corney, Jamie Fiess, Samuel Nolan, and Murray Olsen. This work was supported by Australian Research Council (ARC) Discovery Project No. DE130100575 and the ARC Centre of Excellence for Engineered Quantum Systems (Project No. CE110001013).

\begin{appendix}

\section{Derivation of Eqs.~(\ref{Fs_for_example})} \label{sec_appendix}
Here we derive the QFI $\mathcal{F}_{AB}$ for a MZ interferometer with the following entangled input:
\begin{equation}
	|\Psi_{AB}(t)\rangle = e^{ -i gt \hat{J}_z \hat{N}_b } |\theta, \varphi \rangle \otimes | \beta \rangle,
\end{equation}
where $\hat{N}_b = \hat{b}^\dag \hat{b}$, system $B$ is initially in a coherent state $|\beta\rangle$, and system $A$ is initially in a spin coherent state $ |\theta, \varphi \rangle$. Any spin coherent state can be defined by rotating the maximal Dicke state on the top pole of the Bloch sphere an angle $\theta$ about the $J_y$ axis and an angle $\varphi$ about the $J_z$ axis:
\begin{equation}
	| \theta, \varphi \rangle \equiv \hat{R}(\theta, \varphi) |j, j \rangle = e^{-i \varphi \hat{J}_z} e^{-i \theta \hat{J}_y} |j, j\rangle.
\end{equation}
Recall that $j = N/2$, where $N$ is the total number of system $A$ particles.

The QFI is
\begin{equation}
	\mathcal{F}_{AB} = 4 \text{Var}\left( e^{ i gt \hat{J}_z \hat{N}_b } \hat{J}_y e^{ -i gt \hat{J}_z \hat{N}_b } \right),
\end{equation}
where the expectations in the variance are taken with respect to the initial separable state, $|\theta, \varphi \rangle \otimes |\beta \rangle$. 

By application of the Baker-Campbell-Hausdorff formula
\begin{align}
	e^{\hat{A}} \hat{B} e^{-\hat{A}} 	&= \hat{B} + [\hat{A}, \hat{B}] + \frac{1}{2!}\left[ \hat{A}, [\hat{A}, \hat{B}] \right] \notag \\
							&+ \frac{1}{3!}\left[ \hat{A}, \left[ \hat{A}, [\hat{A}, \hat{B}] \right] \right] + \cdots
\end{align}
it can be shown that
\begin{equation}
	e^{ i gt \hat{J}_z \hat{N}_b } \hat{J}_y e^{ -i gt \hat{J}_z\hat{N}_b } = \sin ( g t \hat{N}_b ) \hat{J}_x + \cos ( g t \hat{N}_b ) \hat{J}_y.
\end{equation}
Therefore, since the initial state is separable, we obtain
\begin{align}
	\mathcal{F}_{AB} 	&= \langle \sin^2( g t \hat{N}_b) \rangle \langle \hat{J}_x^2 \rangle + \langle \cos^2(g t \hat{N}_b) \rangle \langle \hat{J}_y^2 \rangle \notag \\
					&+ \langle \sin(g t \hat{N}_b) \cos( g t \hat{N}_b )\rangle \langle \hat{J}_x \hat{J}_y + \hat{J}_y \hat{J}_x \rangle \notag \\
			& - \left( \langle \sin(g t \hat{N}_b )\rangle \langle\hat{J}_x\rangle + \langle\cos(g t \hat{N}_b )\rangle \langle \hat{J}_y\rangle \right)^2. \label{F_AB_exps_IC}
\end{align}

The system $A$ expectations are more easily computed by rotating the operators by $\hat{R}(\theta, \varphi)$ and then taking expectations with respect to the Dicke state $|j, j\rangle$. Specifically, by virtue of
\begin{subequations}
\label{eqns_rotations}
\begin{align}
	\hat{R}^\dag(\theta,\varphi) \hat{J}_x \hat{R}(\theta,\varphi) 	&= \cos \theta \cos \varphi \hat{J}_x + \cos \theta \sin \varphi \hat{J}_y \notag \\
													&- \sin \theta \hat{J}_z, \\
	 \hat{R}^\dag(\theta,\varphi) \hat{J}_y \hat{R}(\theta,\varphi) 	&= - \sin \varphi \hat{J}_x + \cos \varphi \hat{J}_y,
\end{align}
\end{subequations}
and the application of $\hat{J}_\pm = \hat{J}_x \pm i \hat{J}_y$ with
\begin{equation}
 	\hat{J}_\pm |j,m \rangle = \sqrt{j(j+1)-(m \pm 1)} |j,m \pm 1 \rangle,
\end{equation} 
we obtain
\begin{subequations}
\begin{align}
	\langle \hat{J}_x \rangle	&= j \sin \theta \cos \varphi, \\
	\langle \hat{J}_y \rangle	&= j \sin \theta \sin \varphi, \\
	\langle \hat{J}_x^2 \rangle	&= \frac{j}{2}\left( 1 + (2j-1) \sin^2 \theta \cos^2 \varphi \right),\\
	\langle \hat{J}_y^2 \rangle	&= \frac{j}{2}\left( 1 + (2j-1) \sin^2 \theta \sin^2 \varphi \right), \\
	\langle \hat{J}_x \hat{J}_y + \hat{J}_y \hat{J}_x \rangle	&= j(2j-1) \sin^2 \theta \sin \varphi \cos \varphi.
\end{align}
\end{subequations}
With some simplification this gives 
\begin{align}
	\mathcal{F}_{AB}	&= 2j \Big( 1 + \sin^2 \theta \big[(2j-1)\langle \sin^2(g t\hat{N}_b  + \varphi)\rangle \notag \\
					&- 2j \langle \sin(g t \hat{N}_b + \varphi)\rangle^2 \big] \Big). \label{F_AB_almost}
\end{align}
Incidentally, by setting $t = 0$ we can see that the QFI for a spin coherent state input never exceeds the standard quantum limit:
\begin{equation}
	\mathcal{F}[\hat{J}_y, |\theta, \varphi\rangle] = 2 j \left( 1 - \sin^2 \theta \sin^2 \varphi \right) \leq N.
\end{equation}

In order to compute the system $B$ expectations, note that
\begin{subequations}
\label{coh_avg_exp}
\begin{align}
	\langle \sin(g t \hat{N}_b + \varphi) \rangle	&= -\frac{i}{2}\left( \langle e^{i (g t \hat{N}_b + \varphi)}\rangle - \langle e^{-i (g t \hat{N}_b + \varphi)}\rangle\right) \\
	\langle \sin^2(g t \hat{N}_b + \varphi) \rangle	&= \frac{ 2- \langle e^{2 i (g t \hat{N}_b + \varphi)}\rangle - \langle e^{-2 i ( g t \hat{N}_b + \varphi)}\rangle}{4}.
\end{align}
\end{subequations}
Furthermore, for any $m$,
\begin{align}
	\langle e^{i m \left( g t \hat{N}_b + \varphi \right)} \rangle	&= e^{i m \varphi } \langle \beta | \beta e^{i m g t} \rangle \notag \\
												&= \exp \left[ - |\beta|^2 \left( 1 - e^{i m g t}\right) + 2 m \varphi \right]. \label{coh_overlap}
\end{align}
Substituting Eqs.~(\ref{coh_avg_exp}) and Eq.~(\ref{coh_overlap}) into Eq.~(\ref{F_AB_almost}) gives
\begin{equation}
	\mathcal{F}_{AB} = \mathcal{F}_0 + \mathcal{F}_1 + \mathcal{F}_2, 
\end{equation}
with the expressions for $\mathcal{F}_0$, $\mathcal{F}_1$, and $\mathcal{F}_2$ listed in Eqs.~(\ref{Fs_for_example}).
\end{appendix}

\bibliography{info_fisher_bib}

\begin{thebibliography}{87}%
\makeatletter
\providecommand \@ifxundefined [1]{%
 \@ifx{#1\undefined}
}%
\providecommand \@ifnum [1]{%
 \ifnum #1\expandafter \@firstoftwo
 \else \expandafter \@secondoftwo
 \fi
}%
\providecommand \@ifx [1]{%
 \ifx #1\expandafter \@firstoftwo
 \else \expandafter \@secondoftwo
 \fi
}%
\providecommand \natexlab [1]{#1}%
\providecommand \enquote  [1]{``#1''}%
\providecommand \bibnamefont  [1]{#1}%
\providecommand \bibfnamefont [1]{#1}%
\providecommand \citenamefont [1]{#1}%
\providecommand \href@noop [0]{\@secondoftwo}%
\providecommand \href [0]{\begingroup \@sanitize@url \@href}%
\providecommand \@href[1]{\@@startlink{#1}\@@href}%
\providecommand \@@href[1]{\endgroup#1\@@endlink}%
\providecommand \@sanitize@url [0]{\catcode `\\12\catcode `\$12\catcode
  `\&12\catcode `\#12\catcode `\^12\catcode `\_12\catcode `\%12\relax}%
\providecommand \@@startlink[1]{}%
\providecommand \@@endlink[0]{}%
\providecommand \url  [0]{\begingroup\@sanitize@url \@url }%
\providecommand \@url [1]{\endgroup\@href {#1}{\urlprefix }}%
\providecommand \urlprefix  [0]{URL }%
\providecommand \Eprint [0]{\href }%
\providecommand \doibase [0]{http://dx.doi.org/}%
\providecommand \selectlanguage [0]{\@gobble}%
\providecommand \bibinfo  [0]{\@secondoftwo}%
\providecommand \bibfield  [0]{\@secondoftwo}%
\providecommand \translation [1]{[#1]}%
\providecommand \BibitemOpen [0]{}%
\providecommand \bibitemStop [0]{}%
\providecommand \bibitemNoStop [0]{.\EOS\space}%
\providecommand \EOS [0]{\spacefactor3000\relax}%
\providecommand \BibitemShut  [1]{\csname bibitem#1\endcsname}%
\let\auto@bib@innerbib\@empty
\bibitem [{\citenamefont {Holland}\ and\ \citenamefont
  {Burnett}(1993)}]{Holland:1993}%
  \BibitemOpen
  \bibfield  {author} {\bibinfo {author} {\bibfnamefont {M.~J.}\ \bibnamefont
  {Holland}}\ and\ \bibinfo {author} {\bibfnamefont {K.}~\bibnamefont
  {Burnett}},\ }\bibfield  {title} {\enquote {\bibinfo {title} {Interferometric
  detection of optical phase shifts at the {Heisenberg} limit},}\ }\href
  {\doibase 10.1103/PhysRevLett.71.1355} {\bibfield  {journal} {\bibinfo
  {journal} {Phys. Rev. Lett.}\ }\textbf {\bibinfo {volume} {71}},\ \bibinfo
  {pages} {1355--1358} (\bibinfo {year} {1993})}\BibitemShut {NoStop}%
\bibitem [{\citenamefont {Giovannetti}\ \emph {et~al.}(2006)\citenamefont
  {Giovannetti}, \citenamefont {Lloyd},\ and\ \citenamefont
  {Maccone}}]{Giovannetti:2006}%
  \BibitemOpen
  \bibfield  {author} {\bibinfo {author} {\bibfnamefont {Vittorio}\
  \bibnamefont {Giovannetti}}, \bibinfo {author} {\bibfnamefont {Seth}\
  \bibnamefont {Lloyd}}, \ and\ \bibinfo {author} {\bibfnamefont {Lorenzo}\
  \bibnamefont {Maccone}},\ }\bibfield  {title} {\enquote {\bibinfo {title}
  {Quantum metrology},}\ }\href {\doibase 10.1103/PhysRevLett.96.010401}
  {\bibfield  {journal} {\bibinfo  {journal} {Phys. Rev. Lett.}\ }\textbf
  {\bibinfo {volume} {96}},\ \bibinfo {pages} {010401} (\bibinfo {year}
  {2006})}\BibitemShut {NoStop}%
\bibitem [{\citenamefont {L{\"u}cke}\ \emph {et~al.}(2011)\citenamefont
  {L{\"u}cke}, \citenamefont {Scherer}, \citenamefont {Kruse}, \citenamefont
  {Pezz{\'e}}, \citenamefont {Deuretzbacher}, \citenamefont {Hyllus},
  \citenamefont {Topic}, \citenamefont {Peise}, \citenamefont {Ertmer},
  \citenamefont {Arlt}, \citenamefont {Santos}, \citenamefont {Smerzi},\ and\
  \citenamefont {Klempt}}]{Lucke:2011}%
  \BibitemOpen
  \bibfield  {author} {\bibinfo {author} {\bibfnamefont {B.}~\bibnamefont
  {L{\"u}cke}}, \bibinfo {author} {\bibfnamefont {M.}~\bibnamefont {Scherer}},
  \bibinfo {author} {\bibfnamefont {J.}~\bibnamefont {Kruse}}, \bibinfo
  {author} {\bibfnamefont {L.}~\bibnamefont {Pezz{\'e}}}, \bibinfo {author}
  {\bibfnamefont {F.}~\bibnamefont {Deuretzbacher}}, \bibinfo {author}
  {\bibfnamefont {P.}~\bibnamefont {Hyllus}}, \bibinfo {author} {\bibfnamefont
  {O.}~\bibnamefont {Topic}}, \bibinfo {author} {\bibfnamefont
  {J.}~\bibnamefont {Peise}}, \bibinfo {author} {\bibfnamefont
  {W.}~\bibnamefont {Ertmer}}, \bibinfo {author} {\bibfnamefont
  {J.}~\bibnamefont {Arlt}}, \bibinfo {author} {\bibfnamefont {L.}~\bibnamefont
  {Santos}}, \bibinfo {author} {\bibfnamefont {A.}~\bibnamefont {Smerzi}}, \
  and\ \bibinfo {author} {\bibfnamefont {C.}~\bibnamefont {Klempt}},\
  }\bibfield  {title} {\enquote {\bibinfo {title} {Twin matter waves for
  interferometry beyond the classical limit},}\ }\href@noop {} {\bibfield
  {journal} {\bibinfo  {journal} {Science}\ }\textbf {\bibinfo {volume}
  {334}},\ \bibinfo {pages} {773--776} (\bibinfo {year} {2011})}\BibitemShut
  {NoStop}%
\bibitem [{\citenamefont {Sanders}\ and\ \citenamefont
  {Milburn}(1995)}]{Sanders:1995}%
  \BibitemOpen
  \bibfield  {author} {\bibinfo {author} {\bibfnamefont {B.~C.}\ \bibnamefont
  {Sanders}}\ and\ \bibinfo {author} {\bibfnamefont {G.~J.}\ \bibnamefont
  {Milburn}},\ }\bibfield  {title} {\enquote {\bibinfo {title} {Optimal quantum
  measurements for phase estimation},}\ }\href {\doibase
  10.1103/PhysRevLett.75.2944} {\bibfield  {journal} {\bibinfo  {journal}
  {Phys. Rev. Lett.}\ }\textbf {\bibinfo {volume} {75}},\ \bibinfo {pages}
  {2944--2947} (\bibinfo {year} {1995})}\BibitemShut {NoStop}%
\bibitem [{\citenamefont {Pezz\'e}\ and\ \citenamefont
  {Smerzi}(2013)}]{Pezze:2013}%
  \BibitemOpen
  \bibfield  {author} {\bibinfo {author} {\bibfnamefont {Luca}\ \bibnamefont
  {Pezz\'e}}\ and\ \bibinfo {author} {\bibfnamefont {Augusto}\ \bibnamefont
  {Smerzi}},\ }\bibfield  {title} {\enquote {\bibinfo {title} {Ultrasensitive
  two-mode interferometry with single-mode number squeezing},}\ }\href
  {\doibase 10.1103/PhysRevLett.110.163604} {\bibfield  {journal} {\bibinfo
  {journal} {Phys. Rev. Lett.}\ }\textbf {\bibinfo {volume} {110}},\ \bibinfo
  {pages} {163604} (\bibinfo {year} {2013})}\BibitemShut {NoStop}%
\bibitem [{\citenamefont {Demkowicz-Dobrzanski}\ \emph
  {et~al.}(2012)\citenamefont {Demkowicz-Dobrzanski}, \citenamefont
  {Kolodynski},\ and\ \citenamefont {Guta}}]{Demkowicz-Dobrzanski:2012}%
  \BibitemOpen
  \bibfield  {author} {\bibinfo {author} {\bibfnamefont {Rafal}\ \bibnamefont
  {Demkowicz-Dobrzanski}}, \bibinfo {author} {\bibfnamefont {Jan}\ \bibnamefont
  {Kolodynski}}, \ and\ \bibinfo {author} {\bibfnamefont {Madalin}\
  \bibnamefont {Guta}},\ }\bibfield  {title} {\enquote {\bibinfo {title} {The
  elusive {Heisenberg} limit in quantum-enhanced metrology},}\ }\href
  {http://dx.doi.org/10.1038/ncomms2067} {\bibfield  {journal} {\bibinfo
  {journal} {Nat Commun}\ }\textbf {\bibinfo {volume} {3}},\ \bibinfo {pages}
  {1063} (\bibinfo {year} {2012})}\BibitemShut {NoStop}%
\bibitem [{\citenamefont {Leibfried}\ \emph {et~al.}(2005)\citenamefont
  {Leibfried}, \citenamefont {Knill}, \citenamefont {Seidelin}, \citenamefont
  {Britton}, \citenamefont {Blakestad}, \citenamefont {Chiaverini},
  \citenamefont {Hume}, \citenamefont {Itano}, \citenamefont {Jost},
  \citenamefont {Langer}, \citenamefont {Ozeri}, \citenamefont {Reichle},\ and\
  \citenamefont {Wineland}}]{Leibfried:2005}%
  \BibitemOpen
  \bibfield  {author} {\bibinfo {author} {\bibfnamefont {D.}~\bibnamefont
  {Leibfried}}, \bibinfo {author} {\bibfnamefont {E.}~\bibnamefont {Knill}},
  \bibinfo {author} {\bibfnamefont {S.}~\bibnamefont {Seidelin}}, \bibinfo
  {author} {\bibfnamefont {J.}~\bibnamefont {Britton}}, \bibinfo {author}
  {\bibfnamefont {R.~B.}\ \bibnamefont {Blakestad}}, \bibinfo {author}
  {\bibfnamefont {J.}~\bibnamefont {Chiaverini}}, \bibinfo {author}
  {\bibfnamefont {D.~B.}\ \bibnamefont {Hume}}, \bibinfo {author}
  {\bibfnamefont {W.~M.}\ \bibnamefont {Itano}}, \bibinfo {author}
  {\bibfnamefont {J.~D.}\ \bibnamefont {Jost}}, \bibinfo {author}
  {\bibfnamefont {C.}~\bibnamefont {Langer}}, \bibinfo {author} {\bibfnamefont
  {R.}~\bibnamefont {Ozeri}}, \bibinfo {author} {\bibfnamefont
  {R.}~\bibnamefont {Reichle}}, \ and\ \bibinfo {author} {\bibfnamefont
  {D.~J.}\ \bibnamefont {Wineland}},\ }\bibfield  {title} {\enquote {\bibinfo
  {title} {Creation of a six-atom `{Schrodinger} cat' state},}\ }\href
  {http://dx.doi.org/10.1038/nature04251} {\bibfield  {journal} {\bibinfo
  {journal} {Nature}\ }\textbf {\bibinfo {volume} {438}},\ \bibinfo {pages}
  {639--642} (\bibinfo {year} {2005})}\BibitemShut {NoStop}%
\bibitem [{\citenamefont {Wieczorek}\ \emph {et~al.}(2009)\citenamefont
  {Wieczorek}, \citenamefont {Krischek}, \citenamefont {Kiesel}, \citenamefont
  {Michelberger}, \citenamefont {T\'oth},\ and\ \citenamefont
  {Weinfurter}}]{Wieczorek:2009}%
  \BibitemOpen
  \bibfield  {author} {\bibinfo {author} {\bibfnamefont {Witlef}\ \bibnamefont
  {Wieczorek}}, \bibinfo {author} {\bibfnamefont {Roland}\ \bibnamefont
  {Krischek}}, \bibinfo {author} {\bibfnamefont {Nikolai}\ \bibnamefont
  {Kiesel}}, \bibinfo {author} {\bibfnamefont {Patrick}\ \bibnamefont
  {Michelberger}}, \bibinfo {author} {\bibfnamefont {G\'eza}\ \bibnamefont
  {T\'oth}}, \ and\ \bibinfo {author} {\bibfnamefont {Harald}\ \bibnamefont
  {Weinfurter}},\ }\bibfield  {title} {\enquote {\bibinfo {title} {Experimental
  entanglement of a six-photon symmetric {Dicke} state},}\ }\href {\doibase
  10.1103/PhysRevLett.103.020504} {\bibfield  {journal} {\bibinfo  {journal}
  {Phys. Rev. Lett.}\ }\textbf {\bibinfo {volume} {103}},\ \bibinfo {pages}
  {020504} (\bibinfo {year} {2009})}\BibitemShut {NoStop}%
\bibitem [{\citenamefont {Gao}\ \emph {et~al.}(2010)\citenamefont {Gao},
  \citenamefont {Lu}, \citenamefont {Yao}, \citenamefont {Xu}, \citenamefont
  {Guhne}, \citenamefont {Goebel}, \citenamefont {Chen}, \citenamefont {Peng},
  \citenamefont {Chen},\ and\ \citenamefont {Pan}}]{Gao:2010}%
  \BibitemOpen
  \bibfield  {author} {\bibinfo {author} {\bibfnamefont {Wei-Bo}\ \bibnamefont
  {Gao}}, \bibinfo {author} {\bibfnamefont {Chao-Yang}\ \bibnamefont {Lu}},
  \bibinfo {author} {\bibfnamefont {Xing-Can}\ \bibnamefont {Yao}}, \bibinfo
  {author} {\bibfnamefont {Ping}\ \bibnamefont {Xu}}, \bibinfo {author}
  {\bibfnamefont {Otfried}\ \bibnamefont {Guhne}}, \bibinfo {author}
  {\bibfnamefont {Alexander}\ \bibnamefont {Goebel}}, \bibinfo {author}
  {\bibfnamefont {Yu-Ao}\ \bibnamefont {Chen}}, \bibinfo {author}
  {\bibfnamefont {Cheng-Zhi}\ \bibnamefont {Peng}}, \bibinfo {author}
  {\bibfnamefont {Zeng-Bing}\ \bibnamefont {Chen}}, \ and\ \bibinfo {author}
  {\bibfnamefont {Jian-Wei}\ \bibnamefont {Pan}},\ }\bibfield  {title}
  {\enquote {\bibinfo {title} {Experimental demonstration of a hyper-entangled
  ten-qubit {Schrodinger} cat state},}\ }\href
  {http://dx.doi.org/10.1038/nphys1603} {\bibfield  {journal} {\bibinfo
  {journal} {Nat Phys}\ }\textbf {\bibinfo {volume} {6}},\ \bibinfo {pages}
  {331--335} (\bibinfo {year} {2010})}\BibitemShut {NoStop}%
\bibitem [{\citenamefont {Vlastakis}\ \emph {et~al.}(2013)\citenamefont
  {Vlastakis}, \citenamefont {Kirchmair}, \citenamefont {Leghtas},
  \citenamefont {Nigg}, \citenamefont {Frunzio}, \citenamefont {Girvin},
  \citenamefont {Mirrahimi}, \citenamefont {Devoret},\ and\ \citenamefont
  {Schoelkopf}}]{Vlastakis:2013}%
  \BibitemOpen
  \bibfield  {author} {\bibinfo {author} {\bibfnamefont {Brian}\ \bibnamefont
  {Vlastakis}}, \bibinfo {author} {\bibfnamefont {Gerhard}\ \bibnamefont
  {Kirchmair}}, \bibinfo {author} {\bibfnamefont {Zaki}\ \bibnamefont
  {Leghtas}}, \bibinfo {author} {\bibfnamefont {Simon~E.}\ \bibnamefont
  {Nigg}}, \bibinfo {author} {\bibfnamefont {Luigi}\ \bibnamefont {Frunzio}},
  \bibinfo {author} {\bibfnamefont {S.~M.}\ \bibnamefont {Girvin}}, \bibinfo
  {author} {\bibfnamefont {Mazyar}\ \bibnamefont {Mirrahimi}}, \bibinfo
  {author} {\bibfnamefont {M.~H.}\ \bibnamefont {Devoret}}, \ and\ \bibinfo
  {author} {\bibfnamefont {R.~J.}\ \bibnamefont {Schoelkopf}},\ }\bibfield
  {title} {\enquote {\bibinfo {title} {Deterministically encoding quantum
  information using 100-photon {S}chr{\"o}dinger cat states},}\ }\href
  {\doibase 10.1126/science.1243289} {\bibfield  {journal} {\bibinfo  {journal}
  {Science}\ }\textbf {\bibinfo {volume} {342}},\ \bibinfo {pages} {607--610}
  (\bibinfo {year} {2013})}\BibitemShut {NoStop}%
\bibitem [{\citenamefont {Signoles}\ \emph {et~al.}(2014)\citenamefont
  {Signoles}, \citenamefont {Facon}, \citenamefont {Grosso}, \citenamefont
  {Dotsenko}, \citenamefont {Haroche}, \citenamefont {Raimond}, \citenamefont
  {Brune},\ and\ \citenamefont {Gleyzes}}]{Signoles:2014}%
  \BibitemOpen
  \bibfield  {author} {\bibinfo {author} {\bibfnamefont {Adrien}\ \bibnamefont
  {Signoles}}, \bibinfo {author} {\bibfnamefont {Adrien}\ \bibnamefont
  {Facon}}, \bibinfo {author} {\bibfnamefont {Dorian}\ \bibnamefont {Grosso}},
  \bibinfo {author} {\bibfnamefont {Igor}\ \bibnamefont {Dotsenko}}, \bibinfo
  {author} {\bibfnamefont {Serge}\ \bibnamefont {Haroche}}, \bibinfo {author}
  {\bibfnamefont {Jean-Michel}\ \bibnamefont {Raimond}}, \bibinfo {author}
  {\bibfnamefont {Michel}\ \bibnamefont {Brune}}, \ and\ \bibinfo {author}
  {\bibfnamefont {Sebastien}\ \bibnamefont {Gleyzes}},\ }\bibfield  {title}
  {\enquote {\bibinfo {title} {Confined quantum {Zeno} dynamics of a watched
  atomic arrow},}\ }\href {http://dx.doi.org/10.1038/nphys3076} {\bibfield
  {journal} {\bibinfo  {journal} {Nat Phys}\ }\textbf {\bibinfo {volume}
  {10}},\ \bibinfo {pages} {715--719} (\bibinfo {year} {2014})}\BibitemShut
  {NoStop}%
\bibitem [{\citenamefont {Jeong}\ \emph {et~al.}(2004)\citenamefont {Jeong},
  \citenamefont {Kim}, \citenamefont {Ralph},\ and\ \citenamefont
  {Ham}}]{Jeong:2004}%
  \BibitemOpen
  \bibfield  {author} {\bibinfo {author} {\bibfnamefont {H.}~\bibnamefont
  {Jeong}}, \bibinfo {author} {\bibfnamefont {M.~S.}\ \bibnamefont {Kim}},
  \bibinfo {author} {\bibfnamefont {T.~C.}\ \bibnamefont {Ralph}}, \ and\
  \bibinfo {author} {\bibfnamefont {B.~S.}\ \bibnamefont {Ham}},\ }\bibfield
  {title} {\enquote {\bibinfo {title} {Generation of macroscopic superposition
  states with small nonlinearity},}\ }\href {\doibase
  10.1103/PhysRevA.70.061801} {\bibfield  {journal} {\bibinfo  {journal} {Phys.
  Rev. A}\ }\textbf {\bibinfo {volume} {70}},\ \bibinfo {pages} {061801}
  (\bibinfo {year} {2004})}\BibitemShut {NoStop}%
\bibitem [{\citenamefont {Rebi\'c}\ \emph {et~al.}(2009)\citenamefont
  {Rebi\'c}, \citenamefont {Twamley},\ and\ \citenamefont
  {Milburn}}]{Rebic:2009}%
  \BibitemOpen
  \bibfield  {author} {\bibinfo {author} {\bibfnamefont {Stojan}\ \bibnamefont
  {Rebi\'c}}, \bibinfo {author} {\bibfnamefont {Jason}\ \bibnamefont
  {Twamley}}, \ and\ \bibinfo {author} {\bibfnamefont {Gerard~J.}\ \bibnamefont
  {Milburn}},\ }\bibfield  {title} {\enquote {\bibinfo {title} {Giant {Kerr}
  nonlinearities in circuit quantum electrodynamics},}\ }\href {\doibase
  10.1103/PhysRevLett.103.150503} {\bibfield  {journal} {\bibinfo  {journal}
  {Phys. Rev. Lett.}\ }\textbf {\bibinfo {volume} {103}},\ \bibinfo {pages}
  {150503} (\bibinfo {year} {2009})}\BibitemShut {NoStop}%
\bibitem [{\citenamefont {L\"u}\ \emph {et~al.}(2013)\citenamefont {L\"u},
  \citenamefont {Zhang}, \citenamefont {Ashhab}, \citenamefont {Wu},\ and\
  \citenamefont {Nori}}]{Lu:2013}%
  \BibitemOpen
  \bibfield  {author} {\bibinfo {author} {\bibfnamefont {Xin-You}\ \bibnamefont
  {L\"u}, \bibfnamefont {Xin-You}}, \bibinfo {author} {\bibfnamefont {Wei-Min}\
  \bibnamefont {Zhang}}, \bibinfo {author} {\bibfnamefont {Sahel}\ \bibnamefont
  {Ashhab}}, \bibinfo {author} {\bibfnamefont {Ying}\ \bibnamefont {Wu}}, \
  and\ \bibinfo {author} {\bibfnamefont {Franco}\ \bibnamefont {Nori}},\
  }\bibfield  {title} {\enquote {\bibinfo {title} {Quantum-criticality-induced
  strong {Kerr} nonlinearities in optomechanical systems},}\ }\href
  {http://dx.doi.org/10.1038/srep02943} {\bibfield  {journal} {\bibinfo
  {journal} {Sci. Rep.}\ }\textbf {\bibinfo {volume} {3}} (\bibinfo {year}
  {2013})}\BibitemShut {NoStop}%
\bibitem [{\citenamefont {Debs}\ \emph {et~al.}(2011)\citenamefont {Debs},
  \citenamefont {Altin}, \citenamefont {Barter}, \citenamefont {D\"oring},
  \citenamefont {Dennis}, \citenamefont {McDonald}, \citenamefont {Anderson},
  \citenamefont {Close},\ and\ \citenamefont {Robins}}]{Debs:2011}%
  \BibitemOpen
  \bibfield  {author} {\bibinfo {author} {\bibfnamefont {J.~E.}\ \bibnamefont
  {Debs}}, \bibinfo {author} {\bibfnamefont {P.~A.}\ \bibnamefont {Altin}},
  \bibinfo {author} {\bibfnamefont {T.~H.}\ \bibnamefont {Barter}}, \bibinfo
  {author} {\bibfnamefont {D.}~\bibnamefont {D\"oring}}, \bibinfo {author}
  {\bibfnamefont {G.~R.}\ \bibnamefont {Dennis}}, \bibinfo {author}
  {\bibfnamefont {G.}~\bibnamefont {McDonald}}, \bibinfo {author}
  {\bibfnamefont {R.~P.}\ \bibnamefont {Anderson}}, \bibinfo {author}
  {\bibfnamefont {J.~D.}\ \bibnamefont {Close}}, \ and\ \bibinfo {author}
  {\bibfnamefont {N.~P.}\ \bibnamefont {Robins}},\ }\bibfield  {title}
  {\enquote {\bibinfo {title} {Cold-atom gravimetry with a {B}ose-{E}instein
  condensate},}\ }\href {\doibase 10.1103/PhysRevA.84.033610} {\bibfield
  {journal} {\bibinfo  {journal} {Phys. Rev. A}\ }\textbf {\bibinfo {volume}
  {84}},\ \bibinfo {pages} {033610} (\bibinfo {year} {2011})}\BibitemShut
  {NoStop}%
\bibitem [{\citenamefont {Szigeti}\ \emph {et~al.}(2012)\citenamefont
  {Szigeti}, \citenamefont {Debs}, \citenamefont {Hope}, \citenamefont
  {Robins},\ and\ \citenamefont {Close}}]{Szigeti:2012}%
  \BibitemOpen
  \bibfield  {author} {\bibinfo {author} {\bibfnamefont {S~S}\ \bibnamefont
  {Szigeti}}, \bibinfo {author} {\bibfnamefont {J~E}\ \bibnamefont {Debs}},
  \bibinfo {author} {\bibfnamefont {J~J}\ \bibnamefont {Hope}}, \bibinfo
  {author} {\bibfnamefont {N~P}\ \bibnamefont {Robins}}, \ and\ \bibinfo
  {author} {\bibfnamefont {J~D}\ \bibnamefont {Close}},\ }\bibfield  {title}
  {\enquote {\bibinfo {title} {Why momentum width matters for atom
  interferometry with {Bragg} pulses},}\ }\href
  {http://stacks.iop.org/1367-2630/14/i=2/a=023009} {\bibfield  {journal}
  {\bibinfo  {journal} {New Journal of Physics}\ }\textbf {\bibinfo {volume}
  {14}},\ \bibinfo {pages} {023009} (\bibinfo {year} {2012})}\BibitemShut
  {NoStop}%
\bibitem [{\citenamefont {Robins}\ \emph {et~al.}(2013)\citenamefont {Robins},
  \citenamefont {Altin}, \citenamefont {Debs},\ and\ \citenamefont
  {Close}}]{Robins:2013}%
  \BibitemOpen
  \bibfield  {author} {\bibinfo {author} {\bibfnamefont {N.~P.}\ \bibnamefont
  {Robins}}, \bibinfo {author} {\bibfnamefont {P.~A.}\ \bibnamefont {Altin}},
  \bibinfo {author} {\bibfnamefont {J.~E.}\ \bibnamefont {Debs}}, \ and\
  \bibinfo {author} {\bibfnamefont {J.~D.}\ \bibnamefont {Close}},\ }\bibfield
  {title} {\enquote {\bibinfo {title} {Atom lasers: Production, properties and
  prospects for precision inertial measurement},}\ }\href {\doibase
  http://dx.doi.org/10.1016/j.physrep.2013.03.006} {\bibfield  {journal}
  {\bibinfo  {journal} {Physics Reports}\ }\textbf {\bibinfo {volume} {529}},\
  \bibinfo {pages} {265--296} (\bibinfo {year} {2013})}\BibitemShut {NoStop}%
\bibitem [{\citenamefont {Kheruntsyan}\ \emph {et~al.}(2005)\citenamefont
  {Kheruntsyan}, \citenamefont {Olsen},\ and\ \citenamefont
  {Drummond}}]{Kheruntsyan:2005b}%
  \BibitemOpen
  \bibfield  {author} {\bibinfo {author} {\bibfnamefont {K.~V.}\ \bibnamefont
  {Kheruntsyan}}, \bibinfo {author} {\bibfnamefont {M.~K.}\ \bibnamefont
  {Olsen}}, \ and\ \bibinfo {author} {\bibfnamefont {P.~D.}\ \bibnamefont
  {Drummond}},\ }\bibfield  {title} {\enquote {\bibinfo {title}
  {Einstein-{P}odolsky-{R}osen correlations via dissociation of a molecular
  {B}ose-{E}instein condensate},}\ }\href {\doibase
  10.1103/PhysRevLett.95.150405} {\bibfield  {journal} {\bibinfo  {journal}
  {Phys. Rev. Lett.}\ }\textbf {\bibinfo {volume} {95}},\ \bibinfo {pages}
  {150405} (\bibinfo {year} {2005})}\BibitemShut {NoStop}%
\bibitem [{\citenamefont {Pu}\ and\ \citenamefont {Meystre}(2000)}]{Pu:2000}%
  \BibitemOpen
  \bibfield  {author} {\bibinfo {author} {\bibfnamefont {H.}~\bibnamefont
  {Pu}}\ and\ \bibinfo {author} {\bibfnamefont {P.}~\bibnamefont {Meystre}},\
  }\bibfield  {title} {\enquote {\bibinfo {title} {Creating macroscopic atomic
  {Einstein-Podolsky-Rosen} states from {Bose-Einstein} condensates},}\ }\href
  {\doibase 10.1103/PhysRevLett.85.3987} {\bibfield  {journal} {\bibinfo
  {journal} {Phys. Rev. Lett.}\ }\textbf {\bibinfo {volume} {85}},\ \bibinfo
  {pages} {3987--3990} (\bibinfo {year} {2000})}\BibitemShut {NoStop}%
\bibitem [{\citenamefont {Gross}\ \emph {et~al.}(2011)\citenamefont {Gross},
  \citenamefont {Strobel}, \citenamefont {Nicklas}, \citenamefont {Zibold},
  \citenamefont {Bar-Gill}, \citenamefont {Kurizki},\ and\ \citenamefont
  {Oberthaler}}]{Gross:2011}%
  \BibitemOpen
  \bibfield  {author} {\bibinfo {author} {\bibfnamefont {C.}~\bibnamefont
  {Gross}}, \bibinfo {author} {\bibfnamefont {H.}~\bibnamefont {Strobel}},
  \bibinfo {author} {\bibfnamefont {E.}~\bibnamefont {Nicklas}}, \bibinfo
  {author} {\bibfnamefont {T.}~\bibnamefont {Zibold}}, \bibinfo {author}
  {\bibfnamefont {N.}~\bibnamefont {Bar-Gill}}, \bibinfo {author}
  {\bibfnamefont {G.}~\bibnamefont {Kurizki}}, \ and\ \bibinfo {author}
  {\bibfnamefont {M.~K.}\ \bibnamefont {Oberthaler}},\ }\bibfield  {title}
  {\enquote {\bibinfo {title} {Atomic homodyne detection of continuous-variable
  entangled twin-atom states},}\ }\href {http://dx.doi.org/10.1038/nature10654}
  {\bibfield  {journal} {\bibinfo  {journal} {Nature}\ }\textbf {\bibinfo
  {volume} {480}},\ \bibinfo {pages} {219--223} (\bibinfo {year}
  {2011})}\BibitemShut {NoStop}%
\bibitem [{\citenamefont {Jaskula}\ \emph {et~al.}(2010)\citenamefont
  {Jaskula}, \citenamefont {Bonneau}, \citenamefont {Partridge}, \citenamefont
  {Krachmalnicoff}, \citenamefont {Deuar}, \citenamefont {Kheruntsyan},
  \citenamefont {Aspect}, \citenamefont {Boiron},\ and\ \citenamefont
  {Westbrook}}]{Jaskula:2010}%
  \BibitemOpen
  \bibfield  {author} {\bibinfo {author} {\bibfnamefont {J.-C.}\ \bibnamefont
  {Jaskula}}, \bibinfo {author} {\bibfnamefont {M.}~\bibnamefont {Bonneau}},
  \bibinfo {author} {\bibfnamefont {G.~B.}\ \bibnamefont {Partridge}}, \bibinfo
  {author} {\bibfnamefont {V.}~\bibnamefont {Krachmalnicoff}}, \bibinfo
  {author} {\bibfnamefont {P.}~\bibnamefont {Deuar}}, \bibinfo {author}
  {\bibfnamefont {K.~V.}\ \bibnamefont {Kheruntsyan}}, \bibinfo {author}
  {\bibfnamefont {A.}~\bibnamefont {Aspect}}, \bibinfo {author} {\bibfnamefont
  {D.}~\bibnamefont {Boiron}}, \ and\ \bibinfo {author} {\bibfnamefont {C.~I.}\
  \bibnamefont {Westbrook}},\ }\bibfield  {title} {\enquote {\bibinfo {title}
  {Sub-poissonian number differences in four-wave mixing of matter waves},}\
  }\href {\doibase 10.1103/PhysRevLett.105.190402} {\bibfield  {journal}
  {\bibinfo  {journal} {Phys. Rev. Lett.}\ }\textbf {\bibinfo {volume} {105}},\
  \bibinfo {pages} {190402} (\bibinfo {year} {2010})}\BibitemShut {NoStop}%
\bibitem [{\citenamefont {Bucker}\ \emph {et~al.}(2011)\citenamefont {Bucker},
  \citenamefont {Grond}, \citenamefont {Manz}, \citenamefont {Berrada},
  \citenamefont {Betz}, \citenamefont {Koller}, \citenamefont {Hohenester},
  \citenamefont {Schumm}, \citenamefont {Perrin},\ and\ \citenamefont
  {Schmiedmayer}}]{Bucker:2011}%
  \BibitemOpen
  \bibfield  {author} {\bibinfo {author} {\bibfnamefont {Robert}\ \bibnamefont
  {Bucker}}, \bibinfo {author} {\bibfnamefont {Julian}\ \bibnamefont {Grond}},
  \bibinfo {author} {\bibfnamefont {Stephanie}\ \bibnamefont {Manz}}, \bibinfo
  {author} {\bibfnamefont {Tarik}\ \bibnamefont {Berrada}}, \bibinfo {author}
  {\bibfnamefont {Thomas}\ \bibnamefont {Betz}}, \bibinfo {author}
  {\bibfnamefont {Christian}\ \bibnamefont {Koller}}, \bibinfo {author}
  {\bibfnamefont {Ulrich}\ \bibnamefont {Hohenester}}, \bibinfo {author}
  {\bibfnamefont {Thorsten}\ \bibnamefont {Schumm}}, \bibinfo {author}
  {\bibfnamefont {Aurelien}\ \bibnamefont {Perrin}}, \ and\ \bibinfo {author}
  {\bibfnamefont {Jorg}\ \bibnamefont {Schmiedmayer}},\ }\bibfield  {title}
  {\enquote {\bibinfo {title} {Twin-atom beams},}\ }\href
  {http://dx.doi.org/10.1038/nphys1992} {\bibfield  {journal} {\bibinfo
  {journal} {Nat Phys}\ }\textbf {\bibinfo {volume} {7}},\ \bibinfo {pages}
  {608--611} (\bibinfo {year} {2011})}\BibitemShut {NoStop}%
\bibitem [{\citenamefont {Haine}\ and\ \citenamefont
  {Ferris}(2011)}]{Haine:2011}%
  \BibitemOpen
  \bibfield  {author} {\bibinfo {author} {\bibfnamefont {S.~A.}\ \bibnamefont
  {Haine}}\ and\ \bibinfo {author} {\bibfnamefont {A.~J.}\ \bibnamefont
  {Ferris}},\ }\bibfield  {title} {\enquote {\bibinfo {title} {Surpassing the
  standard quantum limit in an atom interferometer with four-mode entanglement
  produced from four-wave mixing},}\ }\href {\doibase
  10.1103/PhysRevA.84.043624} {\bibfield  {journal} {\bibinfo  {journal} {Phys.
  Rev. A}\ }\textbf {\bibinfo {volume} {84}},\ \bibinfo {pages} {043624}
  (\bibinfo {year} {2011})}\BibitemShut {NoStop}%
\bibitem [{\citenamefont {Lewis-Swan}\ and\ \citenamefont
  {Kheruntsyan}(2014)}]{Lewis-Swan:2014}%
  \BibitemOpen
  \bibfield  {author} {\bibinfo {author} {\bibfnamefont {R.~J.}\ \bibnamefont
  {Lewis-Swan}}\ and\ \bibinfo {author} {\bibfnamefont {K.~V.}\ \bibnamefont
  {Kheruntsyan}},\ }\bibfield  {title} {\enquote {\bibinfo {title} {Proposal
  for demonstrating the {Hong--Ou--Mandel} effect with matter waves},}\ }\href
  {http://dx.doi.org/10.1038/ncomms4752} {\bibfield  {journal} {\bibinfo
  {journal} {Nat Commun}\ }\textbf {\bibinfo {volume} {5}} (\bibinfo {year}
  {2014})}\BibitemShut {NoStop}%
\bibitem [{\citenamefont {Kitagawa}\ and\ \citenamefont
  {Ueda}(1993)}]{Kitagawa:1993}%
  \BibitemOpen
  \bibfield  {author} {\bibinfo {author} {\bibfnamefont {Masahiro}\
  \bibnamefont {Kitagawa}}\ and\ \bibinfo {author} {\bibfnamefont {Masahito}\
  \bibnamefont {Ueda}},\ }\bibfield  {title} {\enquote {\bibinfo {title}
  {Squeezed spin states},}\ }\href {\doibase 10.1103/PhysRevA.47.5138}
  {\bibfield  {journal} {\bibinfo  {journal} {Phys. Rev. A}\ }\textbf {\bibinfo
  {volume} {47}},\ \bibinfo {pages} {5138--5143} (\bibinfo {year}
  {1993})}\BibitemShut {NoStop}%
\bibitem [{\citenamefont {S\o{}ndberg~S\o{}rensen}(2002)}]{Sorensen:2002}%
  \BibitemOpen
  \bibfield  {author} {\bibinfo {author} {\bibfnamefont {Anders}\ \bibnamefont
  {S\o{}ndberg~S\o{}rensen}},\ }\bibfield  {title} {\enquote {\bibinfo {title}
  {Bogoliubov theory of entanglement in a {Bose-Einstein} condensate},}\ }\href
  {\doibase 10.1103/PhysRevA.65.043610} {\bibfield  {journal} {\bibinfo
  {journal} {Phys. Rev. A}\ }\textbf {\bibinfo {volume} {65}},\ \bibinfo
  {pages} {043610} (\bibinfo {year} {2002})}\BibitemShut {NoStop}%
\bibitem [{\citenamefont {Gross}\ \emph {et~al.}(2010)\citenamefont {Gross},
  \citenamefont {Zibold}, \citenamefont {Nicklas}, \citenamefont {Est{\`e}ve},\
  and\ \citenamefont {Oberthaler}}]{Gross:2010}%
  \BibitemOpen
  \bibfield  {author} {\bibinfo {author} {\bibfnamefont {C.}~\bibnamefont
  {Gross}}, \bibinfo {author} {\bibfnamefont {T.}~\bibnamefont {Zibold}},
  \bibinfo {author} {\bibfnamefont {E.}~\bibnamefont {Nicklas}}, \bibinfo
  {author} {\bibfnamefont {J.}~\bibnamefont {Est{\`e}ve}}, \ and\ \bibinfo
  {author} {\bibfnamefont {M.~K.}\ \bibnamefont {Oberthaler}},\ }\bibfield
  {title} {\enquote {\bibinfo {title} {Nonlinear atom interferometer surpasses
  classical precision limit},}\ }\href {http://dx.doi.org/10.1038/nature08919}
  {\bibfield  {journal} {\bibinfo  {journal} {Nature}\ }\textbf {\bibinfo
  {volume} {464}},\ \bibinfo {pages} {1165--1169} (\bibinfo {year}
  {2010})}\BibitemShut {NoStop}%
\bibitem [{\citenamefont {Riedel}\ \emph {et~al.}(2010)\citenamefont {Riedel},
  \citenamefont {B{\"o}hi}, \citenamefont {Li}, \citenamefont {H{\"a}nsch},
  \citenamefont {Sinatra},\ and\ \citenamefont {Treutlein}}]{Riedel:2010}%
  \BibitemOpen
  \bibfield  {author} {\bibinfo {author} {\bibfnamefont {Max~F.}\ \bibnamefont
  {Riedel}}, \bibinfo {author} {\bibfnamefont {Pascal}\ \bibnamefont
  {B{\"o}hi}}, \bibinfo {author} {\bibfnamefont {Yun}\ \bibnamefont {Li}},
  \bibinfo {author} {\bibfnamefont {Theodor~W.}\ \bibnamefont {H{\"a}nsch}},
  \bibinfo {author} {\bibfnamefont {Alice}\ \bibnamefont {Sinatra}}, \ and\
  \bibinfo {author} {\bibfnamefont {Philipp}\ \bibnamefont {Treutlein}},\
  }\bibfield  {title} {\enquote {\bibinfo {title} {Atom-chip-based generation
  of entanglement for quantum metrology},}\ }\href
  {http://dx.doi.org/10.1038/nature08988} {\bibfield  {journal} {\bibinfo
  {journal} {Nature}\ }\textbf {\bibinfo {volume} {464}},\ \bibinfo {pages}
  {1170--1173} (\bibinfo {year} {2010})}\BibitemShut {NoStop}%
\bibitem [{\citenamefont {Johnsson}\ and\ \citenamefont
  {Haine}(2007)}]{Johnsson:2007a}%
  \BibitemOpen
  \bibfield  {author} {\bibinfo {author} {\bibfnamefont {Mattias~T.}\
  \bibnamefont {Johnsson}}\ and\ \bibinfo {author} {\bibfnamefont {Simon~A.}\
  \bibnamefont {Haine}},\ }\bibfield  {title} {\enquote {\bibinfo {title}
  {Generating quadrature squeezing in an atom laser through
  self-interaction},}\ }\href {\doibase 10.1103/PhysRevLett.99.010401}
  {\bibfield  {journal} {\bibinfo  {journal} {Phys. Rev. Lett.}\ }\textbf
  {\bibinfo {volume} {99}},\ \bibinfo {pages} {010401} (\bibinfo {year}
  {2007})}\BibitemShut {NoStop}%
\bibitem [{\citenamefont {Haine}\ and\ \citenamefont
  {Johnsson}(2009)}]{Haine:2009}%
  \BibitemOpen
  \bibfield  {author} {\bibinfo {author} {\bibfnamefont {Simon~A.}\
  \bibnamefont {Haine}}\ and\ \bibinfo {author} {\bibfnamefont {Mattias~T.}\
  \bibnamefont {Johnsson}},\ }\bibfield  {title} {\enquote {\bibinfo {title}
  {Dynamic scheme for generating number squeezing in {B}ose-{E}instein
  condensates through nonlinear interactions},}\ }\href {\doibase
  10.1103/PhysRevA.80.023611} {\bibfield  {journal} {\bibinfo  {journal} {Phys.
  Rev. A}\ }\textbf {\bibinfo {volume} {80}},\ \bibinfo {pages} {023611}
  (\bibinfo {year} {2009})}\BibitemShut {NoStop}%
\bibitem [{\citenamefont {Haine}\ \emph {et~al.}(2014)\citenamefont {Haine},
  \citenamefont {Lau}, \citenamefont {Anderson},\ and\ \citenamefont
  {Johnsson}}]{Haine:2014}%
  \BibitemOpen
  \bibfield  {author} {\bibinfo {author} {\bibfnamefont {S.~A.}\ \bibnamefont
  {Haine}}, \bibinfo {author} {\bibfnamefont {J.}~\bibnamefont {Lau}}, \bibinfo
  {author} {\bibfnamefont {R.~P.}\ \bibnamefont {Anderson}}, \ and\ \bibinfo
  {author} {\bibfnamefont {M.~T.}\ \bibnamefont {Johnsson}},\ }\bibfield
  {title} {\enquote {\bibinfo {title} {Self-induced spatial dynamics to enhance
  spin squeezing via one-axis twisting in a two-component {Bose}-{Einstein}
  condensate},}\ }\href {\doibase 10.1103/PhysRevA.90.023613} {\bibfield
  {journal} {\bibinfo  {journal} {Phys. Rev. A}\ }\textbf {\bibinfo {volume}
  {90}},\ \bibinfo {pages} {023613} (\bibinfo {year} {2014})}\BibitemShut
  {NoStop}%
\bibitem [{\citenamefont {Jing}\ \emph {et~al.}(2000)\citenamefont {Jing},
  \citenamefont {Chen},\ and\ \citenamefont {Ge}}]{Jing:2000}%
  \BibitemOpen
  \bibfield  {author} {\bibinfo {author} {\bibfnamefont {Hui}\ \bibnamefont
  {Jing}}, \bibinfo {author} {\bibfnamefont {Jing-Ling}\ \bibnamefont {Chen}},
  \ and\ \bibinfo {author} {\bibfnamefont {Mo-Lin}\ \bibnamefont {Ge}},\
  }\bibfield  {title} {\enquote {\bibinfo {title} {Quantum-dynamical theory for
  squeezing the output of a {Bose-Einstein} condensate},}\ }\href {\doibase
  10.1103/PhysRevA.63.015601} {\bibfield  {journal} {\bibinfo  {journal} {Phys.
  Rev. A}\ }\textbf {\bibinfo {volume} {63}},\ \bibinfo {pages} {015601}
  (\bibinfo {year} {2000})}\BibitemShut {NoStop}%
\bibitem [{\citenamefont {Fleischhauer}\ and\ \citenamefont
  {Gong}(2002)}]{Fleischhauer:2002b}%
  \BibitemOpen
  \bibfield  {author} {\bibinfo {author} {\bibfnamefont {Michael}\ \bibnamefont
  {Fleischhauer}}\ and\ \bibinfo {author} {\bibfnamefont {Shangqing}\
  \bibnamefont {Gong}},\ }\bibfield  {title} {\enquote {\bibinfo {title}
  {Stationary source of nonclassical or entangled atoms},}\ }\href {\doibase
  10.1103/PhysRevLett.88.070404} {\bibfield  {journal} {\bibinfo  {journal}
  {Phys. Rev. Lett.}\ }\textbf {\bibinfo {volume} {88}},\ \bibinfo {pages}
  {070404} (\bibinfo {year} {2002})}\BibitemShut {NoStop}%
\bibitem [{\citenamefont {Haine}\ and\ \citenamefont
  {Hope}(2005{\natexlab{a}})}]{Haine:2005}%
  \BibitemOpen
  \bibfield  {author} {\bibinfo {author} {\bibfnamefont {S.~A.}\ \bibnamefont
  {Haine}}\ and\ \bibinfo {author} {\bibfnamefont {J.~J.}\ \bibnamefont
  {Hope}},\ }\bibfield  {title} {\enquote {\bibinfo {title} {Outcoupling from a
  bose-einstein condensate with squeezed light to produce entangled-atom laser
  beams},}\ }\href {\doibase 10.1103/PhysRevA.72.033601} {\bibfield  {journal}
  {\bibinfo  {journal} {Phys. Rev. A}\ }\textbf {\bibinfo {volume} {72}},\
  \bibinfo {pages} {033601} (\bibinfo {year} {2005}{\natexlab{a}})}\BibitemShut
  {NoStop}%
\bibitem [{\citenamefont {Haine}\ and\ \citenamefont
  {Hope}(2005{\natexlab{b}})}]{Haine:2005b}%
  \BibitemOpen
  \bibfield  {author} {\bibinfo {author} {\bibfnamefont {S.~A.}\ \bibnamefont
  {Haine}}\ and\ \bibinfo {author} {\bibfnamefont {J.~J.}\ \bibnamefont
  {Hope}},\ }\bibfield  {title} {\enquote {\bibinfo {title} {A multi-mode model
  of a non-classical atom laser produced by outcoupling from a bose-einstein
  condensate with squeezed light},}\ }\href {\doibase 10.1002/lapl.200510052}
  {\bibfield  {journal} {\bibinfo  {journal} {Laser Physics Letters}\ }\textbf
  {\bibinfo {volume} {2}},\ \bibinfo {pages} {597--602} (\bibinfo {year}
  {2005}{\natexlab{b}})}\BibitemShut {NoStop}%
\bibitem [{\citenamefont {Haine}\ \emph {et~al.}(2006)\citenamefont {Haine},
  \citenamefont {Olsen},\ and\ \citenamefont {Hope}}]{Haine:2006b}%
  \BibitemOpen
  \bibfield  {author} {\bibinfo {author} {\bibfnamefont {S.~A.}\ \bibnamefont
  {Haine}}, \bibinfo {author} {\bibfnamefont {M.~K.}\ \bibnamefont {Olsen}}, \
  and\ \bibinfo {author} {\bibfnamefont {J.~J.}\ \bibnamefont {Hope}},\
  }\bibfield  {title} {\enquote {\bibinfo {title} {Generating controllable
  atom-light entanglement with a raman atom laser system},}\ }\href {\doibase
  10.1103/PhysRevLett.96.133601} {\bibfield  {journal} {\bibinfo  {journal}
  {Phys. Rev. Lett.}\ }\textbf {\bibinfo {volume} {96}},\ \bibinfo {pages}
  {133601} (\bibinfo {year} {2006})}\BibitemShut {NoStop}%
\bibitem [{\citenamefont {Hammerer}\ \emph {et~al.}(2010)\citenamefont
  {Hammerer}, \citenamefont {S\o{}rensen},\ and\ \citenamefont
  {Polzik}}]{Hammerer:2010}%
  \BibitemOpen
  \bibfield  {author} {\bibinfo {author} {\bibfnamefont {Klemens}\ \bibnamefont
  {Hammerer}}, \bibinfo {author} {\bibfnamefont {Anders~S.}\ \bibnamefont
  {S\o{}rensen}}, \ and\ \bibinfo {author} {\bibfnamefont {Eugene~S.}\
  \bibnamefont {Polzik}},\ }\bibfield  {title} {\enquote {\bibinfo {title}
  {Quantum interface between light and atomic ensembles},}\ }\href {\doibase
  10.1103/RevModPhys.82.1041} {\bibfield  {journal} {\bibinfo  {journal} {Rev.
  Mod. Phys.}\ }\textbf {\bibinfo {volume} {82}},\ \bibinfo {pages}
  {1041--1093} (\bibinfo {year} {2010})}\BibitemShut {NoStop}%
\bibitem [{\citenamefont {Haine}(2013)}]{Haine:2013}%
  \BibitemOpen
  \bibfield  {author} {\bibinfo {author} {\bibfnamefont {S.~A.}\ \bibnamefont
  {Haine}},\ }\bibfield  {title} {\enquote {\bibinfo {title}
  {Information-recycling beam splitters for quantum enhanced atom
  interferometry},}\ }\href {\doibase 10.1103/PhysRevLett.110.053002}
  {\bibfield  {journal} {\bibinfo  {journal} {Phys. Rev. Lett.}\ }\textbf
  {\bibinfo {volume} {110}},\ \bibinfo {pages} {053002} (\bibinfo {year}
  {2013})}\BibitemShut {NoStop}%
\bibitem [{\citenamefont {Szigeti}\ \emph
  {et~al.}(2014{\natexlab{a}})\citenamefont {Szigeti}, \citenamefont
  {Tonekaboni}, \citenamefont {Lau}, \citenamefont {Hood},\ and\ \citenamefont
  {Haine}}]{Szigeti:2014b}%
  \BibitemOpen
  \bibfield  {author} {\bibinfo {author} {\bibfnamefont {Stuart~S.}\
  \bibnamefont {Szigeti}}, \bibinfo {author} {\bibfnamefont {Behnam}\
  \bibnamefont {Tonekaboni}}, \bibinfo {author} {\bibfnamefont {Wing Yung~S.}\
  \bibnamefont {Lau}}, \bibinfo {author} {\bibfnamefont {Samantha~N.}\
  \bibnamefont {Hood}}, \ and\ \bibinfo {author} {\bibfnamefont {Simon~A.}\
  \bibnamefont {Haine}},\ }\bibfield  {title} {\enquote {\bibinfo {title}
  {Squeezed-light-enhanced atom interferometry below the standard quantum
  limit},}\ }\href {\doibase 10.1103/PhysRevA.90.063630} {\bibfield  {journal}
  {\bibinfo  {journal} {Phys. Rev. A}\ }\textbf {\bibinfo {volume} {90}},\
  \bibinfo {pages} {063630} (\bibinfo {year} {2014}{\natexlab{a}})}\BibitemShut
  {NoStop}%
\bibitem [{\citenamefont {Tonekaboni}\ \emph {et~al.}(2015)\citenamefont
  {Tonekaboni}, \citenamefont {Haine},\ and\ \citenamefont
  {Szigeti}}]{Tonekaboni:2015}%
  \BibitemOpen
  \bibfield  {author} {\bibinfo {author} {\bibfnamefont {Behnam}\ \bibnamefont
  {Tonekaboni}}, \bibinfo {author} {\bibfnamefont {Simon~A.}\ \bibnamefont
  {Haine}}, \ and\ \bibinfo {author} {\bibfnamefont {Stuart~S.}\ \bibnamefont
  {Szigeti}},\ }\bibfield  {title} {\enquote {\bibinfo {title}
  {Heisenberg-limited metrology with a squeezed vacuum state, three-mode
  mixing, and information recycling},}\ }\href {\doibase
  10.1103/PhysRevA.91.033616} {\bibfield  {journal} {\bibinfo  {journal} {Phys.
  Rev. A}\ }\textbf {\bibinfo {volume} {91}},\ \bibinfo {pages} {033616}
  (\bibinfo {year} {2015})}\BibitemShut {NoStop}%
\bibitem [{\citenamefont {Haine}\ \emph {et~al.}(2015)\citenamefont {Haine},
  \citenamefont {Szigeti}, \citenamefont {Lang},\ and\ \citenamefont
  {Caves}}]{Haine:2015}%
  \BibitemOpen
  \bibfield  {author} {\bibinfo {author} {\bibfnamefont {Simon~A.}\
  \bibnamefont {Haine}}, \bibinfo {author} {\bibfnamefont {Stuart~S.}\
  \bibnamefont {Szigeti}}, \bibinfo {author} {\bibfnamefont {Matthias~D.}\
  \bibnamefont {Lang}}, \ and\ \bibinfo {author} {\bibfnamefont {Carlton~M.}\
  \bibnamefont {Caves}},\ }\bibfield  {title} {\enquote {\bibinfo {title}
  {Heisenberg-limited metrology with information recycling},}\ }\href {\doibase
  10.1103/PhysRevA.91.041802} {\bibfield  {journal} {\bibinfo  {journal} {Phys.
  Rev. A}\ }\textbf {\bibinfo {volume} {91}},\ \bibinfo {pages} {041802}
  (\bibinfo {year} {2015})}\BibitemShut {NoStop}%
\bibitem [{\citenamefont {Braunstein}\ and\ \citenamefont
  {Caves}(1994)}]{Braunstein:1994}%
  \BibitemOpen
  \bibfield  {author} {\bibinfo {author} {\bibfnamefont {Samuel~L.}\
  \bibnamefont {Braunstein}}\ and\ \bibinfo {author} {\bibfnamefont
  {Carlton~M.}\ \bibnamefont {Caves}},\ }\bibfield  {title} {\enquote {\bibinfo
  {title} {Statistical distance and the geometry of quantum states},}\ }\href
  {\doibase 10.1103/PhysRevLett.72.3439} {\bibfield  {journal} {\bibinfo
  {journal} {Phys. Rev. Lett.}\ }\textbf {\bibinfo {volume} {72}},\ \bibinfo
  {pages} {3439--3443} (\bibinfo {year} {1994})}\BibitemShut {NoStop}%
\bibitem [{\citenamefont {Paris}(2009)}]{Paris:2009}%
  \BibitemOpen
  \bibfield  {author} {\bibinfo {author} {\bibfnamefont {Matteo G.~A.}\
  \bibnamefont {Paris}},\ }\bibfield  {title} {\enquote {\bibinfo {title}
  {Quantum estimation for quantum technology},}\ }\href {\doibase
  10.1142/S0219749909004839} {\bibfield  {journal} {\bibinfo  {journal}
  {International Journal of Quantum Information}\ }\textbf {\bibinfo {volume}
  {07}},\ \bibinfo {pages} {125--137} (\bibinfo {year} {2009})}\BibitemShut
  {NoStop}%
\bibitem [{\citenamefont {Demkowicz-Dobrza{\'n}ski}\ \emph
  {et~al.}(2014)\citenamefont {Demkowicz-Dobrza{\'n}ski}, \citenamefont
  {Jarzyna},\ and\ \citenamefont
  {Ko{\l}ody{\'n}ski}}]{Demkowicz-Dobrzanski:2014}%
  \BibitemOpen
  \bibfield  {author} {\bibinfo {author} {\bibfnamefont {Rafa{\l}}\
  \bibnamefont {Demkowicz-Dobrza{\'n}ski}}, \bibinfo {author} {\bibfnamefont
  {M.}~\bibnamefont {Jarzyna}}, \ and\ \bibinfo {author} {\bibfnamefont
  {J.}~\bibnamefont {Ko{\l}ody{\'n}ski}},\ }\href@noop {} {\enquote {\bibinfo
  {title} {Quantum limits in optical interferometry},}\ } (\bibinfo {year}
  {2014}),\ \bibinfo {note} {to be published in Progress in Optics},\ \Eprint
  {http://arxiv.org/abs/arXiv:1405.7703 [quant-ph]} {arXiv:1405.7703
  [quant-ph]} \BibitemShut {NoStop}%
\bibitem [{\citenamefont {T\'oth}\ and\ \citenamefont
  {Apellaniz}(2014)}]{Toth:2014}%
  \BibitemOpen
  \bibfield  {author} {\bibinfo {author} {\bibfnamefont {G\'eza}\ \bibnamefont
  {T\'oth}}\ and\ \bibinfo {author} {\bibfnamefont {Iagoba}\ \bibnamefont
  {Apellaniz}},\ }\bibfield  {title} {\enquote {\bibinfo {title} {Quantum
  metrology from a quantum information science perspective},}\ }\href
  {http://stacks.iop.org/1751-8121/47/i=42/a=424006} {\bibfield  {journal}
  {\bibinfo  {journal} {Journal of Physics A: Mathematical and Theoretical}\
  }\textbf {\bibinfo {volume} {47}},\ \bibinfo {pages} {424006} (\bibinfo
  {year} {2014})}\BibitemShut {NoStop}%
\bibitem [{\citenamefont {Yurke}\ \emph {et~al.}(1986)\citenamefont {Yurke},
  \citenamefont {McCall},\ and\ \citenamefont {Klauder}}]{Yurke:1986}%
  \BibitemOpen
  \bibfield  {author} {\bibinfo {author} {\bibfnamefont {Bernard}\ \bibnamefont
  {Yurke}}, \bibinfo {author} {\bibfnamefont {Samuel~L.}\ \bibnamefont
  {McCall}}, \ and\ \bibinfo {author} {\bibfnamefont {John~R.}\ \bibnamefont
  {Klauder}},\ }\bibfield  {title} {\enquote {\bibinfo {title} {{SU}(2) and
  {SU}(1,1) interferometers},}\ }\href {\doibase 10.1103/PhysRevA.33.4033}
  {\bibfield  {journal} {\bibinfo  {journal} {Phys. Rev. A}\ }\textbf {\bibinfo
  {volume} {33}},\ \bibinfo {pages} {4033--4054} (\bibinfo {year}
  {1986})}\BibitemShut {NoStop}%
\bibitem [{\citenamefont {Arecchi}\ \emph {et~al.}(1972)\citenamefont
  {Arecchi}, \citenamefont {Courtens}, \citenamefont {Gilmore},\ and\
  \citenamefont {Thomas}}]{Arecchi:1972}%
  \BibitemOpen
  \bibfield  {author} {\bibinfo {author} {\bibfnamefont {F.~T.}\ \bibnamefont
  {Arecchi}}, \bibinfo {author} {\bibfnamefont {Eric}\ \bibnamefont
  {Courtens}}, \bibinfo {author} {\bibfnamefont {Robert}\ \bibnamefont
  {Gilmore}}, \ and\ \bibinfo {author} {\bibfnamefont {Harry}\ \bibnamefont
  {Thomas}},\ }\bibfield  {title} {\enquote {\bibinfo {title} {Atomic coherent
  states in quantum optics},}\ }\href {\doibase 10.1103/PhysRevA.6.2211}
  {\bibfield  {journal} {\bibinfo  {journal} {Phys. Rev. A}\ }\textbf {\bibinfo
  {volume} {6}},\ \bibinfo {pages} {2211--2237} (\bibinfo {year}
  {1972})}\BibitemShut {NoStop}%
\bibitem [{\citenamefont {Agarwal}(1998)}]{Agarwal:1998}%
  \BibitemOpen
  \bibfield  {author} {\bibinfo {author} {\bibfnamefont {G.~S.}\ \bibnamefont
  {Agarwal}},\ }\bibfield  {title} {\enquote {\bibinfo {title} {State
  reconstruction for a collection of two-level systems},}\ }\href {\doibase
  10.1103/PhysRevA.57.671} {\bibfield  {journal} {\bibinfo  {journal} {Phys.
  Rev. A}\ }\textbf {\bibinfo {volume} {57}},\ \bibinfo {pages} {671--673}
  (\bibinfo {year} {1998})}\BibitemShut {NoStop}%
\bibitem [{\citenamefont {Mondal}(2015)}]{Mondal:2015}%
  \BibitemOpen
  \bibfield  {author} {\bibinfo {author} {\bibfnamefont {Debasis}\ \bibnamefont
  {Mondal}},\ }\bibfield  {title} {\enquote {\bibinfo {title} {{G}eneralized
  {F}ubini-study metric and {F}isher information metric},}\ }\href@noop {}
  {\bibfield  {journal} {\bibinfo  {journal} {arXiv:1503.04146}\ } (\bibinfo
  {year} {2015})}\BibitemShut {NoStop}%
\bibitem [{\citenamefont {Pezz\'e}\ and\ \citenamefont
  {Smerzi}(2009)}]{Pezze:2009}%
  \BibitemOpen
  \bibfield  {author} {\bibinfo {author} {\bibfnamefont {Luca}\ \bibnamefont
  {Pezz\'e}}\ and\ \bibinfo {author} {\bibfnamefont {Augusto}\ \bibnamefont
  {Smerzi}},\ }\bibfield  {title} {\enquote {\bibinfo {title} {Entanglement,
  nonlinear dynamics, and the {Heisenberg} limit},}\ }\href {\doibase
  10.1103/PhysRevLett.102.100401} {\bibfield  {journal} {\bibinfo  {journal}
  {Phys. Rev. Lett.}\ }\textbf {\bibinfo {volume} {102}},\ \bibinfo {pages}
  {100401} (\bibinfo {year} {2009})}\BibitemShut {NoStop}%
\bibitem [{Note1()}]{Note1}%
  \BibitemOpen
  \bibinfo {note} {If we restricted measurements to system $A$, then this is
  equivalent to tracing out system $B$, in which case the QFI is $\protect
  \mathcal {F}_A$.}\BibitemShut {Stop}%
\bibitem [{\citenamefont {Wasilewski}\ \emph {et~al.}(2010)\citenamefont
  {Wasilewski}, \citenamefont {Jensen}, \citenamefont {Krauter}, \citenamefont
  {Renema}, \citenamefont {Balabas},\ and\ \citenamefont
  {Polzik}}]{Wasilewski:2010}%
  \BibitemOpen
  \bibfield  {author} {\bibinfo {author} {\bibfnamefont {W.}~\bibnamefont
  {Wasilewski}}, \bibinfo {author} {\bibfnamefont {K.}~\bibnamefont {Jensen}},
  \bibinfo {author} {\bibfnamefont {H.}~\bibnamefont {Krauter}}, \bibinfo
  {author} {\bibfnamefont {J.~J.}\ \bibnamefont {Renema}}, \bibinfo {author}
  {\bibfnamefont {M.~V.}\ \bibnamefont {Balabas}}, \ and\ \bibinfo {author}
  {\bibfnamefont {E.~S.}\ \bibnamefont {Polzik}},\ }\bibfield  {title}
  {\enquote {\bibinfo {title} {Quantum noise limited and entanglement-assisted
  magnetometry},}\ }\href {\doibase 10.1103/PhysRevLett.104.133601} {\bibfield
  {journal} {\bibinfo  {journal} {Phys. Rev. Lett.}\ }\textbf {\bibinfo
  {volume} {104}},\ \bibinfo {pages} {133601} (\bibinfo {year}
  {2010})}\BibitemShut {NoStop}%
\bibitem [{\citenamefont {Schleier-Smith}\ \emph {et~al.}(2010)\citenamefont
  {Schleier-Smith}, \citenamefont {Leroux},\ and\ \citenamefont
  {Vuleti\ifmmode~\acute{c}\else \'{c}\fi{}}}]{Schleier-Smith:2010}%
  \BibitemOpen
  \bibfield  {author} {\bibinfo {author} {\bibfnamefont {Monika~H.}\
  \bibnamefont {Schleier-Smith}}, \bibinfo {author} {\bibfnamefont {Ian~D.}\
  \bibnamefont {Leroux}}, \ and\ \bibinfo {author} {\bibfnamefont {Vladan}\
  \bibnamefont {Vuleti\ifmmode~\acute{c}\else \'{c}\fi{}}},\ }\bibfield
  {title} {\enquote {\bibinfo {title} {Squeezing the collective spin of a
  dilute atomic ensemble by cavity feedback},}\ }\href {\doibase
  10.1103/PhysRevA.81.021804} {\bibfield  {journal} {\bibinfo  {journal} {Phys.
  Rev. A}\ }\textbf {\bibinfo {volume} {81}},\ \bibinfo {pages} {021804}
  (\bibinfo {year} {2010})}\BibitemShut {NoStop}%
\bibitem [{\citenamefont {Chen}\ \emph {et~al.}(2011)\citenamefont {Chen},
  \citenamefont {Bohnet}, \citenamefont {Sankar}, \citenamefont {Dai},\ and\
  \citenamefont {Thompson}}]{Chen:2011}%
  \BibitemOpen
  \bibfield  {author} {\bibinfo {author} {\bibfnamefont {Zilong}\ \bibnamefont
  {Chen}}, \bibinfo {author} {\bibfnamefont {Justin~G.}\ \bibnamefont
  {Bohnet}}, \bibinfo {author} {\bibfnamefont {Shannon~R.}\ \bibnamefont
  {Sankar}}, \bibinfo {author} {\bibfnamefont {Jiayan}\ \bibnamefont {Dai}}, \
  and\ \bibinfo {author} {\bibfnamefont {James~K.}\ \bibnamefont {Thompson}},\
  }\bibfield  {title} {\enquote {\bibinfo {title} {Conditional spin squeezing
  of a large ensemble via the vacuum rabi splitting},}\ }\href {\doibase
  10.1103/PhysRevLett.106.133601} {\bibfield  {journal} {\bibinfo  {journal}
  {Phys. Rev. Lett.}\ }\textbf {\bibinfo {volume} {106}},\ \bibinfo {pages}
  {133601} (\bibinfo {year} {2011})}\BibitemShut {NoStop}%
\bibitem [{\citenamefont {Szigeti}\ \emph {et~al.}(2009)\citenamefont
  {Szigeti}, \citenamefont {Hush}, \citenamefont {Carvalho},\ and\
  \citenamefont {Hope}}]{Szigeti:2009}%
  \BibitemOpen
  \bibfield  {author} {\bibinfo {author} {\bibfnamefont {S.~S.}\ \bibnamefont
  {Szigeti}}, \bibinfo {author} {\bibfnamefont {M.~R.}\ \bibnamefont {Hush}},
  \bibinfo {author} {\bibfnamefont {A.~R.~R.}\ \bibnamefont {Carvalho}}, \ and\
  \bibinfo {author} {\bibfnamefont {J.~J.}\ \bibnamefont {Hope}},\ }\bibfield
  {title} {\enquote {\bibinfo {title} {Continuous measurement feedback control
  of a {Bose-Einstein} condensate using phase-contrast imaging},}\ }\href
  {\doibase 10.1103/PhysRevA.80.013614} {\bibfield  {journal} {\bibinfo
  {journal} {Physical Review A (Atomic, Molecular, and Optical Physics)}\
  }\textbf {\bibinfo {volume} {80}},\ \bibinfo {eid} {013614} (\bibinfo {year}
  {2009})}\BibitemShut {NoStop}%
\bibitem [{\citenamefont {Szigeti}\ \emph {et~al.}(2010)\citenamefont
  {Szigeti}, \citenamefont {Hush}, \citenamefont {Carvalho},\ and\
  \citenamefont {Hope}}]{Szigeti:2010}%
  \BibitemOpen
  \bibfield  {author} {\bibinfo {author} {\bibfnamefont {S.~S.}\ \bibnamefont
  {Szigeti}}, \bibinfo {author} {\bibfnamefont {M.~R.}\ \bibnamefont {Hush}},
  \bibinfo {author} {\bibfnamefont {A.~R.~R.}\ \bibnamefont {Carvalho}}, \ and\
  \bibinfo {author} {\bibfnamefont {J.~J.}\ \bibnamefont {Hope}},\ }\bibfield
  {title} {\enquote {\bibinfo {title} {Feedback control of an interacting
  {Bose-Einstein} condensate using phase-contrast imaging},}\ }\href {\doibase
  10.1103/PhysRevA.82.043632} {\bibfield  {journal} {\bibinfo  {journal} {Phys.
  Rev. A}\ }\textbf {\bibinfo {volume} {82}},\ \bibinfo {pages} {043632}
  (\bibinfo {year} {2010})}\BibitemShut {NoStop}%
\bibitem [{\citenamefont {Vanderbruggen}\ \emph {et~al.}(2011)\citenamefont
  {Vanderbruggen}, \citenamefont {Bernon}, \citenamefont {Bertoldi},
  \citenamefont {Landragin},\ and\ \citenamefont
  {Bouyer}}]{vanderbruggen2011spin}%
  \BibitemOpen
  \bibfield  {author} {\bibinfo {author} {\bibfnamefont {Thomas}\ \bibnamefont
  {Vanderbruggen}}, \bibinfo {author} {\bibfnamefont {Simon}\ \bibnamefont
  {Bernon}}, \bibinfo {author} {\bibfnamefont {Andrea}\ \bibnamefont
  {Bertoldi}}, \bibinfo {author} {\bibfnamefont {Arnaud}\ \bibnamefont
  {Landragin}}, \ and\ \bibinfo {author} {\bibfnamefont {Philippe}\
  \bibnamefont {Bouyer}},\ }\bibfield  {title} {\enquote {\bibinfo {title}
  {Spin-squeezing and {Dicke}-state preparation by heterodyne measurement},}\
  }\href@noop {} {\bibfield  {journal} {\bibinfo  {journal} {Physical Review
  A}\ }\textbf {\bibinfo {volume} {83}},\ \bibinfo {pages} {013821} (\bibinfo
  {year} {2011})}\BibitemShut {NoStop}%
\bibitem [{\citenamefont {Bernon}\ \emph {et~al.}(2011)\citenamefont {Bernon},
  \citenamefont {Vanderbruggen}, \citenamefont {Kohlhaas}, \citenamefont
  {Bertoldi}, \citenamefont {Landragin},\ and\ \citenamefont
  {Bouyer}}]{bernon2011heterodyne}%
  \BibitemOpen
  \bibfield  {author} {\bibinfo {author} {\bibfnamefont {Simon}\ \bibnamefont
  {Bernon}}, \bibinfo {author} {\bibfnamefont {Thomas}\ \bibnamefont
  {Vanderbruggen}}, \bibinfo {author} {\bibfnamefont {Ralf}\ \bibnamefont
  {Kohlhaas}}, \bibinfo {author} {\bibfnamefont {Andrea}\ \bibnamefont
  {Bertoldi}}, \bibinfo {author} {\bibfnamefont {Arnaud}\ \bibnamefont
  {Landragin}}, \ and\ \bibinfo {author} {\bibfnamefont {Philippe}\
  \bibnamefont {Bouyer}},\ }\bibfield  {title} {\enquote {\bibinfo {title}
  {Heterodyne non-demolition measurements on cold atomic samples: towards the
  preparation of non-classical states for atom interferometry},}\ }\href@noop
  {} {\bibfield  {journal} {\bibinfo  {journal} {New Journal of Physics}\
  }\textbf {\bibinfo {volume} {13}},\ \bibinfo {pages} {065021} (\bibinfo
  {year} {2011})}\BibitemShut {NoStop}%
\bibitem [{\citenamefont {Brahms}\ \emph {et~al.}(2012)\citenamefont {Brahms},
  \citenamefont {Botter}, \citenamefont {Schreppler}, \citenamefont {Brooks},\
  and\ \citenamefont {Stamper-Kurn}}]{Brahms:2012}%
  \BibitemOpen
  \bibfield  {author} {\bibinfo {author} {\bibfnamefont {Nathan}\ \bibnamefont
  {Brahms}}, \bibinfo {author} {\bibfnamefont {Thierry}\ \bibnamefont
  {Botter}}, \bibinfo {author} {\bibfnamefont {Sydney}\ \bibnamefont
  {Schreppler}}, \bibinfo {author} {\bibfnamefont {Daniel W.~C.}\ \bibnamefont
  {Brooks}}, \ and\ \bibinfo {author} {\bibfnamefont {Dan~M.}\ \bibnamefont
  {Stamper-Kurn}},\ }\bibfield  {title} {\enquote {\bibinfo {title} {Optical
  detection of the quantization of collective atomic motion},}\ }\href
  {\doibase 10.1103/PhysRevLett.108.133601} {\bibfield  {journal} {\bibinfo
  {journal} {Phys. Rev. Lett.}\ }\textbf {\bibinfo {volume} {108}},\ \bibinfo
  {pages} {133601} (\bibinfo {year} {2012})}\BibitemShut {NoStop}%
\bibitem [{\citenamefont {Bohnet}\ \emph {et~al.}(2014)\citenamefont {Bohnet},
  \citenamefont {Cox}, \citenamefont {Norcia}, \citenamefont {Weiner},
  \citenamefont {Chen},\ and\ \citenamefont {Thompson}}]{Bohnet:2014}%
  \BibitemOpen
  \bibfield  {author} {\bibinfo {author} {\bibfnamefont {J.~G.}\ \bibnamefont
  {Bohnet}}, \bibinfo {author} {\bibfnamefont {K.~C.}\ \bibnamefont {Cox}},
  \bibinfo {author} {\bibfnamefont {M.~A.}\ \bibnamefont {Norcia}}, \bibinfo
  {author} {\bibfnamefont {J.~M.}\ \bibnamefont {Weiner}}, \bibinfo {author}
  {\bibfnamefont {Z.}~\bibnamefont {Chen}}, \ and\ \bibinfo {author}
  {\bibfnamefont {J.~K.}\ \bibnamefont {Thompson}},\ }\bibfield  {title}
  {\enquote {\bibinfo {title} {Reduced spin measurement back-action for a phase
  sensitivity ten times beyond the standard quantum limit},}\ }\href@noop {}
  {\bibfield  {journal} {\bibinfo  {journal} {Nature Photonics}\ }\textbf
  {\bibinfo {volume} {8}},\ \bibinfo {pages} {731} (\bibinfo {year}
  {2014})}\BibitemShut {NoStop}%
\bibitem [{\citenamefont {Cox}\ \emph {et~al.}(2015)\citenamefont {Cox},
  \citenamefont {Weiner}, \citenamefont {Greve},\ and\ \citenamefont
  {Thompson}}]{Cox:2015}%
  \BibitemOpen
  \bibfield  {author} {\bibinfo {author} {\bibfnamefont {Kevin~C.}\
  \bibnamefont {Cox}}, \bibinfo {author} {\bibfnamefont {Joshua~M.}\
  \bibnamefont {Weiner}}, \bibinfo {author} {\bibfnamefont {Graham~P.}\
  \bibnamefont {Greve}}, \ and\ \bibinfo {author} {\bibfnamefont {James~K.}\
  \bibnamefont {Thompson}},\ }\bibfield  {title} {\enquote {\bibinfo {title}
  {Generating entanglement between atomic spins with low-noise probing of an
  optical cavity},}\ }\href@noop {} {\bibfield  {journal} {\bibinfo  {journal}
  {arXiv:1504.05160v1}\ } (\bibinfo {year} {2015})}\BibitemShut {NoStop}%
\bibitem [{\citenamefont {Wallraff}\ \emph {et~al.}(2004)\citenamefont
  {Wallraff}, \citenamefont {Schuster}, \citenamefont {Blais}, \citenamefont
  {Frunzio}, \citenamefont {Huang}, \citenamefont {Majer}, \citenamefont
  {Kumar}, \citenamefont {Girvin},\ and\ \citenamefont
  {Schoelkopf}}]{Wallraff:2004}%
  \BibitemOpen
  \bibfield  {author} {\bibinfo {author} {\bibfnamefont {A.}~\bibnamefont
  {Wallraff}}, \bibinfo {author} {\bibfnamefont {D.~I.}\ \bibnamefont
  {Schuster}}, \bibinfo {author} {\bibfnamefont {A.}~\bibnamefont {Blais}},
  \bibinfo {author} {\bibfnamefont {L.}~\bibnamefont {Frunzio}}, \bibinfo
  {author} {\bibfnamefont {R.~S.}\ \bibnamefont {Huang}}, \bibinfo {author}
  {\bibfnamefont {J.}~\bibnamefont {Majer}}, \bibinfo {author} {\bibfnamefont
  {S.}~\bibnamefont {Kumar}}, \bibinfo {author} {\bibfnamefont {S.~M.}\
  \bibnamefont {Girvin}}, \ and\ \bibinfo {author} {\bibfnamefont {R.~J.}\
  \bibnamefont {Schoelkopf}},\ }\bibfield  {title} {\enquote {\bibinfo {title}
  {Strong coupling of a single photon to a superconducting qubit using circuit
  quantum electrodynamics},}\ }\href {http://dx.doi.org/10.1038/nature02851}
  {\bibfield  {journal} {\bibinfo  {journal} {Nature}\ }\textbf {\bibinfo
  {volume} {431}},\ \bibinfo {pages} {162--167} (\bibinfo {year}
  {2004})}\BibitemShut {NoStop}%
\bibitem [{\citenamefont {Schuster}\ \emph {et~al.}(2005)\citenamefont
  {Schuster}, \citenamefont {Wallraff}, \citenamefont {Blais}, \citenamefont
  {Frunzio}, \citenamefont {Huang}, \citenamefont {Majer}, \citenamefont
  {Girvin},\ and\ \citenamefont {Schoelkopf}}]{Schuster:2005}%
  \BibitemOpen
  \bibfield  {author} {\bibinfo {author} {\bibfnamefont {D.~I.}\ \bibnamefont
  {Schuster}}, \bibinfo {author} {\bibfnamefont {A.}~\bibnamefont {Wallraff}},
  \bibinfo {author} {\bibfnamefont {A.}~\bibnamefont {Blais}}, \bibinfo
  {author} {\bibfnamefont {L.}~\bibnamefont {Frunzio}}, \bibinfo {author}
  {\bibfnamefont {R.-S.}\ \bibnamefont {Huang}}, \bibinfo {author}
  {\bibfnamefont {J.}~\bibnamefont {Majer}}, \bibinfo {author} {\bibfnamefont
  {S.~M.}\ \bibnamefont {Girvin}}, \ and\ \bibinfo {author} {\bibfnamefont
  {R.~J.}\ \bibnamefont {Schoelkopf}},\ }\bibfield  {title} {\enquote {\bibinfo
  {title} {ac stark shift and dephasing of a superconducting qubit strongly
  coupled to a cavity field},}\ }\href {\doibase 10.1103/PhysRevLett.94.123602}
  {\bibfield  {journal} {\bibinfo  {journal} {Phys. Rev. Lett.}\ }\textbf
  {\bibinfo {volume} {94}},\ \bibinfo {pages} {123602} (\bibinfo {year}
  {2005})}\BibitemShut {NoStop}%
\bibitem [{\citenamefont {Haigh}\ \emph {et~al.}(2015)\citenamefont {Haigh},
  \citenamefont {Lambert}, \citenamefont {Doherty},\ and\ \citenamefont
  {Ferguson}}]{Haigh:2015}%
  \BibitemOpen
  \bibfield  {author} {\bibinfo {author} {\bibfnamefont {J.~A.}\ \bibnamefont
  {Haigh}}, \bibinfo {author} {\bibfnamefont {N.~J.}\ \bibnamefont {Lambert}},
  \bibinfo {author} {\bibfnamefont {A.~C.}\ \bibnamefont {Doherty}}, \ and\
  \bibinfo {author} {\bibfnamefont {A.~J.}\ \bibnamefont {Ferguson}},\
  }\bibfield  {title} {\enquote {\bibinfo {title} {Dispersive readout of
  ferromagnetic resonance for strongly coupled magnons and microwave
  photons},}\ }\href {\doibase 10.1103/PhysRevB.91.104410} {\bibfield
  {journal} {\bibinfo  {journal} {Phys. Rev. B}\ }\textbf {\bibinfo {volume}
  {91}},\ \bibinfo {pages} {104410} (\bibinfo {year} {2015})}\BibitemShut
  {NoStop}%
\bibitem [{\citenamefont {Walls}\ and\ \citenamefont
  {Milburn}(2008)}]{Walls:2008}%
  \BibitemOpen
  \bibfield  {author} {\bibinfo {author} {\bibfnamefont {D.~F.}\ \bibnamefont
  {Walls}}\ and\ \bibinfo {author} {\bibfnamefont {G.~J.}\ \bibnamefont
  {Milburn}},\ }\href@noop {} {\emph {\bibinfo {title} {Quantum Optics}}},\
  \bibinfo {edition} {2nd}\ ed.\ (\bibinfo  {publisher} {Springer-Verlag},\
  \bibinfo {address} {Berlin and Heidelberg},\ \bibinfo {year}
  {2008})\BibitemShut {NoStop}%
\bibitem [{\citenamefont {Moore}\ and\ \citenamefont
  {Meystre}(2000)}]{Moore:2000}%
  \BibitemOpen
  \bibfield  {author} {\bibinfo {author} {\bibfnamefont {M.~G.}\ \bibnamefont
  {Moore}}\ and\ \bibinfo {author} {\bibfnamefont {P.}~\bibnamefont
  {Meystre}},\ }\bibfield  {title} {\enquote {\bibinfo {title} {Generating
  entangled atom-photon pairs from {Bose-Einstein} condensates},}\ }\href
  {\doibase 10.1103/PhysRevLett.85.5026} {\bibfield  {journal} {\bibinfo
  {journal} {Phys. Rev. Lett.}\ }\textbf {\bibinfo {volume} {85}},\ \bibinfo
  {pages} {5026--5029} (\bibinfo {year} {2000})}\BibitemShut {NoStop}%
\bibitem [{\citenamefont {Bouyer}\ and\ \citenamefont
  {Kasevich}(1997)}]{Bouyer:1997}%
  \BibitemOpen
  \bibfield  {author} {\bibinfo {author} {\bibfnamefont {P.}~\bibnamefont
  {Bouyer}}\ and\ \bibinfo {author} {\bibfnamefont {M.~A.}\ \bibnamefont
  {Kasevich}},\ }\bibfield  {title} {\enquote {\bibinfo {title}
  {Heisenberg-limited spectroscopy with degenerate bose-einstein gases},}\
  }\href {\doibase 10.1103/PhysRevA.56.R1083} {\bibfield  {journal} {\bibinfo
  {journal} {Phys. Rev. A}\ }\textbf {\bibinfo {volume} {56}},\ \bibinfo
  {pages} {R1083--R1086} (\bibinfo {year} {1997})}\BibitemShut {NoStop}%
\bibitem [{\citenamefont {Kim}\ \emph {et~al.}(1998)\citenamefont {Kim},
  \citenamefont {Pfister}, \citenamefont {Holland}, \citenamefont {Noh},\ and\
  \citenamefont {Hall}}]{Kim:1998}%
  \BibitemOpen
  \bibfield  {author} {\bibinfo {author} {\bibfnamefont {Taesoo}\ \bibnamefont
  {Kim}}, \bibinfo {author} {\bibfnamefont {Olivier}\ \bibnamefont {Pfister}},
  \bibinfo {author} {\bibfnamefont {Murray~J.}\ \bibnamefont {Holland}},
  \bibinfo {author} {\bibfnamefont {Jaewoo}\ \bibnamefont {Noh}}, \ and\
  \bibinfo {author} {\bibfnamefont {John~L.}\ \bibnamefont {Hall}},\ }\bibfield
   {title} {\enquote {\bibinfo {title} {Influence of decorrelation on
  {Heisenberg}-limited interferometry with quantum correlated photons},}\
  }\href {\doibase 10.1103/PhysRevA.57.4004} {\bibfield  {journal} {\bibinfo
  {journal} {Phys. Rev. A}\ }\textbf {\bibinfo {volume} {57}},\ \bibinfo
  {pages} {4004--4013} (\bibinfo {year} {1998})}\BibitemShut {NoStop}%
\bibitem [{\citenamefont {Bollinger}\ \emph {et~al.}(1996)\citenamefont
  {Bollinger}, \citenamefont {Itano}, \citenamefont {Wineland},\ and\
  \citenamefont {Heinzen}}]{Bollinger:1996}%
  \BibitemOpen
  \bibfield  {author} {\bibinfo {author} {\bibfnamefont {J.~J~.}\ \bibnamefont
  {Bollinger}}, \bibinfo {author} {\bibfnamefont {Wayne~M.}\ \bibnamefont
  {Itano}}, \bibinfo {author} {\bibfnamefont {D.~J.}\ \bibnamefont {Wineland}},
  \ and\ \bibinfo {author} {\bibfnamefont {D.~J.}\ \bibnamefont {Heinzen}},\
  }\bibfield  {title} {\enquote {\bibinfo {title} {Optimal frequency
  measurements with maximally correlated states},}\ }\href {\doibase
  10.1103/PhysRevA.54.R4649} {\bibfield  {journal} {\bibinfo  {journal} {Phys.
  Rev. A}\ }\textbf {\bibinfo {volume} {54}},\ \bibinfo {pages} {R4649--R4652}
  (\bibinfo {year} {1996})}\BibitemShut {NoStop}%
\bibitem [{\citenamefont {Bachor}\ and\ \citenamefont
  {Ralph}(2004)}]{Bachor:2004}%
  \BibitemOpen
  \bibfield  {author} {\bibinfo {author} {\bibfnamefont {Hans~A.}\ \bibnamefont
  {Bachor}}\ and\ \bibinfo {author} {\bibfnamefont {Timothy~C.}\ \bibnamefont
  {Ralph}},\ }\href@noop {} {\emph {\bibinfo {title} {A Guide to Experiments in
  Quantum Optics}}},\ \bibinfo {edition} {2nd}\ ed.\ (\bibinfo  {publisher}
  {Wiley},\ \bibinfo {year} {2004})\BibitemShut {NoStop}%
\bibitem [{\citenamefont {Bloch}\ \emph {et~al.}(2008)\citenamefont {Bloch},
  \citenamefont {Dalibard},\ and\ \citenamefont {Zwerger}}]{Bloch:2008}%
  \BibitemOpen
  \bibfield  {author} {\bibinfo {author} {\bibfnamefont {Immanuel}\
  \bibnamefont {Bloch}}, \bibinfo {author} {\bibfnamefont {Jean}\ \bibnamefont
  {Dalibard}}, \ and\ \bibinfo {author} {\bibfnamefont {Wilhelm}\ \bibnamefont
  {Zwerger}},\ }\bibfield  {title} {\enquote {\bibinfo {title} {Many-body
  physics with ultracold gases},}\ }\href {\doibase 10.1103/RevModPhys.80.885}
  {\bibfield  {journal} {\bibinfo  {journal} {Rev. Mod. Phys.}\ }\textbf
  {\bibinfo {volume} {80}},\ \bibinfo {pages} {885--964} (\bibinfo {year}
  {2008})}\BibitemShut {NoStop}%
\bibitem [{\citenamefont {Cronin}\ \emph {et~al.}(2009)\citenamefont {Cronin},
  \citenamefont {Schmiedmayer},\ and\ \citenamefont {Pritchard}}]{Cronin:2009}%
  \BibitemOpen
  \bibfield  {author} {\bibinfo {author} {\bibfnamefont {Alexander~D.}\
  \bibnamefont {Cronin}}, \bibinfo {author} {\bibfnamefont {J\"org}\
  \bibnamefont {Schmiedmayer}}, \ and\ \bibinfo {author} {\bibfnamefont
  {David~E.}\ \bibnamefont {Pritchard}},\ }\bibfield  {title} {\enquote
  {\bibinfo {title} {Optics and interferometry with atoms and molecules},}\
  }\href {\doibase 10.1103/RevModPhys.81.1051} {\bibfield  {journal} {\bibinfo
  {journal} {Rev. Mod. Phys.}\ }\textbf {\bibinfo {volume} {81}},\ \bibinfo
  {pages} {1051--1129} (\bibinfo {year} {2009})}\BibitemShut {NoStop}%
\bibitem [{\citenamefont {Li}\ \emph {et~al.}(2013)\citenamefont {Li},
  \citenamefont {Dudin},\ and\ \citenamefont {Kuzmich}}]{Li:2013}%
  \BibitemOpen
  \bibfield  {author} {\bibinfo {author} {\bibfnamefont {L.}~\bibnamefont
  {Li}}, \bibinfo {author} {\bibfnamefont {Y.~O.}\ \bibnamefont {Dudin}}, \
  and\ \bibinfo {author} {\bibfnamefont {A.}~\bibnamefont {Kuzmich}},\
  }\bibfield  {title} {\enquote {\bibinfo {title} {Entanglement between light
  and an optical atomic excitation},}\ }\href
  {http://dx.doi.org/10.1038/nature12227} {\bibfield  {journal} {\bibinfo
  {journal} {Nature}\ }\textbf {\bibinfo {volume} {498}},\ \bibinfo {pages}
  {466--469} (\bibinfo {year} {2013})}\BibitemShut {NoStop}%
\bibitem [{\citenamefont {Paik}\ \emph {et~al.}(2011)\citenamefont {Paik},
  \citenamefont {Schuster}, \citenamefont {Bishop}, \citenamefont {Kirchmair},
  \citenamefont {Catelani}, \citenamefont {Sears}, \citenamefont {Johnson},
  \citenamefont {Reagor}, \citenamefont {Frunzio}, \citenamefont {Glazman},
  \citenamefont {Girvin}, \citenamefont {Devoret},\ and\ \citenamefont
  {Schoelkopf}}]{Paik:2011}%
  \BibitemOpen
  \bibfield  {author} {\bibinfo {author} {\bibfnamefont {Hanhee}\ \bibnamefont
  {Paik}}, \bibinfo {author} {\bibfnamefont {D.~I.}\ \bibnamefont {Schuster}},
  \bibinfo {author} {\bibfnamefont {Lev~S.}\ \bibnamefont {Bishop}}, \bibinfo
  {author} {\bibfnamefont {G.}~\bibnamefont {Kirchmair}}, \bibinfo {author}
  {\bibfnamefont {G.}~\bibnamefont {Catelani}}, \bibinfo {author}
  {\bibfnamefont {A.~P.}\ \bibnamefont {Sears}}, \bibinfo {author}
  {\bibfnamefont {B.~R.}\ \bibnamefont {Johnson}}, \bibinfo {author}
  {\bibfnamefont {M.~J.}\ \bibnamefont {Reagor}}, \bibinfo {author}
  {\bibfnamefont {L.}~\bibnamefont {Frunzio}}, \bibinfo {author} {\bibfnamefont
  {L.~I.}\ \bibnamefont {Glazman}}, \bibinfo {author} {\bibfnamefont {S.~M.}\
  \bibnamefont {Girvin}}, \bibinfo {author} {\bibfnamefont {M.~H.}\
  \bibnamefont {Devoret}}, \ and\ \bibinfo {author} {\bibfnamefont {R.~J.}\
  \bibnamefont {Schoelkopf}},\ }\bibfield  {title} {\enquote {\bibinfo {title}
  {Observation of high coherence in {Josephson} junction qubits measured in a
  three-dimensional circuit {QED} architecture},}\ }\href {\doibase
  10.1103/PhysRevLett.107.240501} {\bibfield  {journal} {\bibinfo  {journal}
  {Phys. Rev. Lett.}\ }\textbf {\bibinfo {volume} {107}},\ \bibinfo {pages}
  {240501} (\bibinfo {year} {2011})}\BibitemShut {NoStop}%
\bibitem [{\citenamefont {Xiang}\ \emph {et~al.}(2013)\citenamefont {Xiang},
  \citenamefont {Ashhab}, \citenamefont {You},\ and\ \citenamefont
  {Nori}}]{Xiang:2013}%
  \BibitemOpen
  \bibfield  {author} {\bibinfo {author} {\bibfnamefont {Ze-Liang}\
  \bibnamefont {Xiang}}, \bibinfo {author} {\bibfnamefont {Sahel}\ \bibnamefont
  {Ashhab}}, \bibinfo {author} {\bibfnamefont {J.~Q.}\ \bibnamefont {You}}, \
  and\ \bibinfo {author} {\bibfnamefont {Franco}\ \bibnamefont {Nori}},\
  }\bibfield  {title} {\enquote {\bibinfo {title} {Hybrid quantum circuits:
  Superconducting circuits interacting with other quantum systems},}\ }\href
  {\doibase 10.1103/RevModPhys.85.623} {\bibfield  {journal} {\bibinfo
  {journal} {Rev. Mod. Phys.}\ }\textbf {\bibinfo {volume} {85}},\ \bibinfo
  {pages} {623--653} (\bibinfo {year} {2013})}\BibitemShut {NoStop}%
\bibitem [{\citenamefont {Aspelmeyer}\ \emph {et~al.}(2014)\citenamefont
  {Aspelmeyer}, \citenamefont {Kippenberg},\ and\ \citenamefont
  {Marquardt}}]{Aspelmeyer:2014}%
  \BibitemOpen
  \bibfield  {author} {\bibinfo {author} {\bibfnamefont {Markus}\ \bibnamefont
  {Aspelmeyer}}, \bibinfo {author} {\bibfnamefont {Tobias~J.}\ \bibnamefont
  {Kippenberg}}, \ and\ \bibinfo {author} {\bibfnamefont {Florian}\
  \bibnamefont {Marquardt}},\ }\bibfield  {title} {\enquote {\bibinfo {title}
  {Cavity optomechanics},}\ }\href {\doibase 10.1103/RevModPhys.86.1391}
  {\bibfield  {journal} {\bibinfo  {journal} {Rev. Mod. Phys.}\ }\textbf
  {\bibinfo {volume} {86}},\ \bibinfo {pages} {1391--1452} (\bibinfo {year}
  {2014})}\BibitemShut {NoStop}%
\bibitem [{\citenamefont {Leibfried}\ \emph {et~al.}(2003)\citenamefont
  {Leibfried}, \citenamefont {Blatt}, \citenamefont {Monroe},\ and\
  \citenamefont {Wineland}}]{Leibfried:2003}%
  \BibitemOpen
  \bibfield  {author} {\bibinfo {author} {\bibfnamefont {D.}~\bibnamefont
  {Leibfried}}, \bibinfo {author} {\bibfnamefont {R.}~\bibnamefont {Blatt}},
  \bibinfo {author} {\bibfnamefont {C.}~\bibnamefont {Monroe}}, \ and\ \bibinfo
  {author} {\bibfnamefont {D.}~\bibnamefont {Wineland}},\ }\bibfield  {title}
  {\enquote {\bibinfo {title} {Quantum dynamics of single trapped ions},}\
  }\href {\doibase 10.1103/RevModPhys.75.281} {\bibfield  {journal} {\bibinfo
  {journal} {Rev. Mod. Phys.}\ }\textbf {\bibinfo {volume} {75}},\ \bibinfo
  {pages} {281--324} (\bibinfo {year} {2003})}\BibitemShut {NoStop}%
\bibitem [{\citenamefont {Wineland}(2013)}]{Wineland:2013}%
  \BibitemOpen
  \bibfield  {author} {\bibinfo {author} {\bibfnamefont {David~J.}\
  \bibnamefont {Wineland}},\ }\bibfield  {title} {\enquote {\bibinfo {title}
  {Nobel lecture: Superposition, entanglement, and raising {S}chr\"odinger's
  cat*},}\ }\href {\doibase 10.1103/RevModPhys.85.1103} {\bibfield  {journal}
  {\bibinfo  {journal} {Rev. Mod. Phys.}\ }\textbf {\bibinfo {volume} {85}},\
  \bibinfo {pages} {1103--1114} (\bibinfo {year} {2013})}\BibitemShut {NoStop}%
\bibitem [{\citenamefont {Stute}\ \emph {et~al.}(2013)\citenamefont {Stute},
  \citenamefont {Casabone}, \citenamefont {Brandstatter}, \citenamefont
  {Friebe}, \citenamefont {Northup},\ and\ \citenamefont {Blatt}}]{Stute:2013}%
  \BibitemOpen
  \bibfield  {author} {\bibinfo {author} {\bibfnamefont {A.}~\bibnamefont
  {Stute}}, \bibinfo {author} {\bibfnamefont {B.}~\bibnamefont {Casabone}},
  \bibinfo {author} {\bibfnamefont {B.}~\bibnamefont {Brandstatter}}, \bibinfo
  {author} {\bibfnamefont {K.}~\bibnamefont {Friebe}}, \bibinfo {author}
  {\bibfnamefont {T.~E.}\ \bibnamefont {Northup}}, \ and\ \bibinfo {author}
  {\bibfnamefont {R.}~\bibnamefont {Blatt}},\ }\bibfield  {title} {\enquote
  {\bibinfo {title} {Quantum-state transfer from an ion to a photon},}\ }\href
  {http://dx.doi.org/10.1038/nphoton.2012.358} {\bibfield  {journal} {\bibinfo
  {journal} {Nat Photon}\ }\textbf {\bibinfo {volume} {7}},\ \bibinfo {pages}
  {219--222} (\bibinfo {year} {2013})}\BibitemShut {NoStop}%
\bibitem [{\citenamefont {Rowe}\ \emph {et~al.}(2001)\citenamefont {Rowe},
  \citenamefont {Kielpinski}, \citenamefont {Meyer}, \citenamefont {Sackett},
  \citenamefont {Itano}, \citenamefont {Monroe},\ and\ \citenamefont
  {Wineland}}]{Rowe:2001}%
  \BibitemOpen
  \bibfield  {author} {\bibinfo {author} {\bibfnamefont {M.~A.}\ \bibnamefont
  {Rowe}}, \bibinfo {author} {\bibfnamefont {D.}~\bibnamefont {Kielpinski}},
  \bibinfo {author} {\bibfnamefont {V.}~\bibnamefont {Meyer}}, \bibinfo
  {author} {\bibfnamefont {C.~A.}\ \bibnamefont {Sackett}}, \bibinfo {author}
  {\bibfnamefont {W.~M.}\ \bibnamefont {Itano}}, \bibinfo {author}
  {\bibfnamefont {C.}~\bibnamefont {Monroe}}, \ and\ \bibinfo {author}
  {\bibfnamefont {D.~J.}\ \bibnamefont {Wineland}},\ }\bibfield  {title}
  {\enquote {\bibinfo {title} {Experimental violation of a {Bell's} inequality
  with efficient detection},}\ }\href {http://dx.doi.org/10.1038/35057215}
  {\bibfield  {journal} {\bibinfo  {journal} {Nature}\ }\textbf {\bibinfo
  {volume} {409}},\ \bibinfo {pages} {791--794} (\bibinfo {year}
  {2001})}\BibitemShut {NoStop}%
\bibitem [{\citenamefont {Smith}\ \emph {et~al.}(2012)\citenamefont {Smith},
  \citenamefont {Gillett}, \citenamefont {de~Almeida}, \citenamefont
  {Branciard}, \citenamefont {Fedrizzi}, \citenamefont {Weinhold},
  \citenamefont {Lita}, \citenamefont {Calkins}, \citenamefont {Gerrits},
  \citenamefont {Wiseman}, \citenamefont {Nam},\ and\ \citenamefont
  {White}}]{Smith:2012}%
  \BibitemOpen
  \bibfield  {author} {\bibinfo {author} {\bibfnamefont {Devin~H.}\
  \bibnamefont {Smith}}, \bibinfo {author} {\bibfnamefont {Geoff}\ \bibnamefont
  {Gillett}}, \bibinfo {author} {\bibfnamefont {Marcelo~P.}\ \bibnamefont
  {de~Almeida}}, \bibinfo {author} {\bibfnamefont {Cyril}\ \bibnamefont
  {Branciard}}, \bibinfo {author} {\bibfnamefont {Alessandro}\ \bibnamefont
  {Fedrizzi}}, \bibinfo {author} {\bibfnamefont {Till~J.}\ \bibnamefont
  {Weinhold}}, \bibinfo {author} {\bibfnamefont {Adriana}\ \bibnamefont
  {Lita}}, \bibinfo {author} {\bibfnamefont {Brice}\ \bibnamefont {Calkins}},
  \bibinfo {author} {\bibfnamefont {Thomas}\ \bibnamefont {Gerrits}}, \bibinfo
  {author} {\bibfnamefont {Howard~M.}\ \bibnamefont {Wiseman}}, \bibinfo
  {author} {\bibfnamefont {Sae~Woo}\ \bibnamefont {Nam}}, \ and\ \bibinfo
  {author} {\bibfnamefont {Andrew~G.}\ \bibnamefont {White}},\ }\bibfield
  {title} {\enquote {\bibinfo {title} {Conclusive quantum steering with
  superconducting transition-edge sensors},}\ }\href
  {http://dx.doi.org/10.1038/ncomms1628} {\bibfield  {journal} {\bibinfo
  {journal} {Nat Commun}\ }\textbf {\bibinfo {volume} {3}},\ \bibinfo {pages}
  {625} (\bibinfo {year} {2012})}\BibitemShut {NoStop}%
\bibitem [{\citenamefont {Guerlin}\ \emph {et~al.}(2007)\citenamefont
  {Guerlin}, \citenamefont {Bernu}, \citenamefont {Deleglise}, \citenamefont
  {Sayrin}, \citenamefont {Gleyzes}, \citenamefont {Kuhr}, \citenamefont
  {Brune}, \citenamefont {Raimond},\ and\ \citenamefont
  {Haroche}}]{Guerlin:2007}%
  \BibitemOpen
  \bibfield  {author} {\bibinfo {author} {\bibfnamefont {Christine}\
  \bibnamefont {Guerlin}}, \bibinfo {author} {\bibfnamefont {Julien}\
  \bibnamefont {Bernu}}, \bibinfo {author} {\bibfnamefont {Samuel}\
  \bibnamefont {Deleglise}}, \bibinfo {author} {\bibfnamefont {Clement}\
  \bibnamefont {Sayrin}}, \bibinfo {author} {\bibfnamefont {Sebastien}\
  \bibnamefont {Gleyzes}}, \bibinfo {author} {\bibfnamefont {Stefan}\
  \bibnamefont {Kuhr}}, \bibinfo {author} {\bibfnamefont {Michel}\ \bibnamefont
  {Brune}}, \bibinfo {author} {\bibfnamefont {Jean-Michel}\ \bibnamefont
  {Raimond}}, \ and\ \bibinfo {author} {\bibfnamefont {Serge}\ \bibnamefont
  {Haroche}},\ }\bibfield  {title} {\enquote {\bibinfo {title} {Progressive
  field-state collapse and quantum non-demolition photon counting},}\ }\href
  {http://dx.doi.org/10.1038/nature06057} {\bibfield  {journal} {\bibinfo
  {journal} {Nature}\ }\textbf {\bibinfo {volume} {448}},\ \bibinfo {pages}
  {889--893} (\bibinfo {year} {2007})}\BibitemShut {NoStop}%
\bibitem [{\citenamefont {Groen}\ \emph {et~al.}(2013)\citenamefont {Groen},
  \citenamefont {Rist\`e}, \citenamefont {Tornberg}, \citenamefont {Cramer},
  \citenamefont {de~Groot}, \citenamefont {Picot}, \citenamefont {Johansson},\
  and\ \citenamefont {DiCarlo}}]{Groen:2013}%
  \BibitemOpen
  \bibfield  {author} {\bibinfo {author} {\bibfnamefont {J.~P.}\ \bibnamefont
  {Groen}}, \bibinfo {author} {\bibfnamefont {D.}~\bibnamefont {Rist\`e}},
  \bibinfo {author} {\bibfnamefont {L.}~\bibnamefont {Tornberg}}, \bibinfo
  {author} {\bibfnamefont {J.}~\bibnamefont {Cramer}}, \bibinfo {author}
  {\bibfnamefont {P.~C.}\ \bibnamefont {de~Groot}}, \bibinfo {author}
  {\bibfnamefont {T.}~\bibnamefont {Picot}}, \bibinfo {author} {\bibfnamefont
  {G.}~\bibnamefont {Johansson}}, \ and\ \bibinfo {author} {\bibfnamefont
  {L.}~\bibnamefont {DiCarlo}},\ }\bibfield  {title} {\enquote {\bibinfo
  {title} {Partial-measurement backaction and nonclassical weak values in a
  superconducting circuit},}\ }\href {\doibase 10.1103/PhysRevLett.111.090506}
  {\bibfield  {journal} {\bibinfo  {journal} {Phys. Rev. Lett.}\ }\textbf
  {\bibinfo {volume} {111}},\ \bibinfo {pages} {090506} (\bibinfo {year}
  {2013})}\BibitemShut {NoStop}%
\bibitem [{\citenamefont {Murch}\ \emph {et~al.}(2013)\citenamefont {Murch},
  \citenamefont {Weber}, \citenamefont {Macklin},\ and\ \citenamefont
  {Siddiqi}}]{Murch:2013}%
  \BibitemOpen
  \bibfield  {author} {\bibinfo {author} {\bibfnamefont {K.~W.}\ \bibnamefont
  {Murch}}, \bibinfo {author} {\bibfnamefont {S.~J.}\ \bibnamefont {Weber}},
  \bibinfo {author} {\bibfnamefont {C.}~\bibnamefont {Macklin}}, \ and\
  \bibinfo {author} {\bibfnamefont {I.}~\bibnamefont {Siddiqi}},\ }\bibfield
  {title} {\enquote {\bibinfo {title} {Observing single quantum trajectories of
  a superconducting quantum bit},}\ }\href
  {http://dx.doi.org/10.1038/nature12539} {\bibfield  {journal} {\bibinfo
  {journal} {Nature}\ }\textbf {\bibinfo {volume} {502}},\ \bibinfo {pages}
  {211--214} (\bibinfo {year} {2013})}\BibitemShut {NoStop}%
\bibitem [{\citenamefont {Cohen}\ \emph {et~al.}(2013)\citenamefont {Cohen},
  \citenamefont {Meenehan},\ and\ \citenamefont {Painter}}]{Cohen:2013}%
  \BibitemOpen
  \bibfield  {author} {\bibinfo {author} {\bibfnamefont {Justin~D.}\
  \bibnamefont {Cohen}}, \bibinfo {author} {\bibfnamefont {Se\'{a}n~M.}\
  \bibnamefont {Meenehan}}, \ and\ \bibinfo {author} {\bibfnamefont {Oskar}\
  \bibnamefont {Painter}},\ }\bibfield  {title} {\enquote {\bibinfo {title}
  {Optical coupling to nanoscale optomechanical cavities for near
  quantum-limited motion transduction},}\ }\href {\doibase
  10.1364/OE.21.011227} {\bibfield  {journal} {\bibinfo  {journal} {Opt.
  Express}\ }\textbf {\bibinfo {volume} {21}},\ \bibinfo {pages} {11227--11236}
  (\bibinfo {year} {2013})}\BibitemShut {NoStop}%
\bibitem [{\citenamefont {Kessler}\ \emph {et~al.}(2014)\citenamefont
  {Kessler}, \citenamefont {Lovchinsky}, \citenamefont {Sushkov},\ and\
  \citenamefont {Lukin}}]{Kessler:2014}%
  \BibitemOpen
  \bibfield  {author} {\bibinfo {author} {\bibfnamefont {E.~M.}\ \bibnamefont
  {Kessler}}, \bibinfo {author} {\bibfnamefont {I.}~\bibnamefont {Lovchinsky}},
  \bibinfo {author} {\bibfnamefont {A.~O.}\ \bibnamefont {Sushkov}}, \ and\
  \bibinfo {author} {\bibfnamefont {M.~D.}\ \bibnamefont {Lukin}},\ }\bibfield
  {title} {\enquote {\bibinfo {title} {Quantum error correction for
  metrology},}\ }\href {\doibase 10.1103/PhysRevLett.112.150802} {\bibfield
  {journal} {\bibinfo  {journal} {Phys. Rev. Lett.}\ }\textbf {\bibinfo
  {volume} {112}},\ \bibinfo {pages} {150802} (\bibinfo {year}
  {2014})}\BibitemShut {NoStop}%
\bibitem [{\citenamefont {Szigeti}\ \emph
  {et~al.}(2014{\natexlab{b}})\citenamefont {Szigeti}, \citenamefont
  {Carvalho}, \citenamefont {Morley},\ and\ \citenamefont
  {Hush}}]{Szigeti:2014}%
  \BibitemOpen
  \bibfield  {author} {\bibinfo {author} {\bibfnamefont {Stuart~S.}\
  \bibnamefont {Szigeti}}, \bibinfo {author} {\bibfnamefont {Andre R.~R.}\
  \bibnamefont {Carvalho}}, \bibinfo {author} {\bibfnamefont {James~G.}\
  \bibnamefont {Morley}}, \ and\ \bibinfo {author} {\bibfnamefont {Michael~R.}\
  \bibnamefont {Hush}},\ }\bibfield  {title} {\enquote {\bibinfo {title}
  {Ignorance is bliss: General and robust cancellation of decoherence via
  no-knowledge quantum feedback},}\ }\href {\doibase
  10.1103/PhysRevLett.113.020407} {\bibfield  {journal} {\bibinfo  {journal}
  {Phys. Rev. Lett.}\ }\textbf {\bibinfo {volume} {113}},\ \bibinfo {pages}
  {020407} (\bibinfo {year} {2014}{\natexlab{b}})}\BibitemShut {NoStop}%
\end{thebibliography}%

\end{document}